\documentclass[prb,aps,twocolumn,reprint,noeprint,groupedaddress]{revtex4-2}

\usepackage{graphicx}
\usepackage[nointegrals]{wasysym} %
\usepackage[export]{adjustbox}
\usepackage{amsmath,amsfonts,amssymb,latexsym}
\usepackage{hhline}
\usepackage{bm}
\usepackage{verbatim}
\usepackage{enumitem}
\usepackage{mathrsfs}
\usepackage{slashed}
\usepackage{empheq}
\usepackage{physics}

\usepackage{graphicx}
\usepackage{dcolumn}
\usepackage{bm}
\usepackage{multirow}

\usepackage[normalem]{ulem}

\usepackage{amsmath}
\usepackage{amsthm}
\usepackage{amstext}
\usepackage{amssymb}
\usepackage{mathrsfs}
\usepackage{amsfonts}
\usepackage{amsbsy} 

\usepackage{braket}

\usepackage[all,matrix,cmtip]{xy}
\usepackage{mathtools}

\usepackage{xcolor}
\definecolor{red}{rgb}{1,0,0}
\definecolor{blue}{rgb}{0,0,1}
\definecolor{dblue}{rgb}{0,0,0.4}
\definecolor{green}{rgb}{0,1,0}
\definecolor{black}{rgb}{0,0,0}
\definecolor{white}{rgb}{1,1,1}
\definecolor{pastelblue}{RGB}{20,93,160}

\definecolor{brn}{rgb}{.8,.4,.0}
\definecolor{redo}{rgb}{1,.5,.0}
\definecolor{ddgrn}{rgb}{0,0.4,0}
\definecolor{dgrn}{rgb}{0,0.55,0}
\definecolor{dbl}{rgb}{0,0,0.5}

\usepackage[colorlinks,citecolor=pastelblue,linkcolor=pastelblue,urlcolor=pastelblue]{hyperref}

\renewcommand{\Im}{{\rm Im}} 
\renewcommand{\Re}{{\rm Re}}

\newcommand{\bpm}{\begin{pmatrix}}
	\newcommand{\epm}{\end{pmatrix}}
\newcommand{\bmm}{\begin{matrix}}
	\newcommand{\emm}{\end{matrix}}
\newcommand{\bvm}{\begin{vmatrix}}
	\newcommand{\evm}{\end{vmatrix}}

\usepackage{euscript}

\makeatletter
\newsavebox{\@brx}
\newcommand{\llangle}[1][]{\savebox{\@brx}{\(\m@th{#1\langle}\)}%
	\mathopen{\copy\@brx\kern-0.5\wd\@brx\usebox{\@brx}}}
\newcommand{\rrangle}[1][]{\savebox{\@brx}{\(\m@th{#1\rangle}\)}%
	\mathclose{\copy\@brx\kern-0.5\wd\@brx\usebox{\@brx}}}
\makeatother

\newcommand{\bs}{\boldsymbol}

\allowdisplaybreaks

\usepackage{tikz}
\definecolor{myRed}{RGB}{188,0,4} 
\definecolor{myGray}{RGB}{146,146,146} 
\definecolor{myBlue}{RGB}{0,0,133} 

\begin{document}

\title{Composite Bogoliubov Fermi liquid in a half-filled Chern band}%

\author{Zhengyan Darius Shi}%
\email{zhengyanshi@stanford.edu}
\affiliation{
Department of Physics, Stanford University, Stanford, California 94305, USA}

\author{Pavel A. Nosov}
\email{pnosov@fas.harvard.edu}
\affiliation{Department of Physics, Harvard University, Cambridge, MA 02138, USA
}%

\date{\today}

\begin{abstract}
The composite Fermi liquid (CFL) in the half-filled Landau level is a cornerstone of the quantum Hall phase diagram. Recent experiments and numerics indicate that an anomalous composite Fermi liquid (ACFL) can also arise at half filling of a Chern band without any external magnetic field, opening new possibilities for paired states of composite fermions beyond the fully gapped Pfaffian phase. We argue that in inversion-asymmetric Chern bands with lattice rotational symmetry reduced to $C_3$, as realized in experimental platforms where signatures of the ACFL have been observed, composite fermions can form a superconductor with neutral gapless Bogoliubov Fermi surfaces. We term the resulting electronic state {\it the  composite Bogoliubov Fermi liquid (CBFL)}. This phase has a number of remarkable properties that make it distinct from both the ACFL and the fully gapped Pfaffian. For instance, it is incompressible, has quantized Hall conductance, shows no quantum oscillations as a function of magnetic field or doping, and has topological ground state degeneracy on a torus despite the presence of gapless quasiparticles. At the same time, the neutral Bogoliubov Fermi surface yields  metallic $T$-linear specific heat, non-quantized thermal conductance, Landau damping of density fluctuations, and a non-analytic $|\bs{q}|^3$ contribution to the equal-time structure factor $S(\bs{q})$. We also briefly discuss vortex physics and possible fractionalized daughter states induced by doping or external magnetic fields. Our results pave the way for a broader understanding of gapless topological phases arising from paired composite fermions in Chern bands that go beyond the conventional Landau level paradigm.

\end{abstract}

\maketitle

\section{Introduction}

The composite Fermi liquid (CFL) is a strongly correlated metal realized in the half-filled Landau level~\cite{Stormer1999_FQHreview}. Unlike the electronic Fermi liquid, the elementary degrees of freedom of the CFL are neutral ``composite fermions" (CFs), which may be viewed as electrons bound to two units of magnetic flux~\cite{Jain1989_CFframework,Lopez1991_CSGL,Kalmeyer1992_CFL,Halperin1993_HLRtheory,Son2015_DiracCFL}. While electrons see a large magnetic field and move in cyclotron orbits, CFs see zero magnetic field on average and acquire a nontrivial dispersion, allowing them to fill a Fermi sea at finite density. This conceptual picture successfully accounts for a wide range of experimental phenomena at half filling~\cite{Eisenstein1992_CFLevidence,Kang1993_CFLevidence,Willett1993_CFLevidence_FS,Willett1993_CFLevidence_SAW,Goldman1994_CFLevidence,Smet1996_CFLevidence} and provides a unified framework for understanding neighboring fractional quantum Hall (FQH) states at the Jain filling fractions  $\nu = p/(2p+1)$~\cite{Laughlin1983_FQHtheory,Halperin1984_anyonFQH,Jain1989_CFframework}. 

Under attractive interactions, CFs can pair up to form a superconductor. In a Landau level, the CF dispersion enjoys an inversion symmetry, and weak BCS pairing generically produces a fully gapped Bogoliubov spectrum. The simplest pairing channel compatible with Fermi statistics is $p+ip$ pairing, which gives rise to the celebrated Pfaffian phase~\cite{willett1987_GaAsPf,Li2017_graphenePf}--an insulator with quantized Hall conductance and non-Abelian topological order~\cite{Moore1991_nonAbelian,Read1999_pair,Ivanov2000_pwave_nonabelian}. Properties of a fully gapped $p+ip$ CF superconductor have been investigated in great detail \cite{Foster2003_ConductivityMR,Parameswaran2011_Type1SC,Wang2014_HLRpairing}.

Remarkably, recent experimental developments have revealed that the notion of CFs extends well beyond the Landau level setting. In particular, transport measurements in twisted MoTe$_2$ and rhombohedral graphene~\cite{Park2023_FQAH_TMD,Lu2023_FQAHPenta} have provided preliminary evidence for an anomalous composite Fermi liquid (ACFL) at moire lattice filling $\nu = 1/2$, consistent with the numerical predictions in Refs.~\cite{Dong2023_ACFL,Goldman2023_ACFL}. This interpretation is further supported by the observation of Jain states at $\nu = p/(2p+1)$, which naturally arise as daughter states of the ACFL~\cite{Cai2023_FQAHTMD,Park2023_FQAH_TMD,Xu2023_FQAHTMD,Zeng2023_FQAHTMD,Lu2023_FQAHPenta,Lu2025_EQAH}. 

\begin{table}[t!]
  \label{tab:MR_CBFL_CFL}
  \begin{ruledtabular}
    \begin{tabular}{lccc}
      & Pfaffian & CBFL & ACFL \\
      \colrule
      $C_v(T)$            & $\propto T^{\#} e^{-\Delta/T}$ & $\propto T$      & $\propto T^{2/3}$ \\
      $R_{xx}$       & $0$           & $0$            & $ > 0$             \\
      $R_{xy}$       & $2h/e^2$           & $2h/e^2$            & $\approx 2h/e^2$             \\
      $\kappa_{xy}$       & $3\kappa_0/2$   & Unquantized      & Unquantized       \\
      Compressible        & No              & No               & Yes               \\
      Landau damping       & No              & Yes               & Yes               \\
      $S(\bs{q})$ &Analytic       &$|\bs{q}|^3$  &$|\bs{q}|^3 \ln |\bs{q}|$ \\
      Quantum oscillations & No              & No               & Yes               \\
      Torus GSD           & 6               & 2                & 1                 \\
    \end{tabular}
  \end{ruledtabular}
  \caption{Important experimental observables for three candidate states in a half-filled inversion-asymmetric Chern band: (1) The Pfaffian state with non-Abelian topological order; (2) the composite fermion superconductor with Bogoliubov Fermi surface (CBFL); (3) the composite Fermi liquid (CFL) with screened Coulomb interactions. $S(\bs{q})$ denotes the equal-time structure factor at small $\bs{q}$ and GSD is the topological ground state degeneracy.}
\end{table}

The emergence of the ACFL raises a fundamental theoretical question: are there paired states of CFs in the zero-field FQAH setting that go beyond the Pfaffian and its close relatives, such as the anti-Pfaffian or PH-Pfaffian? In this letter, we answer this question in the affirmative. We show that in a half-filled Chern band with broken inversion symmetry, attractive interactions between CFs can stabilize a paired state with a gapless Bogoliubov Fermi surface (BFS)~\footnote{A related paired state of spinons in gapless spin liquids has been considered in Ref.~\cite{Barkeshli2012_spinonBFS}.} We term the resulting electronic phase the composite Bogoliubov Fermi liquid (CBFL). Importantly, broken inversion symmetry is a natural and generic feature of both twisted MoTe$_2$ and rhombohedral graphene, where trigonal warping reduces the $C_6$ lattice rotation symmetry down to $C_3$, which does not contain inversion as a subgroup~\cite{Wu2019_TMD_foundational,Dong2023_FQAHpenta_MIT,Dong2023_FQAHpenta_harvard,Zhou2023_FQAHpenta}.

As summarized in Table~\ref{tab:MR_CBFL_CFL}, universal low-energy properties of the CBFL are sharply distinct from both the ACFL and the Pfaffian state. Like the Pfaffian, the CBFL is an incompressible phase with a quantized DC resistivity and no quantum oscillations. Unlike the Pfaffian, however, the CBFL hosts a gapless BFS, leading to Landau-damped density fluctuations, a non-analytic $|\bs{q}|^3$ term in the equal-time structure factor $S(\bs{q})$, and metallic $T$-linear specific heat, features more commonly associated with metallic phases such as the ACFL. Most surprisingly, we prove that the CBFL phase has a robust two-fold torus ground state degeneracy (GSD) in the thermodynamic limit despite the presence of gapless BFS fluctuations. As a result, the CBFL phase realizes an exotic gapless topological phase distinct from the Pfaffian state with gapped topological order and the ACFL state with no topological order. 

In what follows, we construct a quantum Ginzburg Landau theory for paired states of CFs and use it to derive the physical properties summarized in Table~\ref{tab:MR_CBFL_CFL}. We further comment on transitions out of the CBFL under doping or externally applied magnetic fields and conclude by outlining several open questions.

\begin{figure}[t!]
    \centering
\includegraphics[width=1.0\linewidth]{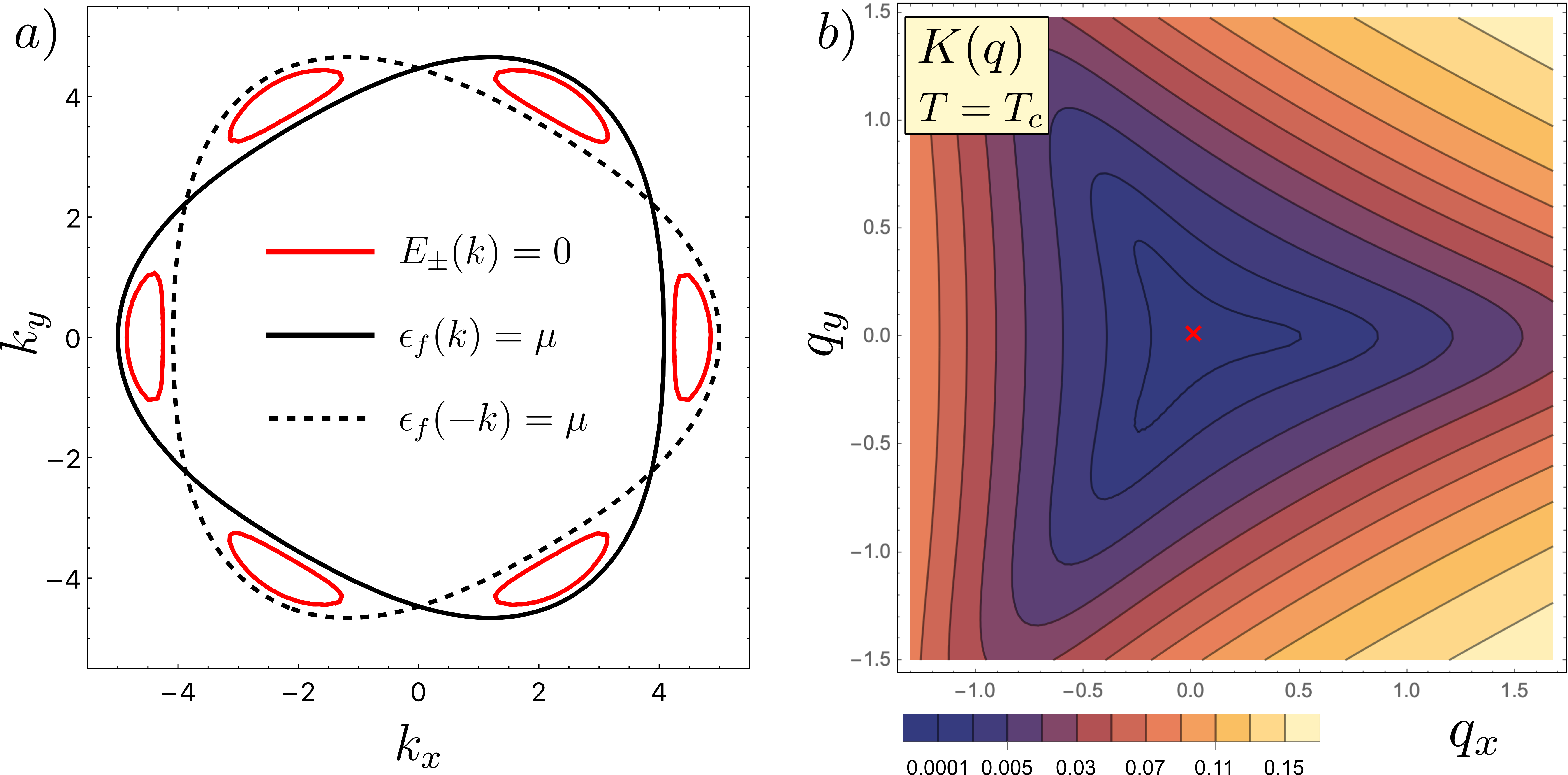}
    \caption{Illustrative solution of the self-consistent mean-field equation for the chiral $p+ip$ SC order parameter with the composite fermion (CF) dispersion $\epsilon_f(\bm{k})=\frac{k^2}{2m}(1+\lambda \cos 3\theta_{\bm{k}})$  and attractive interactions in the $l=1$ channel (see SM Sec.~\ref{subapp:gapequation} for details). a) The CF Fermi surface in a normal state is shown in black. The resulting Bogoliubov Fermi surface at $T=0$ is shown in red. b) Schematic momentum dependence of the quadratic part of the Ginzburg-Landau action at $T=T_c$. The red cross denotes the global minimum at $\bs{q}=0$.} \label{fig:fig1}
\end{figure}
\section{Construction of the CBFL phase}

We begin with a microscopic model described by the following Hamiltonian
\begin{equation}
    H = \sum_{\bs{k}} (\epsilon_{\bs{k}} - \mu) c^{\dagger}_{\bs{k}} c_{\bs{k}} + \sum_{\bs{k}, \bs{p}, \bs{q}} V_{\bs{q}}(\bs{k}, \bs{p}) c^{\dagger}_{\bs{k} + \bs{q}} c^{\dagger}_{\bs{p} - \bs{q}} c_{\bs{p} } c_{\bs{k} }\,, 
\end{equation}
where $c^{\dagger}_{\bs{k}}$ creates band electrons in an isolated Chern band, $\mu$ is the chemical potential, and $V_{\bs{q}}(\bs{k}, \bs{p})$ describes a band-projected screened Coulomb interaction. 

As demonstrated by Ref.~\cite{Dong2023_ACFL,Goldman2023_ACFL}, an appropriate choice of $\epsilon_{\bs{k}}$ and $V_{\bs{q}}(\bs{k}, \bs{p})$ gives rise to a stable ACFL phase at lattice filling $\nu = 1/2$. The effective field theory for the ACFL can be constructed through a parton decomposition $c = f \Phi$ where $f$/$\Phi$ is a fermionic/bosonic parton~\cite{Barkeshli2012_latticeCFL_FL}. We choose $\Phi$ to carry electric charge $+1$ and $f$ to be neutral, such that $c$ carries the correct electric charge $+1$. This decomposition has a $U(1)$ gauge redundancy which can be removed by introducing a dynamical $U(1)$ gauge field $a$ that couples to $f$ and $\Phi$ with opposite signs. The low-energy effective Lagrangian then takes the general form
\begin{equation}\label{eq:parton}
    L_{\rm eff} = L[f, a] + L[\Phi, A-a] \,. 
\end{equation}
The equations of motion for $a$ impose a constraint on the lattice filling $\nu_f = \nu_{\Phi} = 1/2$. Consistent with LSM constraints at this filling, we can place $\Phi$ in a gapped bosonic Laughlin state $U(1)_2$ and $f$ in a Fermi liquid (FL) phase and obtain the ACFL Lagrangian
\begin{equation}\label{eq:ACFL}
    L_{\rm ACFL} = L_{\rm FL}[f, a] - \frac{2i}{4\pi} \alpha d \alpha + \frac{i}{2\pi} \alpha d (A-a)\,.
\end{equation}
We identify $f$ as the emergent CF, and denote its mean-field dispersion by $\epsilon_f(\bs{k})$, which inherits the $C_3$ rotation symmetry of the electron dispersion $\epsilon(\bs{k})$.\footnote{With broken inversion symmetry, the ACFL phase itself exhibits a number of exotic properties that go beyond the Landau level CFL, which we explore in a companion paper~\cite{Nosov2026_C3ACFL}.} 

Now let us introduce an effective attraction to induce pairing of CFs. Unlike in the Landau level context, the absence of inversion symmetry in our model breaks the nesting condition $\epsilon_f(\bs{k}) = \epsilon_f(-\bs{k})$. As a result, a CF pairing condensate $\ev{f^{\dagger}(\bs{k} + \bs{q}) f(-\bs{k})} \neq 0$ does not completely gap out the composite Fermi surface. Instead, there are residual Bogoliubov Fermi pockets that host gapless neutral excitations~\cite{Allais2012_BFS_loopcurrent,Agterberg2017_BFS,Brydon2018_BFS,Lapp2020_BFS,Oh2021_BFS_disorder,Bhattacharya2023_BFS_stability,Pal2024_BFS}. As an illustrative example of the emergence of BFS, we consider an ACFL state with $\epsilon_f(\bs{k})=\frac{k^2}{2m}(1+\lambda \cos 3\theta_{\bs{k}})$, where $|\lambda|<1$ controls the strength of inversion-breaking. By introducing an attractive interaction in the $p+ip$ channel and solving the mean-field gap equation self-consistently (see SM Sec.~\ref{subapp:gapequation} for details), we obtain a typical zero-temperature BFS and the quadratic part of the Ginzburg–Landau free energy, $K(q)$, as a function of total pair momentum $\bs{q}$ at $T_c$ (see Fig.~\ref{fig:fig1}). While the dominant pairing instability occurs at $\bs{q} = 0$ in Fig.~\ref{fig:fig1}, realistic interactions can shift the pairing momentum to a nonzero $\bs{q}$, as illustrated in recent studies of superconductors with electronic Bogoliubov Fermi surfaces~\cite{Yang2025_BFSrhombohedral,Christos2025_rhombohedralPairing,Gaggioli2025_RhombohedralVortex}.
However, since the essential properties of the CBFL do not depend on the value of $\bs{q}$, we will henceforth focus on $\bs{q} = 0$ for notational simplicity.

\section{Low energy properties of the CBFL}

We now construct an effective Ginzburg-Landau theory for the CBFL phase to derive its universal low-energy properties. Following standard procedures (see SM Sec.~\ref{subapp:Derivation_LG}), we introduce a Hubbard-Stratanovich field $\Delta$ in the Cooper pair channel to decouple the four-fermion interactions. The mean-field value of $\Delta$ defines a Bogoliubov-de Gennes (BdG) Hamiltonian whose solution contains a BFS. Expanding around the mean-field and integrating out the gapless Bogoliubov fermions, we land on the effective Lagrangian
\begin{equation}\label{eq:LG_main}
    L_{\rm CBFL} = L[\phi, 2a] - \frac{2i}{4\pi} \alpha d \alpha + \frac{i}{2\pi} \alpha d (A-a) \,,
\end{equation}
where $L[\phi, 2a]$ describes phase fluctuations of $\Delta \sim e^{i\phi}$
\begin{equation}
    L[\phi, 2a] = \frac{1}{8} (\partial_{\mu} \phi - 2 a_{\mu}) \Pi^{\mu\nu}_{\rm BdG} (\partial_{\nu} \phi - 2a_{\nu}) \,,\label{eq:GL_phase+a}
\end{equation}
and the kernel $\Pi^{\mu\nu}_{\rm BdG}$ is given by a set of correlation functions evaluated using the BdG Hamiltonian
\begin{equation*}
    \Pi_{\rm BdG}^{\mu\nu}(\bm{q},\Omega) =\begin{pmatrix}
       G_{\rho_f \rho_f}(\bs{q}, \Omega)  & -i G_{\rho_f J^j_f}(\bs{q}, \Omega)\\
        -i G_{J^i_f\rho_f }(\bs{q}, \Omega) & K_{\rm diam}^{ij} - G_{J^i_f J^j_f}(\bs{q}, \Omega)
    \end{pmatrix}.
\end{equation*}
Here $G_{AB}$ is the Euclidean correlation function between $A$ and $B$, e.g. $G_{\rho_f \rho_f}(\bs{q}, \Omega) =\ev{\rho_f(\bs{q}, \omega) \rho_f(-\bs{q}, - \omega)}_c$, etc., and $K^{ij}_{\rm diam}$ is a diamagnetic term defined in SM Sec.~\ref{subapp:Derivation_LG}. Armed with this Ginzburg-Landau Lagrangian and the functional form of $\Pi^{\mu\nu}_{\rm BdG}$ calculated in SM Sec.~\ref{app:LGtheory}, we can now deduce all the physical properties in Table~\ref{tab:MR_CBFL_CFL}.

\subsection{Single-electron properties and thermodynamics}

We begin with single-electron properties. The Chern-Simons terms in the ACFL Lagrangian \eqref{eq:ACFL} imply that a single electron $c$ is the bound state of $f$ and a monopole operator $\mathcal{M}_a^{\dagger,2}$ that inserts $4\pi$ flux of the emergent gauge field $a$. Following Ref.~\cite{Kim1994_monopole}, we decompose the electron Green's function as
\begin{equation}
    \begin{aligned}
    &G_c(\bs{x}, t) = \ev{f^{\dagger}(\bs{x}, t) \mathcal{M}_a^2(\bs{x}, t) f(0, 0) \mathcal{M}_a^{\dagger,2}(0,0)} \\
    &= \ev{f^{\dagger}(\bs{x}, t) W_{\gamma} f(0, 0) \mathcal{M}_a^{\dagger, 2}(0, 0) W^{\dagger}_{\gamma}\mathcal{M}_a^2(\bs{x},t)} \,,
    \end{aligned}
\end{equation}
where $W_{\gamma}$ is the Wilson line operator of the gauge field $a$ along a straight path $\gamma$ connecting $(0,0)$ and $(\bs{x}, t)$
\begin{equation}
    W_{\gamma} = e^{i \int_{\gamma} a_{\mu}(s) ds^{\mu}} \,. 
\end{equation}
In the limit of weak gauge fluctuations, we can approximately factorize $G_c$ as $G_c(\bs{x},t) = G_f(\bs{x}, t) G_{\mathcal{M}_a^2}(-\bs{x}, -t)$ where 
\begin{equation}
    \begin{aligned}
    G_f(\bs{x}, t) &= \ev{f^{\dagger}(\bs{x}, t) W_{\gamma} f(0, 0)} \,, \\
    G_{\mathcal{M}_a^2}(\bs{x}, t) &= \ev{\mathcal{M}_a^{\dagger, 2}(\bs{x}, t) W^{\dagger}_{\gamma}\mathcal{M}_a^2(0,0)}\,.
    \end{aligned}
\end{equation}
Individually, $f$ and $\mathcal{M}_a^2$ are not gauge-invariant under $a$ and carry equal and opposite gauge charge. However, due to the insertion of Wilson lines, $G_f, G_{\mathcal{M}_a^2}$ are gauge-invariant and can be calculated within perturbation theory. When the $f$ fermions form a BFS, there are gapless $f$ excitations at low energy. Therefore, $G_f(\bs{x}, t)$ decays as a power law at large $|\bs{x}|, t$, whose precise form can be determined from the mean-field BdG Hamiltonian. The monopole correlation function $G_{\mathcal{M}_a^2}$ is more intricate. In SM Sec.~\ref{app:LGtheory}, we show that the transverse gauge field $a_T$ develops a nonzero Higgs mass $\rho_s$ despite the presence of many gapless excitations on the BFS. This Higgs mass leads to a short-range interaction between monopole endpoints. In addition, because a monopole and an antimonopole are connected by a physical vortex line in the Higgs phase, there is also a nonzero line tension that confines the monopoles. This line tension induces an exponentially decaying monopole correlation function $G_{\mathcal{M}_a^2}(\bs{x}, t)$ as we derive in SM Sec.~\ref{subapp:monopoles}. Taking the product of $G_f$ and $G_{\mathcal{M}_a^2}$, we therefore conclude that the electron Green's function $G_c$ also decays exponentially in space and time, implying a nonzero electron energy gap. 

Despite the single-electron energy gap, the thermodynamics of the CBFL state is sensitive to the gapless BFS. In a fully gapped quantum Hall state (e.g. the Pfaffian state), the specific heat at low temperature scales as $C_v(T) \sim e^{-\Delta_0/T}$, where $\Delta_0$ is the energy gap to the cheapest neutral excitation. In contrast, for the ACFL state with screened Coulomb interactions, the specific heat scales as a singular power law $T^{2/3}$ due to gapless gauge field fluctuations~\cite{Halperin1993_HLRtheory,Chen2025_CFLtensor}. The CBFL phase sits in between these two conventional phases. Like the ACFL, the CBFL features an infinite number of gapless fermionic modes that contribute to the low-temperature specific heat. However, unlike in the ACFL, the gauge field $a$ is Higgsed in the CBFL state and its massive fluctuations do not lead to a singular self-energy for $f$. As a result, composite fermions on the Bogoliubov Fermi surface remain well-defined quasiparticles and contribute a Fermi-liquid-like specific heat $C^{\rm BFL}_v(T) \sim \gamma_{\rm BFL} T$. The coefficient $\gamma_{\rm BFL}$ is proportional to the averaged density of states on the BFS, which is non-universal. In addition to the BFS contribution, there is a gauge field contribution to the specific heat which is also linear in $T$ with a different coefficient $C_v^{\rm gauge}(T) \sim \gamma_{\rm gauge} T$ (see SM Sec.~\ref{subapp:thermodynamics} for detailed calculations). The total physical specific heat is therefore $C_v(T) \sim (\gamma_{\rm gauge} + \gamma_{\rm BFL}) T$, which is distinct from both the ACFL and the Pfaffian.

\subsection{Electrical and thermal transport}

Going beyond single-particle properties, we now study two-body correlations encoded in electrical and thermal responses. Despite the presence of a gapless Fermi surface, the electromagnetic response of the CBFL state resembles closely a fully gapped quantum Hall state at the same lattice filling $\nu = 1/2$. To understand this result, let us recall the Ioffe-Larkin rule~\cite{Ioffe1989_rule} which relates the resistivity $R_f$ of the composite fermions to the resistivity $R_c$ of the physical electrons
\begin{equation}
    R_c(\omega) = R_f(\omega) + \begin{pmatrix}
        0 & 2 h/e^2 \\ - 2h/e^2 & 0 
    \end{pmatrix} \,. \label{eq:Ioffe-Larkin}
\end{equation}
In the DC limit ($\omega = 0$), momentum-relaxing scattering processes in the ACFL produce an unquantized $R_f^{xx}$. On the other hand, for a CF superconductor, a nonzero stiffness $\rho_s \equiv \frac{1}{4}\left[K_{\rm diam}^{ii} - G_{J^i_f J^i_f}(\bs{q}{\rightarrow} 0, \Omega{=}0)\right]$ implies dissipationless electrical transport with strictly vanishing $R_f$. As a result, the CBFL is indistinguishable from the gapped Pfaffian state from the perspective of DC transport and exhibits a quantized resistivity tensor with $R_c^{xy} = 2h/e^2$ and $R_c^{xx} = 0$, which is stable to all perturbations at half-filling. Despite the breaking of rotational symmetry down to $C_3$, the $\bs{q} = 0$ resistivity tensor remains isotropic. In contrast, at nonzero frequency, the CBFL bears more resemblance to the ACFL phase, as both have non-vanishing optical weight down to zero frequency. 

Next, we turn to thermal transport. In a fully gapped quantum Hall state, the bulk thermal conductance $\kappa_{xx}$ vanishes identically at $T = 0$ and a quantized thermal Hall conductance $\kappa_{xy}$ directly measures the chiral central charge of the quantum Hall edge state. In the CBFL phase, the gapless composite Bogoliubov fermions do not carry a charge current, but do carry a thermal current. A standard Boltzmann calculation shows that the CBFL contributes a large longitudinal thermal conductance $\kappa_{xx}(T) \sim C_v(T) v_F^2 \tau_{\rm tr}$, where $v_F$ is the Fermi velocity averaged over the BFS and $\tau_{\rm tr}$ is the transport scattering time~\cite{Lapp2020_BFS,Pal2024_thermoelectricBFS}. At weak inversion breaking, the Bogoliubov Fermi pockets are small and Umklapp scattering is not operative. The most efficient scattering mechanism is impurity scattering, which gives a constant $\tau_{\rm tr}$ as $T \rightarrow 0$. As a result, the bulk thermal conductance $\kappa_{xx}(T) \sim T$. The presence of a metallic bulk thermal conductance also destroys the quantization of edge thermal Hall conductance, providing another sharp distinction between the CBFL and the Pfaffian. 

\subsection{Spatially resolved dynamics and Landau damping}

At nonzero momentum, the singular effects of BFS fluctuations become more pronounced. Specifically, in the London limit, where $q$ is much less than the inverse coherence length and $|\Omega| \ll v_F q$, fluctuations of the BFS induce Landau damping in the density correlations of $f$ (see SM Sec.~\ref{subapp:BdG_rhorho}). On the Matsubara axis, we find
\begin{equation}
    G_{\rho_f \rho_f}(\bs{q}, \Omega)  = \tilde{\chi}_0 -  (\Gamma'_{\hat {\bm{q}}}+ i\operatorname{sgn}(\Omega)\Gamma''_{\hat {\bm{q}}} ) \frac{i\Omega}{q} + \ldots \,, \label{eq:density_f_main}
\end{equation}
where $\tilde{\chi}_0$ is a positive constant, and $\Gamma_{\hat{\bm{q}}}=\Gamma'_{\hat{\bm{q}}}+i \Gamma''_{\hat{\bm{q}}}$ is a complex function of the direction $\hat{\bm{q}}$ given in SM Sec.~\ref{subapp:BdG_rhorho}. After performing analytic continuation to real frequency $\omega$, we find $G^R_{\rho_f \rho_f}(\bs{q}, \omega)\approx \tilde{\chi}_0 -\Gamma_{\hat{\bm{q}}} \omega/q$. From Eq.~\eqref{eq:GL_phase+a}, one can see that the  Landau damping term translates to a frequency-dependent damping of the Goldstone boson associated with the CF superconductor.\footnote{A similar Landau damping of Goldstone modes has previously appeared in the context of nematic Fermi fluids~\cite{Oganesyan2001_nematic}.}
When the Goldstone boson is eaten by the gauge field $a$, this $\omega$-dependent damping then appears as an $\omega$-dependent mass term in the effective action for $a$. Through the flux attachment dictionary (SM Sec.~\ref{subapp:density_CBFL}), this damping leads to a singular electron density-density correlation function
\begin{equation}
  G^R_{\rho_c \rho_c}(\bs{q}, \omega) = \frac{q^2}{4(4\pi)^2\rho_s} \Bigg(1 -\frac{\Gamma_{\hat{\bm{q}}}^{TT}}{4\rho_s}\;\frac{\omega}{q} \Bigg)+ \ldots \, ,
\end{equation}
where $\Gamma_{\hat{\bm{q}}}^{TT}$is a complex coefficient associated with the Landau damping term in the composite fermion current-current correlator, $ G_{J^i_f J^j_f}^R(\bs{q},\omega) \approx \tilde \chi_J - \Gamma^{ij}_{\hat{\bs{q}}} \omega/q $, projected onto the transverse subspace.
Since $ G^R_{\rho_c \rho_c}(\bs{q}, \omega=0)\sim q^2$, we conclude that the CBFL state is incompressible, much like a fully gapped quantum Hall state. However, at nonzero $\omega$, $ G^R_{\rho_c \rho_c}(\bs{q}, \omega)$ has an additional imaginary Landau-damping term which is a smoking-gun gauge-invariant manifestation of the gapless BFS fluctuations. 

The coexistence of Landau damping and a finite Higgs mass for the gauge field $a$ leads to a distinctive non-analytic contribution to the small-momentum expansion of the equal-time density–density correlator $S(\bs{q}) \equiv  G_{\rho_c \rho_c}(\bs{q}, \tau=0)$ (the structure factor). Specifically, we find a $\sim q^3$ contribution, whose prefactor diverges as the Higgs mass of the gauge field is taken to zero. The appearance of this non-analytic term should be contrasted with the Pfaffian state, whose structure factor admits a regular expansion in powers of $q^2$. In this respect, the CBFL more closely resembles the conventional CFL, where an RPA analysis yields a $q^3\ln 1/q$ non-analytic term \footnote{The existence of the additional logarithm on top of the $q^3$ dependence beyond RPA was recently questioned in Ref.~\cite{Anakru2025_noLogCFL}.}.

Finally, coming back to real space, the BFS induces Friedel oscillations in $G^R_{\rho_c\rho_c}(\bs{x}, t)$ as a function of $|\bs{x}|$ for every orientation $\hat x$. Since the BFS is generically anisotropic, the oscillation period is also a nontrivial function of $\hat x$ that depends on the structure of the BFS. 

\section{Topological properties of the CBFL}

The physical observables considered so far can be accessed through standard experimental probes that measure few-body correlators. As a result, they miss the topological properties of the CBFL phase that are only encoded in its long-range entanglement patterns. This limitation raises an important conceptual question: given the gapless BFS, is there any sense in which the CBFL phase remains topological? 

To answer this question, we turn to the torus ground state degeneracy (GSD), which is a fundamental diagnostic of topological order. It is known that the ACFL has a unique ground state on the torus, while the Pfaffian state has a six-fold torus GSD with exponentially suppressed finite-size splittings (see SM Sec.~\ref{app:degeneracy} for details). Given the gapless BFS excitations, one may expect that the CBFL is more similar to the ACFL in its topological properties and has no protected torus GSD. However, this expectation turns out to be false. 

To determine the torus GSD of the CBFL, we use an effective Lagrangian (see SM Sec.~\ref{subapp:LG_withBFS} for a derivation within the framework of Wilsonian RG) that retains both the gapless fermions $f$ and the phase fluctuations $\phi$ 
\begin{equation*}
    L_{\rm CBFL} = L_{\rm BFL}[\psi, a] + L[\phi, 2a] - \frac{2i}{4\pi} \alpha d \alpha + \frac{i}{2\pi} \alpha d (A-a)\,, 
\end{equation*}
where $L_{\rm BFL}[\psi, a]$ describes an interacting Bogoliubov Fermi liquid (BFL) written in the Bogoliubov quasiparticle basis $\psi$. After performing a particle-vortex duality on $L[\phi, 2a]$, we can rewrite the Lagrangian as
\begin{equation*}
    L_{\rm CBFL} = L_{\rm BFL}[\psi, a] + \frac{2 i}{2\pi} \beta d a - \frac{2i}{4\pi} \alpha d \alpha + \frac{i}{2\pi} \alpha d (A-a)\,, 
\end{equation*}
where $\beta$ is the hydrodynamic gauge field dual to $\phi$. As a thought experiment, imagine that the BFL sector was gapped. Then we could integrate out $a$ and set $2\beta = \alpha$. The resulting phase would be a gapped $U(1)_8$ topological order described by the Lagrangian
\begin{equation}
    L_{U(1)_8} = - \frac{8i}{4\pi} \beta d \beta + \frac{2i}{2\pi} \beta d A \,. 
\end{equation}
This topological order is generated by a single order-8 anyon $v$ which can be created by a matter field carrying unit gauge charge under $\beta$. By definition, the Bogoliubov fermion $\psi$ can be created by a $\pi$-flux of $\beta$, which carries gauge charge 4 under $\beta$ due to the level-8 Chern-Simons term for $\beta$. This charge assignment implies that $\psi$ sources the anyon $v^4$ in the $U(1)_8$ topological order. 

In the real system, the composite Bogoliubov fermions form a Fermi surface, making the anyon $v^4$ gapless. As a result, all anyons that braid with $v^4$ cease to be well-defined. Since the braiding phase between $v^n$ and $v^m$ is $e^{2\pi i nm/8}$, the only surviving anyons are $\{1, v^2, v^4, v^6\}$. Furthermore, anyons related to each other by fusing with $v^4$ cannot be distinguished when $v^4$ is gapless. Thus, the unique anyon sector distinguishable from the vacuum is the semion $v^2$, which guarantees a two-fold torus GSD. The same result can be obtained from the more formal perspective of emergent 1-form symmetries and their 't Hooft anomalies~\cite{Gaiotto2015_gensym}. In that language, the nontrivial torus GSD originates from a self-anomalous $\mathbb{Z}_4$ 1-form symmetry that is preserved by the gapless BFS (see SM Sec.~\ref{app:topology}). 

The topological phase realized by the CBFL should be distinguished from a more familiar type of gapless phase known as FL*, in which a conventional gapless Fermi liquid is stacked with a gapped topological order. Unlike in FL*, gapped anyons and gapless excitations in the CBFL are intertwined in a nontrivial manner: there is no sense in which the CBFL can be interpreted as a Fermi liquid stacked with a $U(1)_2$ topological order, because a decoupled semion with charge 1/2 cannot emerge in any electronic system in which the microscopic electron carries charge 1. As a final remark, we note that if we instead started with a bosonic ACFL realized by hard-core bosons at lattice filling $\nu = 1$, then the descendent bosonic CBFL would not have this two-fold degeneracy (see SM Sec.~\ref{app:topology}).

\section{Effects of doping and magnetic fields}

The theory of CBFL at $\nu = 1/2$ allows us to study possible evolutions out of the CBFL phase under doping or an external magnetic field $B$. Both processes produce a non-vanishing effective magnetic field $B_{\rm eff}$ seen by the CFs. In the unpaired state with a regular Fermi sea of CFs, this leads to quantum oscillations in transport and thermodynamic quantities. Strikingly, the response of a Bogoliubov Fermi surface to a weak external field is qualitatively different: Bogoliubov quasiparticles of CFs follow the semiclassical equations of motion, with a momentum-dependent effective charge $e_*(\bs{k})$ that changes sign at discrete points on the BFS (see SM Sec.~\ref{app:LGtheory}). As a result, quasiparticles cannot complete closed orbits on the BFS and there is no quantum oscillation~\cite{Allais2012_BFS_loopcurrent}. At slightly larger values of $B_{\rm eff}$, the superfluid stiffness remains nonzero, but the CF superconductor may develop a vortex lattice, {which further screens the effective magnetic field experienced by gapless quasiparticles~\cite{Franz2000_vortex}. In the presence of disorder/spatial boundary, the vortex lattice gets pinned and the DC resistivity remains quantized to its value at half-filling, leading to a resistivity plateau near $B_{\rm eff} = 0$. However, since order parameter fluctuations are Landau-damped by the BFS, vortex dynamics in the CBFL phase may be qualitatively different from a fully gapped CF superconductor. It would be important to analyze these vortex damping effects more systematically in the future. 

At even larger $B_{\rm eff}$, the superfluid stiffness eventually vanishes and the vortex lattice of CFs may melt into a translation-preserving phase. If the parent Chern band is very flat, it is natural to expect the melted daughter state to be a gapped FQH state. For example, starting with the Pfaffian at $\nu = 1/2$, various theoretical approaches predict gapped daughter states at $\nu = 8/17$ and $\nu = 7/13$~\cite{Levin2009_daughter,Yutushui2024_daughter,Zheltonozhskii2024_daughter,Zhang2025_daughter}, which have been observed in various graphene FQH platforms~\cite{Li2017_daughter_exp,Zibrov2017_daughter_exp,Huang2021_daughter_exp,Singh2023_daughter_exp}. In SM Sec.~\ref{app:daughter}, we make some preliminary attempts at generalizing the construction of daughter states to the CBFL context, finding a sequence of gapped topological phases at $\nu = \frac{(p+8n)}{2(p+8n) + 1}$ where $n \neq 0$ and $p$ are integers. If the parent Chern band has appreciable dispersion, then the daughter states could even be itinerant, forming a superconductor or metal of electrons.\footnote{This general route of obtaining itinerant phases from doping FQAH insulators has been explored recently in Refs.~\cite{Shi2024_doping,Kim2024_anyonSC,Divic2024_HofHubb,Kuhlenkamp2025_HofHubb,Shi2025_dopeMR,Nosov2025_plateau,Shi2025_anyon_delocalization,Pichler2025_anyonSC}.} If the Pfaffian state is realized at $\nu = 1/2$, then doping indeed induces charge-2e superconductivity~\cite{Shi2025_dopeMR}. It is tantalizing to conjecture that doping the CBFL still gives a superconductor, but now with an electronic BFS that originates from the CF BFS. We leave a more thorough examination of these doping-induced itinerant phases to future work.

\section{Conclusion and outlook}

In summary, we have shown that half-filled Chern bands with broken inversion symmetry host a new type of gapless topological phase dubbed the composite Bogoliubov Fermi liquid (CBFL), whose unusual physical properties are summarized in Table~\ref{tab:MR_CBFL_CFL}. Specifically, this phase is characterized by a quantized $2h/e^2$ Hall resistivity, two-fold topological ground state degeneracy on a torus, metallic $T$-linear specific heat and non-quantized thermal conductance, incompressible Landau-overdamped density response, equal time structure factor with a $\sim q^3$ non-analytic contribution at small momenta, and a gapped single-electron spectral function.

Our results pave the way for a more detailed exploration of the CBFL phase across different regimes. One important direction concerns the structure and dynamics of the vortex lattice in the presence of Bogoliubov Fermi surfaces, particularly in relation to the Nernst and Seebeck effects \cite{Behnia2016_thermalReview,Silaev2015_thermoelectricTRSB,VanHarlingen1980_thermoelectricity_SC}. More broadly, thermoelectric phenomena have recently attracted significant interest in the context of the half-filled Landau level \cite{Cooper1997_thermoelectric,Potter2016_thermoelectricCFL,Wang2017_CFLthermal}, and it would be interesting to revisit some of these questions in the CBFL setting. Another important question is a more realistic analysis of the dominant pairing channel with proper account for gauge field fluctuations. In particular, the pairing condensate may form at finite center-of-mass momentum $\bm{q}$ (i.e. a CF pair density wave~\cite{Santos2018_CFPDW}), or become strongly peaked at specific “hot spots” on the Fermi surface if trigonal warping effects are sufficiently strong. The latter possibility might be favored due to the fact that the singular current-current interactions mediated by the gauge field need not be purely repulsive if the inversion symmetry is broken~\cite{Nosov2026_C3ACFL}. While our general conclusions are largely insensitive to such details, they may nevertheless play an important role in quantitative comparisons with experiments and numerics. Finally, understanding the effects of disorder on the CBFL phase is especially challenging, as it requires incorporating flux disorder that is tied to the local density variations via flux attachment constrained \cite{Kivelson1997_HallCF,Wang2017_CFLthermal,Kumar2019_HallCF}. The interplay between such disorder and a Bogoliubov Fermi surface remains largely unexplored.\\

\section*{Acknowledgments}

We are grateful to Srinivas Raghu, Bertrand Halperin, Steven Kivelson, Nicholas O'Dea, Pietro Bonetti, and Eslam Khalaf for useful discussions. We also thank Elizaveta Andriyakhina for a related collaboration. ZDS is supported by a Leinweber Institute for Theoretical Physics postdoctoral fellowship at Stanford University. PAN is supported in part by a Harvard Quantum Initiative postdoctoral fellowship at Harvard University.

\bibliography{ACFL_BFS}

\begin{thebibliography}{94}%
\makeatletter
\providecommand \@ifxundefined [1]{%
 \@ifx{#1\undefined}
}%
\providecommand \@ifnum [1]{%
 \ifnum #1\expandafter \@firstoftwo
 \else \expandafter \@secondoftwo
 \fi
}%
\providecommand \@ifx [1]{%
 \ifx #1\expandafter \@firstoftwo
 \else \expandafter \@secondoftwo
 \fi
}%
\providecommand \natexlab [1]{#1}%
\providecommand \enquote  [1]{``#1''}%
\providecommand \bibnamefont  [1]{#1}%
\providecommand \bibfnamefont [1]{#1}%
\providecommand \citenamefont [1]{#1}%
\providecommand \href@noop [0]{\@secondoftwo}%
\providecommand \href [0]{\begingroup \@sanitize@url \@href}%
\providecommand \@href[1]{\@@startlink{#1}\@@href}%
\providecommand \@@href[1]{\endgroup#1\@@endlink}%
\providecommand \@sanitize@url [0]{\catcode `\\12\catcode `\$12\catcode
  `\&12\catcode `\#12\catcode `\^12\catcode `\_12\catcode `\%12\relax}%
\providecommand \@@startlink[1]{}%
\providecommand \@@endlink[0]{}%
\providecommand \url  [0]{\begingroup\@sanitize@url \@url }%
\providecommand \@url [1]{\endgroup\@href {#1}{\urlprefix }}%
\providecommand \urlprefix  [0]{URL }%
\providecommand \Eprint [0]{\href }%
\providecommand \doibase [0]{https://doi.org/}%
\providecommand \selectlanguage [0]{\@gobble}%
\providecommand \bibinfo  [0]{\@secondoftwo}%
\providecommand \bibfield  [0]{\@secondoftwo}%
\providecommand \translation [1]{[#1]}%
\providecommand \BibitemOpen [0]{}%
\providecommand \bibitemStop [0]{}%
\providecommand \bibitemNoStop [0]{.\EOS\space}%
\providecommand \EOS [0]{\spacefactor3000\relax}%
\providecommand \BibitemShut  [1]{\csname bibitem#1\endcsname}%
\let\auto@bib@innerbib\@empty
\bibitem [{\citenamefont {Stormer}\ \emph {et~al.}(1999)\citenamefont
  {Stormer}, \citenamefont {Tsui},\ and\ \citenamefont
  {Gossard}}]{Stormer1999_FQHreview}%
  \BibitemOpen
  \bibfield  {author} {\bibinfo {author} {\bibfnamefont {H.~L.}\ \bibnamefont
  {Stormer}}, \bibinfo {author} {\bibfnamefont {D.~C.}\ \bibnamefont {Tsui}},\
  and\ \bibinfo {author} {\bibfnamefont {A.~C.}\ \bibnamefont {Gossard}},\
  }\bibfield  {title} {\bibinfo {title} {The fractional quantum {H}all
  effect},\ }\href {https://doi.org/10.1103/RevModPhys.71.S298} {\bibfield
  {journal} {\bibinfo  {journal} {Rev. Mod. Phys.}\ }\textbf {\bibinfo {volume}
  {71}},\ \bibinfo {pages} {S298} (\bibinfo {year} {1999})}\BibitemShut
  {NoStop}%
\bibitem [{\citenamefont {Jain}(1989)}]{Jain1989_CFframework}%
  \BibitemOpen
  \bibfield  {author} {\bibinfo {author} {\bibfnamefont {J.~K.}\ \bibnamefont
  {Jain}},\ }\bibfield  {title} {\bibinfo {title} {Composite-fermion approach
  for the fractional quantum {H}all effect},\ }\href
  {https://doi.org/10.1103/PhysRevLett.63.199} {\bibfield  {journal} {\bibinfo
  {journal} {Phys. Rev. Lett.}\ }\textbf {\bibinfo {volume} {63}},\ \bibinfo
  {pages} {199} (\bibinfo {year} {1989})}\BibitemShut {NoStop}%
\bibitem [{\citenamefont {Lopez}\ and\ \citenamefont
  {Fradkin}(1991)}]{Lopez1991_CSGL}%
  \BibitemOpen
  \bibfield  {author} {\bibinfo {author} {\bibfnamefont {A.}~\bibnamefont
  {Lopez}}\ and\ \bibinfo {author} {\bibfnamefont {E.}~\bibnamefont
  {Fradkin}},\ }\bibfield  {title} {\bibinfo {title} {Fractional quantum {H}all
  effect and {C}hern-{S}imons gauge theories},\ }\href
  {https://doi.org/10.1103/PhysRevB.44.5246} {\bibfield  {journal} {\bibinfo
  {journal} {Phys. Rev. B}\ }\textbf {\bibinfo {volume} {44}},\ \bibinfo
  {pages} {5246} (\bibinfo {year} {1991})}\BibitemShut {NoStop}%
\bibitem [{\citenamefont {Kalmeyer}\ and\ \citenamefont
  {Zhang}(1992)}]{Kalmeyer1992_CFL}%
  \BibitemOpen
  \bibfield  {author} {\bibinfo {author} {\bibfnamefont {V.}~\bibnamefont
  {Kalmeyer}}\ and\ \bibinfo {author} {\bibfnamefont {S.-C.}\ \bibnamefont
  {Zhang}},\ }\bibfield  {title} {\bibinfo {title} {Metallic phase of the
  quantum hall system at even-denominator filling fractions},\ }\href
  {https://doi.org/10.1103/PhysRevB.46.9889} {\bibfield  {journal} {\bibinfo
  {journal} {Phys. Rev. B}\ }\textbf {\bibinfo {volume} {46}},\ \bibinfo
  {pages} {9889} (\bibinfo {year} {1992})}\BibitemShut {NoStop}%
\bibitem [{\citenamefont {Halperin}\ \emph {et~al.}(1993)\citenamefont
  {Halperin}, \citenamefont {Lee},\ and\ \citenamefont
  {Read}}]{Halperin1993_HLRtheory}%
  \BibitemOpen
  \bibfield  {author} {\bibinfo {author} {\bibfnamefont {B.~I.}\ \bibnamefont
  {Halperin}}, \bibinfo {author} {\bibfnamefont {P.~A.}\ \bibnamefont {Lee}},\
  and\ \bibinfo {author} {\bibfnamefont {N.}~\bibnamefont {Read}},\ }\bibfield
  {title} {\bibinfo {title} {Theory of the half-filled {L}andau level},\ }\href
  {https://doi.org/10.1103/PhysRevB.47.7312} {\bibfield  {journal} {\bibinfo
  {journal} {Phys. Rev. B}\ }\textbf {\bibinfo {volume} {47}},\ \bibinfo
  {pages} {7312} (\bibinfo {year} {1993})}\BibitemShut {NoStop}%
\bibitem [{\citenamefont {Son}(2015)}]{Son2015_DiracCFL}%
  \BibitemOpen
  \bibfield  {author} {\bibinfo {author} {\bibfnamefont {D.~T.}\ \bibnamefont
  {Son}},\ }\bibfield  {title} {\bibinfo {title} {Is the composite fermion a
  dirac particle?},\ }\href {https://doi.org/10.1103/PhysRevX.5.031027}
  {\bibfield  {journal} {\bibinfo  {journal} {Phys. Rev. X}\ }\textbf {\bibinfo
  {volume} {5}},\ \bibinfo {pages} {031027} (\bibinfo {year}
  {2015})}\BibitemShut {NoStop}%
\bibitem [{\citenamefont {Eisenstein}\ \emph {et~al.}(1992)\citenamefont
  {Eisenstein}, \citenamefont {Pfeiffer},\ and\ \citenamefont
  {West}}]{Eisenstein1992_CFLevidence}%
  \BibitemOpen
  \bibfield  {author} {\bibinfo {author} {\bibfnamefont {J.~P.}\ \bibnamefont
  {Eisenstein}}, \bibinfo {author} {\bibfnamefont {L.~N.}\ \bibnamefont
  {Pfeiffer}},\ and\ \bibinfo {author} {\bibfnamefont {K.~W.}\ \bibnamefont
  {West}},\ }\bibfield  {title} {\bibinfo {title} {Coulomb barrier to tunneling
  between parallel two-dimensional electron systems},\ }\href
  {https://doi.org/10.1103/PhysRevLett.69.3804} {\bibfield  {journal} {\bibinfo
   {journal} {Phys. Rev. Lett.}\ }\textbf {\bibinfo {volume} {69}},\ \bibinfo
  {pages} {3804} (\bibinfo {year} {1992})}\BibitemShut {NoStop}%
\bibitem [{\citenamefont {Kang}\ \emph {et~al.}(1993)\citenamefont {Kang},
  \citenamefont {Stormer}, \citenamefont {Pfeiffer}, \citenamefont {Baldwin},\
  and\ \citenamefont {West}}]{Kang1993_CFLevidence}%
  \BibitemOpen
  \bibfield  {author} {\bibinfo {author} {\bibfnamefont {W.}~\bibnamefont
  {Kang}}, \bibinfo {author} {\bibfnamefont {H.~L.}\ \bibnamefont {Stormer}},
  \bibinfo {author} {\bibfnamefont {L.~N.}\ \bibnamefont {Pfeiffer}}, \bibinfo
  {author} {\bibfnamefont {K.~W.}\ \bibnamefont {Baldwin}},\ and\ \bibinfo
  {author} {\bibfnamefont {K.~W.}\ \bibnamefont {West}},\ }\bibfield  {title}
  {\bibinfo {title} {How real are composite fermions?},\ }\href
  {https://doi.org/10.1103/PhysRevLett.71.3850} {\bibfield  {journal} {\bibinfo
   {journal} {Phys. Rev. Lett.}\ }\textbf {\bibinfo {volume} {71}},\ \bibinfo
  {pages} {3850} (\bibinfo {year} {1993})}\BibitemShut {NoStop}%
\bibitem [{\citenamefont {Willett}\ \emph
  {et~al.}(1993{\natexlab{a}})\citenamefont {Willett}, \citenamefont {Ruel},
  \citenamefont {West},\ and\ \citenamefont
  {Pfeiffer}}]{Willett1993_CFLevidence_FS}%
  \BibitemOpen
  \bibfield  {author} {\bibinfo {author} {\bibfnamefont {R.~L.}\ \bibnamefont
  {Willett}}, \bibinfo {author} {\bibfnamefont {R.~R.}\ \bibnamefont {Ruel}},
  \bibinfo {author} {\bibfnamefont {K.~W.}\ \bibnamefont {West}},\ and\
  \bibinfo {author} {\bibfnamefont {L.~N.}\ \bibnamefont {Pfeiffer}},\
  }\bibfield  {title} {\bibinfo {title} {Experimental demonstration of a
  {F}ermi surface at one-half filling of the lowest {L}andau level},\ }\href
  {https://doi.org/10.1103/PhysRevLett.71.3846} {\bibfield  {journal} {\bibinfo
   {journal} {Phys. Rev. Lett.}\ }\textbf {\bibinfo {volume} {71}},\ \bibinfo
  {pages} {3846} (\bibinfo {year} {1993}{\natexlab{a}})}\BibitemShut {NoStop}%
\bibitem [{\citenamefont {Willett}\ \emph
  {et~al.}(1993{\natexlab{b}})\citenamefont {Willett}, \citenamefont {Ruel},
  \citenamefont {Paalanen}, \citenamefont {West},\ and\ \citenamefont
  {Pfeiffer}}]{Willett1993_CFLevidence_SAW}%
  \BibitemOpen
  \bibfield  {author} {\bibinfo {author} {\bibfnamefont {R.~L.}\ \bibnamefont
  {Willett}}, \bibinfo {author} {\bibfnamefont {R.~R.}\ \bibnamefont {Ruel}},
  \bibinfo {author} {\bibfnamefont {M.~A.}\ \bibnamefont {Paalanen}}, \bibinfo
  {author} {\bibfnamefont {K.~W.}\ \bibnamefont {West}},\ and\ \bibinfo
  {author} {\bibfnamefont {L.~N.}\ \bibnamefont {Pfeiffer}},\ }\bibfield
  {title} {\bibinfo {title} {Enhanced finite-wave-vector conductivity at
  multiple even-denominator filling factors in two-dimensional electron
  systems},\ }\href {https://doi.org/10.1103/PhysRevB.47.7344} {\bibfield
  {journal} {\bibinfo  {journal} {Phys. Rev. B}\ }\textbf {\bibinfo {volume}
  {47}},\ \bibinfo {pages} {7344} (\bibinfo {year}
  {1993}{\natexlab{b}})}\BibitemShut {NoStop}%
\bibitem [{\citenamefont {Goldman}\ \emph {et~al.}(1994)\citenamefont
  {Goldman}, \citenamefont {Su},\ and\ \citenamefont
  {Jain}}]{Goldman1994_CFLevidence}%
  \BibitemOpen
  \bibfield  {author} {\bibinfo {author} {\bibfnamefont {V.~J.}\ \bibnamefont
  {Goldman}}, \bibinfo {author} {\bibfnamefont {B.}~\bibnamefont {Su}},\ and\
  \bibinfo {author} {\bibfnamefont {J.~K.}\ \bibnamefont {Jain}},\ }\bibfield
  {title} {\bibinfo {title} {Detection of composite fermions by magnetic
  focusing},\ }\href {https://doi.org/10.1103/PhysRevLett.72.2065} {\bibfield
  {journal} {\bibinfo  {journal} {Phys. Rev. Lett.}\ }\textbf {\bibinfo
  {volume} {72}},\ \bibinfo {pages} {2065} (\bibinfo {year}
  {1994})}\BibitemShut {NoStop}%
\bibitem [{\citenamefont {Smet}\ \emph {et~al.}(1996)\citenamefont {Smet},
  \citenamefont {Weiss}, \citenamefont {Blick}, \citenamefont {L\"utjering},
  \citenamefont {von Klitzing}, \citenamefont {Fleischmann}, \citenamefont
  {Ketzmerick}, \citenamefont {Geisel},\ and\ \citenamefont
  {Weimann}}]{Smet1996_CFLevidence}%
  \BibitemOpen
  \bibfield  {author} {\bibinfo {author} {\bibfnamefont {J.~H.}\ \bibnamefont
  {Smet}}, \bibinfo {author} {\bibfnamefont {D.}~\bibnamefont {Weiss}},
  \bibinfo {author} {\bibfnamefont {R.~H.}\ \bibnamefont {Blick}}, \bibinfo
  {author} {\bibfnamefont {G.}~\bibnamefont {L\"utjering}}, \bibinfo {author}
  {\bibfnamefont {K.}~\bibnamefont {von Klitzing}}, \bibinfo {author}
  {\bibfnamefont {R.}~\bibnamefont {Fleischmann}}, \bibinfo {author}
  {\bibfnamefont {R.}~\bibnamefont {Ketzmerick}}, \bibinfo {author}
  {\bibfnamefont {T.}~\bibnamefont {Geisel}},\ and\ \bibinfo {author}
  {\bibfnamefont {G.}~\bibnamefont {Weimann}},\ }\bibfield  {title} {\bibinfo
  {title} {Magnetic focusing of composite fermions through arrays of
  cavities},\ }\href {https://doi.org/10.1103/PhysRevLett.77.2272} {\bibfield
  {journal} {\bibinfo  {journal} {Phys. Rev. Lett.}\ }\textbf {\bibinfo
  {volume} {77}},\ \bibinfo {pages} {2272} (\bibinfo {year}
  {1996})}\BibitemShut {NoStop}%
\bibitem [{\citenamefont {Laughlin}(1983)}]{Laughlin1983_FQHtheory}%
  \BibitemOpen
  \bibfield  {author} {\bibinfo {author} {\bibfnamefont {R.~B.}\ \bibnamefont
  {Laughlin}},\ }\bibfield  {title} {\bibinfo {title} {Anomalous quantum {H}all
  effect: An incompressible quantum fluid with fractionally charged
  excitations},\ }\href {https://doi.org/10.1103/PhysRevLett.50.1395}
  {\bibfield  {journal} {\bibinfo  {journal} {Phys. Rev. Lett.}\ }\textbf
  {\bibinfo {volume} {50}},\ \bibinfo {pages} {1395} (\bibinfo {year}
  {1983})}\BibitemShut {NoStop}%
\bibitem [{\citenamefont {Halperin}(1984)}]{Halperin1984_anyonFQH}%
  \BibitemOpen
  \bibfield  {author} {\bibinfo {author} {\bibfnamefont {B.~I.}\ \bibnamefont
  {Halperin}},\ }\bibfield  {title} {\bibinfo {title} {Statistics of
  quasiparticles and the hierarchy of fractional quantized {H}all states},\
  }\href {https://doi.org/10.1103/PhysRevLett.52.1583} {\bibfield  {journal}
  {\bibinfo  {journal} {Phys. Rev. Lett.}\ }\textbf {\bibinfo {volume} {52}},\
  \bibinfo {pages} {1583} (\bibinfo {year} {1984})}\BibitemShut {NoStop}%
\bibitem [{\citenamefont {{Willett}}\ \emph {et~al.}(1987)\citenamefont
  {{Willett}}, \citenamefont {{Eisenstein}}, \citenamefont {{Stormer}},
  \citenamefont {{Gossard}},\ and\ \citenamefont
  {{Tsui}}}]{willett1987_GaAsPf}%
  \BibitemOpen
  \bibfield  {author} {\bibinfo {author} {\bibfnamefont {R.}~\bibnamefont
  {{Willett}}}, \bibinfo {author} {\bibfnamefont {J.~P.}\ \bibnamefont
  {{Eisenstein}}}, \bibinfo {author} {\bibfnamefont {H.~L.}\ \bibnamefont
  {{Stormer}}}, \bibinfo {author} {\bibfnamefont {A.~C.}\ \bibnamefont
  {{Gossard}}},\ and\ \bibinfo {author} {\bibfnamefont {D.~C.}\ \bibnamefont
  {{Tsui}}},\ }\bibfield  {title} {\bibinfo {title} {{Observation of an
  even-denominator quantum number in the fractional quantum {H}all effect}},\
  }\href {https://doi.org/10.1103/PhysRevLett.59.1776} {\bibfield  {journal}
  {\bibinfo  {journal} {\prl}\ }\textbf {\bibinfo {volume} {59}},\ \bibinfo
  {pages} {1776} (\bibinfo {year} {1987})}\BibitemShut {NoStop}%
\bibitem [{\citenamefont {{Li}}\ \emph
  {et~al.}(2017{\natexlab{a}})\citenamefont {{Li}}, \citenamefont {{Tan}},
  \citenamefont {{Chen}}, \citenamefont {{Zeng}}, \citenamefont {{Taniguchi}},
  \citenamefont {{Watanabe}}, \citenamefont {{Hone}},\ and\ \citenamefont
  {{Dean}}}]{Li2017_graphenePf}%
  \BibitemOpen
  \bibfield  {author} {\bibinfo {author} {\bibfnamefont {J.~I.~A.}\
  \bibnamefont {{Li}}}, \bibinfo {author} {\bibfnamefont {C.}~\bibnamefont
  {{Tan}}}, \bibinfo {author} {\bibfnamefont {S.}~\bibnamefont {{Chen}}},
  \bibinfo {author} {\bibfnamefont {Y.}~\bibnamefont {{Zeng}}}, \bibinfo
  {author} {\bibfnamefont {T.}~\bibnamefont {{Taniguchi}}}, \bibinfo {author}
  {\bibfnamefont {K.}~\bibnamefont {{Watanabe}}}, \bibinfo {author}
  {\bibfnamefont {J.}~\bibnamefont {{Hone}}},\ and\ \bibinfo {author}
  {\bibfnamefont {C.~R.}\ \bibnamefont {{Dean}}},\ }\bibfield  {title}
  {\bibinfo {title} {{Even-denominator fractional quantum {H}all states in
  bilayer graphene}},\ }\href {https://doi.org/10.1126/science.aao2521}
  {\bibfield  {journal} {\bibinfo  {journal} {Science}\ }\textbf {\bibinfo
  {volume} {358}},\ \bibinfo {pages} {648} (\bibinfo {year}
  {2017}{\natexlab{a}})}\BibitemShut {NoStop}%
\bibitem [{\citenamefont {{Moore}}\ and\ \citenamefont
  {{Read}}(1991)}]{Moore1991_nonAbelian}%
  \BibitemOpen
  \bibfield  {author} {\bibinfo {author} {\bibfnamefont {G.}~\bibnamefont
  {{Moore}}}\ and\ \bibinfo {author} {\bibfnamefont {N.}~\bibnamefont
  {{Read}}},\ }\bibfield  {title} {\bibinfo {title} {{Nonabelions in the
  fractional quantum {H}all effect}},\ }\href
  {https://doi.org/10.1016/0550-3213(91)90407-O} {\bibfield  {journal}
  {\bibinfo  {journal} {Nuclear Physics B}\ }\textbf {\bibinfo {volume}
  {360}},\ \bibinfo {pages} {362} (\bibinfo {year} {1991})}\BibitemShut
  {NoStop}%
\bibitem [{\citenamefont {{Read}}\ and\ \citenamefont
  {{Green}}(2000)}]{Read1999_pair}%
  \BibitemOpen
  \bibfield  {author} {\bibinfo {author} {\bibfnamefont {N.}~\bibnamefont
  {{Read}}}\ and\ \bibinfo {author} {\bibfnamefont {D.}~\bibnamefont
  {{Green}}},\ }\bibfield  {title} {\bibinfo {title} {{Paired states of
  fermions in two dimensions with breaking of parity and time-reversal
  symmetries and the fractional quantum {H}all effect}},\ }\href
  {https://doi.org/10.1103/PhysRevB.61.10267} {\bibfield  {journal} {\bibinfo
  {journal} {\prb}\ }\textbf {\bibinfo {volume} {61}},\ \bibinfo {pages}
  {10267} (\bibinfo {year} {2000})}\BibitemShut {NoStop}%
\bibitem [{\citenamefont {{Ivanov}}(2001)}]{Ivanov2000_pwave_nonabelian}%
  \BibitemOpen
  \bibfield  {author} {\bibinfo {author} {\bibfnamefont {D.~A.}\ \bibnamefont
  {{Ivanov}}},\ }\bibfield  {title} {\bibinfo {title} {{Non-Abelian Statistics
  of Half-Quantum Vortices in p-Wave Superconductors}},\ }\href
  {https://doi.org/10.1103/PhysRevLett.86.268} {\bibfield  {journal} {\bibinfo
  {journal} {\prl}\ }\textbf {\bibinfo {volume} {86}},\ \bibinfo {pages} {268}
  (\bibinfo {year} {2001})}\BibitemShut {NoStop}%
\bibitem [{\citenamefont {Foster}\ \emph {et~al.}(2003)\citenamefont {Foster},
  \citenamefont {Bonesteel},\ and\ \citenamefont
  {Simon}}]{Foster2003_ConductivityMR}%
  \BibitemOpen
  \bibfield  {author} {\bibinfo {author} {\bibfnamefont {K.~C.}\ \bibnamefont
  {Foster}}, \bibinfo {author} {\bibfnamefont {N.~E.}\ \bibnamefont
  {Bonesteel}},\ and\ \bibinfo {author} {\bibfnamefont {S.~H.}\ \bibnamefont
  {Simon}},\ }\bibfield  {title} {\bibinfo {title} {Conductivity of paired
  composite fermions},\ }\href {https://doi.org/10.1103/PhysRevLett.91.046804}
  {\bibfield  {journal} {\bibinfo  {journal} {Phys. Rev. Lett.}\ }\textbf
  {\bibinfo {volume} {91}},\ \bibinfo {pages} {046804} (\bibinfo {year}
  {2003})}\BibitemShut {NoStop}%
\bibitem [{\citenamefont {Parameswaran}\ \emph {et~al.}(2011)\citenamefont
  {Parameswaran}, \citenamefont {Kivelson}, \citenamefont {Sondhi},\ and\
  \citenamefont {Spivak}}]{Parameswaran2011_Type1SC}%
  \BibitemOpen
  \bibfield  {author} {\bibinfo {author} {\bibfnamefont {S.~A.}\ \bibnamefont
  {Parameswaran}}, \bibinfo {author} {\bibfnamefont {S.~A.}\ \bibnamefont
  {Kivelson}}, \bibinfo {author} {\bibfnamefont {S.~L.}\ \bibnamefont
  {Sondhi}},\ and\ \bibinfo {author} {\bibfnamefont {B.~Z.}\ \bibnamefont
  {Spivak}},\ }\bibfield  {title} {\bibinfo {title} {Weakly coupled pfaffian as
  a type i quantum {H}all liquid},\ }\href
  {https://doi.org/10.1103/PhysRevLett.106.236801} {\bibfield  {journal}
  {\bibinfo  {journal} {Phys. Rev. Lett.}\ }\textbf {\bibinfo {volume} {106}},\
  \bibinfo {pages} {236801} (\bibinfo {year} {2011})}\BibitemShut {NoStop}%
\bibitem [{\citenamefont {Wang}\ \emph {et~al.}(2014)\citenamefont {Wang},
  \citenamefont {Mandal}, \citenamefont {Chung},\ and\ \citenamefont
  {Chakravarty}}]{Wang2014_HLRpairing}%
  \BibitemOpen
  \bibfield  {author} {\bibinfo {author} {\bibfnamefont {Z.}~\bibnamefont
  {Wang}}, \bibinfo {author} {\bibfnamefont {I.}~\bibnamefont {Mandal}},
  \bibinfo {author} {\bibfnamefont {S.~B.}\ \bibnamefont {Chung}},\ and\
  \bibinfo {author} {\bibfnamefont {S.}~\bibnamefont {Chakravarty}},\
  }\bibfield  {title} {\bibinfo {title} {Pairing in half-filled {L}andau
  level},\ }\href {https://doi.org/https://doi.org/10.1016/j.aop.2014.09.021}
  {\bibfield  {journal} {\bibinfo  {journal} {Annals of Physics}\ }\textbf
  {\bibinfo {volume} {351}},\ \bibinfo {pages} {727} (\bibinfo {year}
  {2014})}\BibitemShut {NoStop}%
\bibitem [{\citenamefont {{Park}}\ \emph {et~al.}(2023)\citenamefont {{Park}},
  \citenamefont {{Cai}}, \citenamefont {{Anderson}}, \citenamefont {{Zhang}},
  \citenamefont {{Zhu}}, \citenamefont {{Liu}}, \citenamefont {{Wang}},
  \citenamefont {{Holtzmann}}, \citenamefont {{Hu}}, \citenamefont {{Liu}},
  \citenamefont {{Taniguchi}}, \citenamefont {{Watanabe}}, \citenamefont
  {{Chu}}, \citenamefont {{Cao}}, \citenamefont {{Fu}}, \citenamefont {{Yao}},
  \citenamefont {{Chang}}, \citenamefont {{Cobden}}, \citenamefont {{Xiao}},\
  and\ \citenamefont {{Xu}}}]{Park2023_FQAH_TMD}%
  \BibitemOpen
  \bibfield  {author} {\bibinfo {author} {\bibfnamefont {H.}~\bibnamefont
  {{Park}}}, \bibinfo {author} {\bibfnamefont {J.}~\bibnamefont {{Cai}}},
  \bibinfo {author} {\bibfnamefont {E.}~\bibnamefont {{Anderson}}}, \bibinfo
  {author} {\bibfnamefont {Y.}~\bibnamefont {{Zhang}}}, \bibinfo {author}
  {\bibfnamefont {J.}~\bibnamefont {{Zhu}}}, \bibinfo {author} {\bibfnamefont
  {X.}~\bibnamefont {{Liu}}}, \bibinfo {author} {\bibfnamefont
  {C.}~\bibnamefont {{Wang}}}, \bibinfo {author} {\bibfnamefont
  {W.}~\bibnamefont {{Holtzmann}}}, \bibinfo {author} {\bibfnamefont
  {C.}~\bibnamefont {{Hu}}}, \bibinfo {author} {\bibfnamefont {Z.}~\bibnamefont
  {{Liu}}}, \bibinfo {author} {\bibfnamefont {T.}~\bibnamefont {{Taniguchi}}},
  \bibinfo {author} {\bibfnamefont {K.}~\bibnamefont {{Watanabe}}}, \bibinfo
  {author} {\bibfnamefont {J.-H.}\ \bibnamefont {{Chu}}}, \bibinfo {author}
  {\bibfnamefont {T.}~\bibnamefont {{Cao}}}, \bibinfo {author} {\bibfnamefont
  {L.}~\bibnamefont {{Fu}}}, \bibinfo {author} {\bibfnamefont {W.}~\bibnamefont
  {{Yao}}}, \bibinfo {author} {\bibfnamefont {C.-Z.}\ \bibnamefont {{Chang}}},
  \bibinfo {author} {\bibfnamefont {D.}~\bibnamefont {{Cobden}}}, \bibinfo
  {author} {\bibfnamefont {D.}~\bibnamefont {{Xiao}}},\ and\ \bibinfo {author}
  {\bibfnamefont {X.}~\bibnamefont {{Xu}}},\ }\bibfield  {title} {\bibinfo
  {title} {{Observation of fractionally quantized anomalous {H}all effect}},\
  }\href {https://doi.org/10.1038/s41586-023-06536-0} {\bibfield  {journal}
  {\bibinfo  {journal} {\nat}\ }\textbf {\bibinfo {volume} {622}},\ \bibinfo
  {pages} {74} (\bibinfo {year} {2023})}\BibitemShut {NoStop}%
\bibitem [{\citenamefont {{Lu}}\ \emph {et~al.}(2024)\citenamefont {{Lu}},
  \citenamefont {{Han}}, \citenamefont {{Yao}}, \citenamefont {{Reddy}},
  \citenamefont {{Yang}}, \citenamefont {{Seo}}, \citenamefont {{Watanabe}},
  \citenamefont {{Taniguchi}}, \citenamefont {{Fu}},\ and\ \citenamefont
  {{Ju}}}]{Lu2023_FQAHPenta}%
  \BibitemOpen
  \bibfield  {author} {\bibinfo {author} {\bibfnamefont {Z.}~\bibnamefont
  {{Lu}}}, \bibinfo {author} {\bibfnamefont {T.}~\bibnamefont {{Han}}},
  \bibinfo {author} {\bibfnamefont {Y.}~\bibnamefont {{Yao}}}, \bibinfo
  {author} {\bibfnamefont {A.~P.}\ \bibnamefont {{Reddy}}}, \bibinfo {author}
  {\bibfnamefont {J.}~\bibnamefont {{Yang}}}, \bibinfo {author} {\bibfnamefont
  {J.}~\bibnamefont {{Seo}}}, \bibinfo {author} {\bibfnamefont
  {K.}~\bibnamefont {{Watanabe}}}, \bibinfo {author} {\bibfnamefont
  {T.}~\bibnamefont {{Taniguchi}}}, \bibinfo {author} {\bibfnamefont
  {L.}~\bibnamefont {{Fu}}},\ and\ \bibinfo {author} {\bibfnamefont
  {L.}~\bibnamefont {{Ju}}},\ }\bibfield  {title} {\bibinfo {title}
  {{Fractional quantum anomalous {H}all effect in multilayer graphene}},\
  }\href {https://doi.org/10.1038/s41586-023-07010-7} {\bibfield  {journal}
  {\bibinfo  {journal} {\nat}\ }\textbf {\bibinfo {volume} {626}},\ \bibinfo
  {pages} {759} (\bibinfo {year} {2024})}\BibitemShut {NoStop}%
\bibitem [{\citenamefont {{Dong}}\ \emph {et~al.}(2023)\citenamefont {{Dong}},
  \citenamefont {{Wang}}, \citenamefont {{Ledwith}}, \citenamefont
  {{Vishwanath}},\ and\ \citenamefont {{Parker}}}]{Dong2023_ACFL}%
  \BibitemOpen
  \bibfield  {author} {\bibinfo {author} {\bibfnamefont {J.}~\bibnamefont
  {{Dong}}}, \bibinfo {author} {\bibfnamefont {J.}~\bibnamefont {{Wang}}},
  \bibinfo {author} {\bibfnamefont {P.~J.}\ \bibnamefont {{Ledwith}}}, \bibinfo
  {author} {\bibfnamefont {A.}~\bibnamefont {{Vishwanath}}},\ and\ \bibinfo
  {author} {\bibfnamefont {D.~E.}\ \bibnamefont {{Parker}}},\ }\bibfield
  {title} {\bibinfo {title} {{Composite {F}ermi Liquid at Zero Magnetic Field
  in Twisted {MoTe}$_{2}$}},\ }\href
  {https://doi.org/10.1103/PhysRevLett.131.136502} {\bibfield  {journal}
  {\bibinfo  {journal} {\prl}\ }\textbf {\bibinfo {volume} {131}},\ \bibinfo
  {eid} {136502} (\bibinfo {year} {2023})}\BibitemShut {NoStop}%
\bibitem [{\citenamefont {{Goldman}}\ \emph {et~al.}(2023)\citenamefont
  {{Goldman}}, \citenamefont {{Reddy}}, \citenamefont {{Paul}},\ and\
  \citenamefont {{Fu}}}]{Goldman2023_ACFL}%
  \BibitemOpen
  \bibfield  {author} {\bibinfo {author} {\bibfnamefont {H.}~\bibnamefont
  {{Goldman}}}, \bibinfo {author} {\bibfnamefont {A.~P.}\ \bibnamefont
  {{Reddy}}}, \bibinfo {author} {\bibfnamefont {N.}~\bibnamefont {{Paul}}},\
  and\ \bibinfo {author} {\bibfnamefont {L.}~\bibnamefont {{Fu}}},\ }\bibfield
  {title} {\bibinfo {title} {{Zero-Field Composite {F}ermi Liquid in Twisted
  Semiconductor Bilayers}},\ }\href
  {https://doi.org/10.1103/PhysRevLett.131.136501} {\bibfield  {journal}
  {\bibinfo  {journal} {\prl}\ }\textbf {\bibinfo {volume} {131}},\ \bibinfo
  {eid} {136501} (\bibinfo {year} {2023})}\BibitemShut {NoStop}%
\bibitem [{\citenamefont {{Cai}}\ \emph {et~al.}(2023)\citenamefont {{Cai}},
  \citenamefont {{Anderson}}, \citenamefont {{Wang}}, \citenamefont {{Zhang}},
  \citenamefont {{Liu}}, \citenamefont {{Holtzmann}}, \citenamefont {{Zhang}},
  \citenamefont {{Fan}}, \citenamefont {{Taniguchi}}, \citenamefont
  {{Watanabe}}, \citenamefont {{Ran}}, \citenamefont {{Cao}}, \citenamefont
  {{Fu}}, \citenamefont {{Xiao}}, \citenamefont {{Yao}},\ and\ \citenamefont
  {{Xu}}}]{Cai2023_FQAHTMD}%
  \BibitemOpen
  \bibfield  {author} {\bibinfo {author} {\bibfnamefont {J.}~\bibnamefont
  {{Cai}}}, \bibinfo {author} {\bibfnamefont {E.}~\bibnamefont {{Anderson}}},
  \bibinfo {author} {\bibfnamefont {C.}~\bibnamefont {{Wang}}}, \bibinfo
  {author} {\bibfnamefont {X.}~\bibnamefont {{Zhang}}}, \bibinfo {author}
  {\bibfnamefont {X.}~\bibnamefont {{Liu}}}, \bibinfo {author} {\bibfnamefont
  {W.}~\bibnamefont {{Holtzmann}}}, \bibinfo {author} {\bibfnamefont
  {Y.}~\bibnamefont {{Zhang}}}, \bibinfo {author} {\bibfnamefont
  {F.}~\bibnamefont {{Fan}}}, \bibinfo {author} {\bibfnamefont
  {T.}~\bibnamefont {{Taniguchi}}}, \bibinfo {author} {\bibfnamefont
  {K.}~\bibnamefont {{Watanabe}}}, \bibinfo {author} {\bibfnamefont
  {Y.}~\bibnamefont {{Ran}}}, \bibinfo {author} {\bibfnamefont
  {T.}~\bibnamefont {{Cao}}}, \bibinfo {author} {\bibfnamefont
  {L.}~\bibnamefont {{Fu}}}, \bibinfo {author} {\bibfnamefont {D.}~\bibnamefont
  {{Xiao}}}, \bibinfo {author} {\bibfnamefont {W.}~\bibnamefont {{Yao}}},\ and\
  \bibinfo {author} {\bibfnamefont {X.}~\bibnamefont {{Xu}}},\ }\bibfield
  {title} {\bibinfo {title} {{Signatures of fractional quantum anomalous {H}all
  states in twisted {MoTe}$_{2}$}},\ }\href
  {https://doi.org/10.1038/s41586-023-06289-w} {\bibfield  {journal} {\bibinfo
  {journal} {\nat}\ }\textbf {\bibinfo {volume} {622}},\ \bibinfo {pages} {63}
  (\bibinfo {year} {2023})}\BibitemShut {NoStop}%
\bibitem [{\citenamefont {{Xu}}\ \emph {et~al.}(2023)\citenamefont {{Xu}},
  \citenamefont {{Sun}}, \citenamefont {{Jia}}, \citenamefont {{Liu}},
  \citenamefont {{Xu}}, \citenamefont {{Li}}, \citenamefont {{Gu}},
  \citenamefont {{Watanabe}}, \citenamefont {{Taniguchi}}, \citenamefont
  {{Tong}}, \citenamefont {{Jia}}, \citenamefont {{Shi}}, \citenamefont
  {{Jiang}}, \citenamefont {{Zhang}}, \citenamefont {{Liu}},\ and\
  \citenamefont {{Li}}}]{Xu2023_FQAHTMD}%
  \BibitemOpen
  \bibfield  {author} {\bibinfo {author} {\bibfnamefont {F.}~\bibnamefont
  {{Xu}}}, \bibinfo {author} {\bibfnamefont {Z.}~\bibnamefont {{Sun}}},
  \bibinfo {author} {\bibfnamefont {T.}~\bibnamefont {{Jia}}}, \bibinfo
  {author} {\bibfnamefont {C.}~\bibnamefont {{Liu}}}, \bibinfo {author}
  {\bibfnamefont {C.}~\bibnamefont {{Xu}}}, \bibinfo {author} {\bibfnamefont
  {C.}~\bibnamefont {{Li}}}, \bibinfo {author} {\bibfnamefont {Y.}~\bibnamefont
  {{Gu}}}, \bibinfo {author} {\bibfnamefont {K.}~\bibnamefont {{Watanabe}}},
  \bibinfo {author} {\bibfnamefont {T.}~\bibnamefont {{Taniguchi}}}, \bibinfo
  {author} {\bibfnamefont {B.}~\bibnamefont {{Tong}}}, \bibinfo {author}
  {\bibfnamefont {J.}~\bibnamefont {{Jia}}}, \bibinfo {author} {\bibfnamefont
  {Z.}~\bibnamefont {{Shi}}}, \bibinfo {author} {\bibfnamefont
  {S.}~\bibnamefont {{Jiang}}}, \bibinfo {author} {\bibfnamefont
  {Y.}~\bibnamefont {{Zhang}}}, \bibinfo {author} {\bibfnamefont
  {X.}~\bibnamefont {{Liu}}},\ and\ \bibinfo {author} {\bibfnamefont
  {T.}~\bibnamefont {{Li}}},\ }\bibfield  {title} {\bibinfo {title}
  {{Observation of Integer and Fractional Quantum Anomalous {H}all Effects in
  Twisted Bilayer {MoTe}$_{2}$}},\ }\href
  {https://doi.org/10.1103/PhysRevX.13.031037} {\bibfield  {journal} {\bibinfo
  {journal} {Physical Review X}\ }\textbf {\bibinfo {volume} {13}},\ \bibinfo
  {eid} {031037} (\bibinfo {year} {2023})}\BibitemShut {NoStop}%
\bibitem [{\citenamefont {Zeng}\ \emph {et~al.}(2023)\citenamefont {Zeng},
  \citenamefont {Xia}, \citenamefont {Kang}, \citenamefont {Zhu}, \citenamefont
  {Kn{\"u}ppel}, \citenamefont {Vaswani}, \citenamefont {Watanabe},
  \citenamefont {Taniguchi}, \citenamefont {Mak},\ and\ \citenamefont
  {Shan}}]{Zeng2023_FQAHTMD}%
  \BibitemOpen
  \bibfield  {author} {\bibinfo {author} {\bibfnamefont {Y.}~\bibnamefont
  {Zeng}}, \bibinfo {author} {\bibfnamefont {Z.}~\bibnamefont {Xia}}, \bibinfo
  {author} {\bibfnamefont {K.}~\bibnamefont {Kang}}, \bibinfo {author}
  {\bibfnamefont {J.}~\bibnamefont {Zhu}}, \bibinfo {author} {\bibfnamefont
  {P.}~\bibnamefont {Kn{\"u}ppel}}, \bibinfo {author} {\bibfnamefont
  {C.}~\bibnamefont {Vaswani}}, \bibinfo {author} {\bibfnamefont
  {K.}~\bibnamefont {Watanabe}}, \bibinfo {author} {\bibfnamefont
  {T.}~\bibnamefont {Taniguchi}}, \bibinfo {author} {\bibfnamefont {K.~F.}\
  \bibnamefont {Mak}},\ and\ \bibinfo {author} {\bibfnamefont {J.}~\bibnamefont
  {Shan}},\ }\bibfield  {title} {\bibinfo {title} {Thermodynamic evidence of
  fractional {C}hern insulator in moir{\'e}{MoTe}$_2$},\ }\href
  {https://doi.org/10.1038/s41586-023-06452-3} {\bibfield  {journal} {\bibinfo
  {journal} {Nature}\ }\textbf {\bibinfo {volume} {622}},\ \bibinfo {pages}
  {69} (\bibinfo {year} {2023})}\BibitemShut {NoStop}%
\bibitem [{\citenamefont {{Lu}}\ \emph {et~al.}(2025)\citenamefont {{Lu}},
  \citenamefont {{Han}}, \citenamefont {{Yao}}, \citenamefont {{Hadjri}},
  \citenamefont {{Yang}}, \citenamefont {{Seo}}, \citenamefont {{Shi}},
  \citenamefont {{Ye}}, \citenamefont {{Watanabe}}, \citenamefont
  {{Taniguchi}},\ and\ \citenamefont {{Ju}}}]{Lu2025_EQAH}%
  \BibitemOpen
  \bibfield  {author} {\bibinfo {author} {\bibfnamefont {Z.}~\bibnamefont
  {{Lu}}}, \bibinfo {author} {\bibfnamefont {T.}~\bibnamefont {{Han}}},
  \bibinfo {author} {\bibfnamefont {Y.}~\bibnamefont {{Yao}}}, \bibinfo
  {author} {\bibfnamefont {Z.}~\bibnamefont {{Hadjri}}}, \bibinfo {author}
  {\bibfnamefont {J.}~\bibnamefont {{Yang}}}, \bibinfo {author} {\bibfnamefont
  {J.}~\bibnamefont {{Seo}}}, \bibinfo {author} {\bibfnamefont
  {L.}~\bibnamefont {{Shi}}}, \bibinfo {author} {\bibfnamefont
  {S.}~\bibnamefont {{Ye}}}, \bibinfo {author} {\bibfnamefont {K.}~\bibnamefont
  {{Watanabe}}}, \bibinfo {author} {\bibfnamefont {T.}~\bibnamefont
  {{Taniguchi}}},\ and\ \bibinfo {author} {\bibfnamefont {L.}~\bibnamefont
  {{Ju}}},\ }\bibfield  {title} {\bibinfo {title} {{Extended quantum anomalous
  {H}all states in graphene/hBN moir{\'e} superlattices}},\ }\href
  {https://doi.org/10.1038/s41586-024-08470-1} {\bibfield  {journal} {\bibinfo
  {journal} {\nat}\ }\textbf {\bibinfo {volume} {637}},\ \bibinfo {pages}
  {1090} (\bibinfo {year} {2025})}\BibitemShut {NoStop}%
\bibitem [{Note1()}]{Note1}%
  \BibitemOpen
  \bibinfo {note} {A related paired state of spinons in gapless spin liquids
  has been considered in Ref.~\cite {Barkeshli2012_spinonBFS}.}\BibitemShut
  {Stop}%
\bibitem [{\citenamefont {Wu}\ \emph {et~al.}(2019)\citenamefont {Wu},
  \citenamefont {Lovorn}, \citenamefont {Tutuc}, \citenamefont {Martin},\ and\
  \citenamefont {MacDonald}}]{Wu2019_TMD_foundational}%
  \BibitemOpen
  \bibfield  {author} {\bibinfo {author} {\bibfnamefont {F.}~\bibnamefont
  {Wu}}, \bibinfo {author} {\bibfnamefont {T.}~\bibnamefont {Lovorn}}, \bibinfo
  {author} {\bibfnamefont {E.}~\bibnamefont {Tutuc}}, \bibinfo {author}
  {\bibfnamefont {I.}~\bibnamefont {Martin}},\ and\ \bibinfo {author}
  {\bibfnamefont {A.~H.}\ \bibnamefont {MacDonald}},\ }\bibfield  {title}
  {\bibinfo {title} {Topological insulators in twisted transition metal
  dichalcogenide homobilayers},\ }\href
  {https://doi.org/10.1103/PhysRevLett.122.086402} {\bibfield  {journal}
  {\bibinfo  {journal} {Phys. Rev. Lett.}\ }\textbf {\bibinfo {volume} {122}},\
  \bibinfo {pages} {086402} (\bibinfo {year} {2019})}\BibitemShut {NoStop}%
\bibitem [{\citenamefont {Dong}\ \emph
  {et~al.}(2024{\natexlab{a}})\citenamefont {Dong}, \citenamefont {Patri},\
  and\ \citenamefont {Senthil}}]{Dong2023_FQAHpenta_MIT}%
  \BibitemOpen
  \bibfield  {author} {\bibinfo {author} {\bibfnamefont {Z.}~\bibnamefont
  {Dong}}, \bibinfo {author} {\bibfnamefont {A.~S.}\ \bibnamefont {Patri}},\
  and\ \bibinfo {author} {\bibfnamefont {T.}~\bibnamefont {Senthil}},\
  }\bibfield  {title} {\bibinfo {title} {Theory of quantum anomalous {H}all
  phases in pentalayer rhombohedral graphene moir\'e structures},\ }\href
  {https://doi.org/10.1103/PhysRevLett.133.206502} {\bibfield  {journal}
  {\bibinfo  {journal} {Phys. Rev. Lett.}\ }\textbf {\bibinfo {volume} {133}},\
  \bibinfo {pages} {206502} (\bibinfo {year} {2024}{\natexlab{a}})}\BibitemShut
  {NoStop}%
\bibitem [{\citenamefont {Dong}\ \emph
  {et~al.}(2024{\natexlab{b}})\citenamefont {Dong}, \citenamefont {Wang},
  \citenamefont {Wang}, \citenamefont {Soejima}, \citenamefont {Zaletel},
  \citenamefont {Vishwanath},\ and\ \citenamefont
  {Parker}}]{Dong2023_FQAHpenta_harvard}%
  \BibitemOpen
  \bibfield  {author} {\bibinfo {author} {\bibfnamefont {J.}~\bibnamefont
  {Dong}}, \bibinfo {author} {\bibfnamefont {T.}~\bibnamefont {Wang}}, \bibinfo
  {author} {\bibfnamefont {T.}~\bibnamefont {Wang}}, \bibinfo {author}
  {\bibfnamefont {T.}~\bibnamefont {Soejima}}, \bibinfo {author} {\bibfnamefont
  {M.~P.}\ \bibnamefont {Zaletel}}, \bibinfo {author} {\bibfnamefont
  {A.}~\bibnamefont {Vishwanath}},\ and\ \bibinfo {author} {\bibfnamefont
  {D.~E.}\ \bibnamefont {Parker}},\ }\bibfield  {title} {\bibinfo {title}
  {Anomalous {H}all crystals in rhombohedral multilayer graphene. i.
  interaction-driven {C}hern bands and fractional quantum {H}all states at zero
  magnetic field},\ }\href {https://doi.org/10.1103/PhysRevLett.133.206503}
  {\bibfield  {journal} {\bibinfo  {journal} {Phys. Rev. Lett.}\ }\textbf
  {\bibinfo {volume} {133}},\ \bibinfo {pages} {206503} (\bibinfo {year}
  {2024}{\natexlab{b}})}\BibitemShut {NoStop}%
\bibitem [{\citenamefont {Zhou}\ \emph {et~al.}(2024)\citenamefont {Zhou},
  \citenamefont {Yang},\ and\ \citenamefont {Zhang}}]{Zhou2023_FQAHpenta}%
  \BibitemOpen
  \bibfield  {author} {\bibinfo {author} {\bibfnamefont {B.}~\bibnamefont
  {Zhou}}, \bibinfo {author} {\bibfnamefont {H.}~\bibnamefont {Yang}},\ and\
  \bibinfo {author} {\bibfnamefont {Y.-H.}\ \bibnamefont {Zhang}},\ }\bibfield
  {title} {\bibinfo {title} {Fractional quantum anomalous {H}all effect in
  rhombohedral multilayer graphene in the moir\'eless limit},\ }\href
  {https://doi.org/10.1103/PhysRevLett.133.206504} {\bibfield  {journal}
  {\bibinfo  {journal} {Phys. Rev. Lett.}\ }\textbf {\bibinfo {volume} {133}},\
  \bibinfo {pages} {206504} (\bibinfo {year} {2024})}\BibitemShut {NoStop}%
\bibitem [{\citenamefont {{Barkeshli}}\ and\ \citenamefont
  {{McGreevy}}(2012)}]{Barkeshli2012_latticeCFL_FL}%
  \BibitemOpen
  \bibfield  {author} {\bibinfo {author} {\bibfnamefont {M.}~\bibnamefont
  {{Barkeshli}}}\ and\ \bibinfo {author} {\bibfnamefont {J.}~\bibnamefont
  {{McGreevy}}},\ }\bibfield  {title} {\bibinfo {title} {{Continuous
  transitions between composite {F}ermi liquid and {Landau Fermi} liquid: A
  route to fractionalized {M}ott insulators}},\ }\href
  {https://doi.org/10.1103/PhysRevB.86.075136} {\bibfield  {journal} {\bibinfo
  {journal} {\prb}\ }\textbf {\bibinfo {volume} {86}},\ \bibinfo {eid} {075136}
  (\bibinfo {year} {2012})}\BibitemShut {NoStop}%
\bibitem [{Note2()}]{Note2}%
  \BibitemOpen
  \bibinfo {note} {With broken inversion symmetry, the ACFL phase itself
  exhibits a number of exotic properties that go beyond the Landau level CFL,
  which we explore in a companion paper~\cite {Nosov2026_C3ACFL}.}\BibitemShut
  {Stop}%
\bibitem [{\citenamefont {{Allais}}\ and\ \citenamefont
  {{Senthil}}(2012)}]{Allais2012_BFS_loopcurrent}%
  \BibitemOpen
  \bibfield  {author} {\bibinfo {author} {\bibfnamefont {A.}~\bibnamefont
  {{Allais}}}\ and\ \bibinfo {author} {\bibfnamefont {T.}~\bibnamefont
  {{Senthil}}},\ }\bibfield  {title} {\bibinfo {title} {{Loop current order and
  d-wave superconductivity: Some observable consequences}},\ }\href
  {https://doi.org/10.1103/PhysRevB.86.045118} {\bibfield  {journal} {\bibinfo
  {journal} {\prb}\ }\textbf {\bibinfo {volume} {86}},\ \bibinfo {eid} {045118}
  (\bibinfo {year} {2012})}\BibitemShut {NoStop}%
\bibitem [{\citenamefont {{Agterberg}}\ \emph {et~al.}(2017)\citenamefont
  {{Agterberg}}, \citenamefont {{Brydon}},\ and\ \citenamefont
  {{Timm}}}]{Agterberg2017_BFS}%
  \BibitemOpen
  \bibfield  {author} {\bibinfo {author} {\bibfnamefont {D.~F.}\ \bibnamefont
  {{Agterberg}}}, \bibinfo {author} {\bibfnamefont {P.~M.~R.}\ \bibnamefont
  {{Brydon}}},\ and\ \bibinfo {author} {\bibfnamefont {C.}~\bibnamefont
  {{Timm}}},\ }\bibfield  {title} {\bibinfo {title} {{Bogoliubov {F}ermi
  Surfaces in Superconductors with Broken Time-Reversal Symmetry}},\ }\href
  {https://doi.org/10.1103/PhysRevLett.118.127001} {\bibfield  {journal}
  {\bibinfo  {journal} {\prl}\ }\textbf {\bibinfo {volume} {118}},\ \bibinfo
  {eid} {127001} (\bibinfo {year} {2017})}\BibitemShut {NoStop}%
\bibitem [{\citenamefont {{Brydon}}\ \emph {et~al.}(2018)\citenamefont
  {{Brydon}}, \citenamefont {{Agterberg}}, \citenamefont {{Menke}},\ and\
  \citenamefont {{Timm}}}]{Brydon2018_BFS}%
  \BibitemOpen
  \bibfield  {author} {\bibinfo {author} {\bibfnamefont {P.~M.~R.}\
  \bibnamefont {{Brydon}}}, \bibinfo {author} {\bibfnamefont {D.~F.}\
  \bibnamefont {{Agterberg}}}, \bibinfo {author} {\bibfnamefont
  {H.}~\bibnamefont {{Menke}}},\ and\ \bibinfo {author} {\bibfnamefont
  {C.}~\bibnamefont {{Timm}}},\ }\bibfield  {title} {\bibinfo {title}
  {{Bogoliubov {F}ermi surfaces: General theory, magnetic order, and
  topology}},\ }\href {https://doi.org/10.1103/PhysRevB.98.224509} {\bibfield
  {journal} {\bibinfo  {journal} {\prb}\ }\textbf {\bibinfo {volume} {98}},\
  \bibinfo {eid} {224509} (\bibinfo {year} {2018})}\BibitemShut {NoStop}%
\bibitem [{\citenamefont {{Lapp}}\ \emph {et~al.}(2020)\citenamefont {{Lapp}},
  \citenamefont {{B{\"o}rner}},\ and\ \citenamefont {{Timm}}}]{Lapp2020_BFS}%
  \BibitemOpen
  \bibfield  {author} {\bibinfo {author} {\bibfnamefont {C.~J.}\ \bibnamefont
  {{Lapp}}}, \bibinfo {author} {\bibfnamefont {G.}~\bibnamefont
  {{B{\"o}rner}}},\ and\ \bibinfo {author} {\bibfnamefont {C.}~\bibnamefont
  {{Timm}}},\ }\bibfield  {title} {\bibinfo {title} {{Experimental consequences
  of {B}ogoliubov {F}ermi surfaces}},\ }\href
  {https://doi.org/10.1103/PhysRevB.101.024505} {\bibfield  {journal} {\bibinfo
   {journal} {\prb}\ }\textbf {\bibinfo {volume} {101}},\ \bibinfo {eid}
  {024505} (\bibinfo {year} {2020})}\BibitemShut {NoStop}%
\bibitem [{\citenamefont {{Oh}}\ \emph {et~al.}(2021)\citenamefont {{Oh}},
  \citenamefont {{Agterberg}},\ and\ \citenamefont
  {{Moon}}}]{Oh2021_BFS_disorder}%
  \BibitemOpen
  \bibfield  {author} {\bibinfo {author} {\bibfnamefont {H.}~\bibnamefont
  {{Oh}}}, \bibinfo {author} {\bibfnamefont {D.~F.}\ \bibnamefont
  {{Agterberg}}},\ and\ \bibinfo {author} {\bibfnamefont {E.-G.}\ \bibnamefont
  {{Moon}}},\ }\bibfield  {title} {\bibinfo {title} {{Using Disorder to
  Identify {B}ogoliubov {F}ermi-Surface States}},\ }\href
  {https://doi.org/10.1103/PhysRevLett.127.257002} {\bibfield  {journal}
  {\bibinfo  {journal} {\prl}\ }\textbf {\bibinfo {volume} {127}},\ \bibinfo
  {eid} {257002} (\bibinfo {year} {2021})}\BibitemShut {NoStop}%
\bibitem [{\citenamefont {{Bhattacharya}}\ and\ \citenamefont
  {{Timm}}(2023)}]{Bhattacharya2023_BFS_stability}%
  \BibitemOpen
  \bibfield  {author} {\bibinfo {author} {\bibfnamefont {A.}~\bibnamefont
  {{Bhattacharya}}}\ and\ \bibinfo {author} {\bibfnamefont {C.}~\bibnamefont
  {{Timm}}},\ }\bibfield  {title} {\bibinfo {title} {{Stability of {B}ogoliubov
  {F}ermi surfaces within {BCS} theory}},\ }\href
  {https://doi.org/10.1103/PhysRevB.107.L220501} {\bibfield  {journal}
  {\bibinfo  {journal} {\prb}\ }\textbf {\bibinfo {volume} {107}},\ \bibinfo
  {eid} {L220501} (\bibinfo {year} {2023})}\BibitemShut {NoStop}%
\bibitem [{\citenamefont {{Pal}}\ \emph
  {et~al.}(2024{\natexlab{a}})\citenamefont {{Pal}}, \citenamefont {{Saha}},\
  and\ \citenamefont {{Dutta}}}]{Pal2024_BFS}%
  \BibitemOpen
  \bibfield  {author} {\bibinfo {author} {\bibfnamefont {A.}~\bibnamefont
  {{Pal}}}, \bibinfo {author} {\bibfnamefont {A.}~\bibnamefont {{Saha}}},\ and\
  \bibinfo {author} {\bibfnamefont {P.}~\bibnamefont {{Dutta}}},\ }\bibfield
  {title} {\bibinfo {title} {{Transport signatures of {Bogoliubov Fermi}
  surfaces in normal metal/time-reversal symmetry broken d-wave superconductor
  junctions}},\ }\href {https://doi.org/10.1088/1367-2630/ad481a} {\bibfield
  {journal} {\bibinfo  {journal} {New Journal of Physics}\ }\textbf {\bibinfo
  {volume} {26}},\ \bibinfo {eid} {053027} (\bibinfo {year}
  {2024}{\natexlab{a}})}\BibitemShut {NoStop}%
\bibitem [{\citenamefont {Yang}\ and\ \citenamefont
  {Zhang}(2025)}]{Yang2025_BFSrhombohedral}%
  \BibitemOpen
  \bibfield  {author} {\bibinfo {author} {\bibfnamefont {H.}~\bibnamefont
  {Yang}}\ and\ \bibinfo {author} {\bibfnamefont {Y.-H.}\ \bibnamefont
  {Zhang}},\ }\bibfield  {title} {\bibinfo {title} {Topological incommensurate
  {F}ulde-{F}errell-{L}arkin-{O}vchinnikov superconductor and {B}ogoliubov
  {F}ermi surface in rhombohedral tetralayer graphene},\ }\href
  {https://doi.org/10.1103/k8s3-dgfs} {\bibfield  {journal} {\bibinfo
  {journal} {Phys. Rev. B}\ }\textbf {\bibinfo {volume} {112}},\ \bibinfo
  {pages} {L020506} (\bibinfo {year} {2025})}\BibitemShut {NoStop}%
\bibitem [{\citenamefont {Christos}\ \emph {et~al.}(2025)\citenamefont
  {Christos}, \citenamefont {Bonetti},\ and\ \citenamefont
  {Scheurer}}]{Christos2025_rhombohedralPairing}%
  \BibitemOpen
  \bibfield  {author} {\bibinfo {author} {\bibfnamefont {M.}~\bibnamefont
  {Christos}}, \bibinfo {author} {\bibfnamefont {P.~M.}\ \bibnamefont
  {Bonetti}},\ and\ \bibinfo {author} {\bibfnamefont {M.~S.}\ \bibnamefont
  {Scheurer}},\ }\bibfield  {title} {\bibinfo {title} {Finite-momentum pairing
  and superlattice superconductivity in valley-imbalanced rhombohedral
  graphene},\ }\href@noop {} {\bibfield  {journal} {\bibinfo  {journal} {arXiv
  preprint arXiv:2503.15471}\ } (\bibinfo {year} {2025})}\BibitemShut {NoStop}%
\bibitem [{\citenamefont {Gaggioli}\ \emph {et~al.}(2025)\citenamefont
  {Gaggioli}, \citenamefont {Guerci},\ and\ \citenamefont
  {Fu}}]{Gaggioli2025_RhombohedralVortex}%
  \BibitemOpen
  \bibfield  {author} {\bibinfo {author} {\bibfnamefont {F.}~\bibnamefont
  {Gaggioli}}, \bibinfo {author} {\bibfnamefont {D.}~\bibnamefont {Guerci}},\
  and\ \bibinfo {author} {\bibfnamefont {L.}~\bibnamefont {Fu}},\ }\bibfield
  {title} {\bibinfo {title} {Spontaneous vortex-antivortex lattice and majorana
  fermions in rhombohedral graphene},\ }\href
  {https://doi.org/10.1103/k8sb-rqxf} {\bibfield  {journal} {\bibinfo
  {journal} {Phys. Rev. Lett.}\ }\textbf {\bibinfo {volume} {135}},\ \bibinfo
  {pages} {116001} (\bibinfo {year} {2025})}\BibitemShut {NoStop}%
\bibitem [{\citenamefont {Kim}\ and\ \citenamefont
  {Wen}(1994)}]{Kim1994_monopole}%
  \BibitemOpen
  \bibfield  {author} {\bibinfo {author} {\bibfnamefont {Y.~B.}\ \bibnamefont
  {Kim}}\ and\ \bibinfo {author} {\bibfnamefont {X.-G.}\ \bibnamefont {Wen}},\
  }\bibfield  {title} {\bibinfo {title} {Instantons and the spectral function
  of electrons in the half-filled {L}andau level},\ }\href
  {https://doi.org/10.1103/PhysRevB.50.8078} {\bibfield  {journal} {\bibinfo
  {journal} {Phys. Rev. B}\ }\textbf {\bibinfo {volume} {50}},\ \bibinfo
  {pages} {8078} (\bibinfo {year} {1994})}\BibitemShut {NoStop}%
\bibitem [{\citenamefont {Chen}\ \emph {et~al.}(2025)\citenamefont {Chen},
  \citenamefont {Lu},\ and\ \citenamefont {Meng}}]{Chen2025_CFLtensor}%
  \BibitemOpen
  \bibfield  {author} {\bibinfo {author} {\bibfnamefont {B.-B.}\ \bibnamefont
  {Chen}}, \bibinfo {author} {\bibfnamefont {H.}~\bibnamefont {Lu}},\ and\
  \bibinfo {author} {\bibfnamefont {Z.~Y.}\ \bibnamefont {Meng}},\ }\bibfield
  {title} {\bibinfo {title} {Probing non-{F}ermi-liquid behaviour of composite
  {F}ermi liquid via efficient thermal simulations},\ }\href@noop {} {\bibfield
   {journal} {\bibinfo  {journal} {arXiv preprint arXiv:2509.02218}\ }
  (\bibinfo {year} {2025})}\BibitemShut {NoStop}%
\bibitem [{\citenamefont {Ioffe}\ and\ \citenamefont
  {Larkin}(1989)}]{Ioffe1989_rule}%
  \BibitemOpen
  \bibfield  {author} {\bibinfo {author} {\bibfnamefont {L.~B.}\ \bibnamefont
  {Ioffe}}\ and\ \bibinfo {author} {\bibfnamefont {A.~I.}\ \bibnamefont
  {Larkin}},\ }\bibfield  {title} {\bibinfo {title} {Gapless fermions and gauge
  fields in dielectrics},\ }\href {https://doi.org/10.1103/PhysRevB.39.8988}
  {\bibfield  {journal} {\bibinfo  {journal} {Phys. Rev. B}\ }\textbf {\bibinfo
  {volume} {39}},\ \bibinfo {pages} {8988} (\bibinfo {year}
  {1989})}\BibitemShut {NoStop}%
\bibitem [{\citenamefont {{Pal}}\ \emph
  {et~al.}(2024{\natexlab{b}})\citenamefont {{Pal}}, \citenamefont {{Dutta}},\
  and\ \citenamefont {{Saha}}}]{Pal2024_thermoelectricBFS}%
  \BibitemOpen
  \bibfield  {author} {\bibinfo {author} {\bibfnamefont {A.}~\bibnamefont
  {{Pal}}}, \bibinfo {author} {\bibfnamefont {P.}~\bibnamefont {{Dutta}}},\
  and\ \bibinfo {author} {\bibfnamefont {A.}~\bibnamefont {{Saha}}},\
  }\bibfield  {title} {\bibinfo {title} {{Identifying {Bogoliubov Fermi}
  surfaces via thermoelectric response in a $d$-wave superconductor
  heterostructure}},\ }\href {https://doi.org/10.48550/arXiv.2409.12157}
  {\bibfield  {journal} {\bibinfo  {journal} {arXiv e-prints}\ ,\ \bibinfo
  {eid} {arXiv:2409.12157}} (\bibinfo {year} {2024}{\natexlab{b}})}\BibitemShut
  {NoStop}%
\bibitem [{Note3()}]{Note3}%
  \BibitemOpen
  \bibinfo {note} {A similar Landau damping of Goldstone modes has previously
  appeared in the context of nematic Fermi fluids~\cite
  {Oganesyan2001_nematic}.}\BibitemShut {Stop}%
\bibitem [{Note4()}]{Note4}%
  \BibitemOpen
  \bibinfo {note} {The existence of the additional logarithm on top of the
  $q^3$ dependence beyond RPA was recently questioned in Ref.~\cite
  {Anakru2025_noLogCFL}.}\BibitemShut {Stop}%
\bibitem [{\citenamefont {Franz}\ and\ \citenamefont
  {Tesanovic}(2000)}]{Franz2000_vortex}%
  \BibitemOpen
  \bibfield  {author} {\bibinfo {author} {\bibfnamefont {M.}~\bibnamefont
  {Franz}}\ and\ \bibinfo {author} {\bibfnamefont {Z.}~\bibnamefont
  {Tesanovic}},\ }\bibfield  {title} {\bibinfo {title} {Quasiparticles in the
  vortex lattice of unconventional superconductors: {B}loch waves or {L}andau
  levels?},\ }\href {https://doi.org/10.1103/PhysRevLett.84.554} {\bibfield
  {journal} {\bibinfo  {journal} {Phys. Rev. Lett.}\ }\textbf {\bibinfo
  {volume} {84}},\ \bibinfo {pages} {554} (\bibinfo {year} {2000})}\BibitemShut
  {NoStop}%
\bibitem [{\citenamefont {{Levin}}\ and\ \citenamefont
  {{Halperin}}(2009)}]{Levin2009_daughter}%
  \BibitemOpen
  \bibfield  {author} {\bibinfo {author} {\bibfnamefont {M.}~\bibnamefont
  {{Levin}}}\ and\ \bibinfo {author} {\bibfnamefont {B.~I.}\ \bibnamefont
  {{Halperin}}},\ }\bibfield  {title} {\bibinfo {title} {{Collective states of
  non-Abelian quasiparticles in a magnetic field}},\ }\href
  {https://doi.org/10.1103/PhysRevB.79.205301} {\bibfield  {journal} {\bibinfo
  {journal} {\prb}\ }\textbf {\bibinfo {volume} {79}},\ \bibinfo {eid} {205301}
  (\bibinfo {year} {2009})}\BibitemShut {NoStop}%
\bibitem [{\citenamefont {{Yutushui}}\ \emph {et~al.}(2024)\citenamefont
  {{Yutushui}}, \citenamefont {{Hermanns}},\ and\ \citenamefont
  {{Mross}}}]{Yutushui2024_daughter}%
  \BibitemOpen
  \bibfield  {author} {\bibinfo {author} {\bibfnamefont {M.}~\bibnamefont
  {{Yutushui}}}, \bibinfo {author} {\bibfnamefont {M.}~\bibnamefont
  {{Hermanns}}},\ and\ \bibinfo {author} {\bibfnamefont {D.~F.}\ \bibnamefont
  {{Mross}}},\ }\bibfield  {title} {\bibinfo {title} {{Paired fermions in
  strong magnetic fields and daughters of even-denominator {H}all plateaus}},\
  }\href {https://doi.org/10.1103/PhysRevB.110.165402} {\bibfield  {journal}
  {\bibinfo  {journal} {\prb}\ }\textbf {\bibinfo {volume} {110}},\ \bibinfo
  {eid} {165402} (\bibinfo {year} {2024})}\BibitemShut {NoStop}%
\bibitem [{\citenamefont {{Zheltonozhskii}}\ \emph {et~al.}(2024)\citenamefont
  {{Zheltonozhskii}}, \citenamefont {{Stern}},\ and\ \citenamefont
  {{Lindner}}}]{Zheltonozhskii2024_daughter}%
  \BibitemOpen
  \bibfield  {author} {\bibinfo {author} {\bibfnamefont {E.}~\bibnamefont
  {{Zheltonozhskii}}}, \bibinfo {author} {\bibfnamefont {A.}~\bibnamefont
  {{Stern}}},\ and\ \bibinfo {author} {\bibfnamefont {N.~H.}\ \bibnamefont
  {{Lindner}}},\ }\bibfield  {title} {\bibinfo {title} {{Identifying the
  topological order of quantized half-filled {L}andau levels through their
  daughter states}},\ }\href {https://doi.org/10.1103/PhysRevB.110.245140}
  {\bibfield  {journal} {\bibinfo  {journal} {\prb}\ }\textbf {\bibinfo
  {volume} {110}},\ \bibinfo {eid} {245140} (\bibinfo {year}
  {2024})}\BibitemShut {NoStop}%
\bibitem [{\citenamefont {{Zhang}}\ \emph {et~al.}(2025)\citenamefont
  {{Zhang}}, \citenamefont {{Vishwanath}},\ and\ \citenamefont
  {{Wen}}}]{Zhang2025_daughter}%
  \BibitemOpen
  \bibfield  {author} {\bibinfo {author} {\bibfnamefont {C.}~\bibnamefont
  {{Zhang}}}, \bibinfo {author} {\bibfnamefont {A.}~\bibnamefont
  {{Vishwanath}}},\ and\ \bibinfo {author} {\bibfnamefont {X.-G.}\ \bibnamefont
  {{Wen}}},\ }\bibfield  {title} {\bibinfo {title} {{Hierarchy construction for
  non-Abelian fractional quantum Hall states via anyon condensation}},\ }\href
  {https://doi.org/10.1103/jndb-435f} {\bibfield  {journal} {\bibinfo
  {journal} {\prb}\ }\textbf {\bibinfo {volume} {112}},\ \bibinfo {eid}
  {125116} (\bibinfo {year} {2025})}\BibitemShut {NoStop}%
\bibitem [{\citenamefont {{Li}}\ \emph
  {et~al.}(2017{\natexlab{b}})\citenamefont {{Li}}, \citenamefont {{Tan}},
  \citenamefont {{Chen}}, \citenamefont {{Zeng}}, \citenamefont {{Taniguchi}},
  \citenamefont {{Watanabe}}, \citenamefont {{Hone}},\ and\ \citenamefont
  {{Dean}}}]{Li2017_daughter_exp}%
  \BibitemOpen
  \bibfield  {author} {\bibinfo {author} {\bibfnamefont {J.~I.~A.}\
  \bibnamefont {{Li}}}, \bibinfo {author} {\bibfnamefont {C.}~\bibnamefont
  {{Tan}}}, \bibinfo {author} {\bibfnamefont {S.}~\bibnamefont {{Chen}}},
  \bibinfo {author} {\bibfnamefont {Y.}~\bibnamefont {{Zeng}}}, \bibinfo
  {author} {\bibfnamefont {T.}~\bibnamefont {{Taniguchi}}}, \bibinfo {author}
  {\bibfnamefont {K.}~\bibnamefont {{Watanabe}}}, \bibinfo {author}
  {\bibfnamefont {J.}~\bibnamefont {{Hone}}},\ and\ \bibinfo {author}
  {\bibfnamefont {C.~R.}\ \bibnamefont {{Dean}}},\ }\bibfield  {title}
  {\bibinfo {title} {{Even-denominator fractional quantum {H}all states in
  bilayer graphene}},\ }\href {https://doi.org/10.1126/science.aao2521}
  {\bibfield  {journal} {\bibinfo  {journal} {Science}\ }\textbf {\bibinfo
  {volume} {358}},\ \bibinfo {pages} {648} (\bibinfo {year}
  {2017}{\natexlab{b}})}\BibitemShut {NoStop}%
\bibitem [{\citenamefont {Zibrov}\ \emph {et~al.}(2017)\citenamefont {Zibrov},
  \citenamefont {Kometter}, \citenamefont {Zhou}, \citenamefont {Spanton},
  \citenamefont {Taniguchi}, \citenamefont {Watanabe}, \citenamefont
  {Zaletel},\ and\ \citenamefont {Young}}]{Zibrov2017_daughter_exp}%
  \BibitemOpen
  \bibfield  {author} {\bibinfo {author} {\bibfnamefont {A.~A.}\ \bibnamefont
  {Zibrov}}, \bibinfo {author} {\bibfnamefont {C.}~\bibnamefont {Kometter}},
  \bibinfo {author} {\bibfnamefont {H.}~\bibnamefont {Zhou}}, \bibinfo {author}
  {\bibfnamefont {E.~M.}\ \bibnamefont {Spanton}}, \bibinfo {author}
  {\bibfnamefont {T.}~\bibnamefont {Taniguchi}}, \bibinfo {author}
  {\bibfnamefont {K.}~\bibnamefont {Watanabe}}, \bibinfo {author}
  {\bibfnamefont {M.~P.}\ \bibnamefont {Zaletel}},\ and\ \bibinfo {author}
  {\bibfnamefont {A.~F.}\ \bibnamefont {Young}},\ }\bibfield  {title} {\bibinfo
  {title} {Tunable interacting composite fermion phases in a half-filled
  bilayer-graphene {L}andau level},\ }\href
  {https://doi.org/10.1038/nature23893} {\bibfield  {journal} {\bibinfo
  {journal} {Nature}\ }\textbf {\bibinfo {volume} {549}},\ \bibinfo {pages}
  {360} (\bibinfo {year} {2017})}\BibitemShut {NoStop}%
\bibitem [{\citenamefont {{Huang}}\ \emph {et~al.}(2022)\citenamefont
  {{Huang}}, \citenamefont {{Fu}}, \citenamefont {{Hickey}}, \citenamefont
  {{Alem}}, \citenamefont {{Lin}}, \citenamefont {{Watanabe}}, \citenamefont
  {{Taniguchi}},\ and\ \citenamefont {{Zhu}}}]{Huang2021_daughter_exp}%
  \BibitemOpen
  \bibfield  {author} {\bibinfo {author} {\bibfnamefont {K.}~\bibnamefont
  {{Huang}}}, \bibinfo {author} {\bibfnamefont {H.}~\bibnamefont {{Fu}}},
  \bibinfo {author} {\bibfnamefont {D.~R.}\ \bibnamefont {{Hickey}}}, \bibinfo
  {author} {\bibfnamefont {N.}~\bibnamefont {{Alem}}}, \bibinfo {author}
  {\bibfnamefont {X.}~\bibnamefont {{Lin}}}, \bibinfo {author} {\bibfnamefont
  {K.}~\bibnamefont {{Watanabe}}}, \bibinfo {author} {\bibfnamefont
  {T.}~\bibnamefont {{Taniguchi}}},\ and\ \bibinfo {author} {\bibfnamefont
  {J.}~\bibnamefont {{Zhu}}},\ }\bibfield  {title} {\bibinfo {title} {{Valley
  Isospin Controlled Fractional Quantum {H}all States in Bilayer Graphene}},\
  }\href {https://doi.org/10.1103/PhysRevX.12.031019} {\bibfield  {journal}
  {\bibinfo  {journal} {Physical Review X}\ }\textbf {\bibinfo {volume} {12}},\
  \bibinfo {eid} {031019} (\bibinfo {year} {2022})}\BibitemShut {NoStop}%
\bibitem [{\citenamefont {{Singh}}\ \emph {et~al.}(2024)\citenamefont
  {{Singh}}, \citenamefont {{Wang}}, \citenamefont {{Tai}}, \citenamefont
  {{Calhoun}}, \citenamefont {{Villegas Rosales}}, \citenamefont {{Madathil}},
  \citenamefont {{Gupta}}, \citenamefont {{Baldwin}}, \citenamefont
  {{Pfeiffer}},\ and\ \citenamefont {{Shayegan}}}]{Singh2023_daughter_exp}%
  \BibitemOpen
  \bibfield  {author} {\bibinfo {author} {\bibfnamefont {S.~K.}\ \bibnamefont
  {{Singh}}}, \bibinfo {author} {\bibfnamefont {C.}~\bibnamefont {{Wang}}},
  \bibinfo {author} {\bibfnamefont {C.~T.}\ \bibnamefont {{Tai}}}, \bibinfo
  {author} {\bibfnamefont {C.~S.}\ \bibnamefont {{Calhoun}}}, \bibinfo {author}
  {\bibfnamefont {K.~A.}\ \bibnamefont {{Villegas Rosales}}}, \bibinfo {author}
  {\bibfnamefont {P.~T.}\ \bibnamefont {{Madathil}}}, \bibinfo {author}
  {\bibfnamefont {A.}~\bibnamefont {{Gupta}}}, \bibinfo {author} {\bibfnamefont
  {K.~W.}\ \bibnamefont {{Baldwin}}}, \bibinfo {author} {\bibfnamefont {L.~N.}\
  \bibnamefont {{Pfeiffer}}},\ and\ \bibinfo {author} {\bibfnamefont
  {M.}~\bibnamefont {{Shayegan}}},\ }\bibfield  {title} {\bibinfo {title}
  {{Topological phase transition between {J}ain states and daughter states of
  the {\ensuremath{\nu}} = 1/2 fractional quantum {H}all state}},\ }\href
  {https://doi.org/10.1038/s41567-024-02517-w} {\bibfield  {journal} {\bibinfo
  {journal} {Nature Physics}\ }\textbf {\bibinfo {volume} {20}},\ \bibinfo
  {pages} {1247} (\bibinfo {year} {2024})}\BibitemShut {NoStop}%
\bibitem [{Note5()}]{Note5}%
  \BibitemOpen
  \bibinfo {note} {This general route of obtaining itinerant phases from doping
  FQAH insulators has been explored recently in Refs.~\cite
  {Shi2024_doping,Kim2024_anyonSC,Divic2024_HofHubb,Kuhlenkamp2025_HofHubb,Shi2025_dopeMR,Nosov2025_plateau,Shi2025_anyon_delocalization,Pichler2025_anyonSC}.}\BibitemShut
  {Stop}%
\bibitem [{\citenamefont {{Darius Shi}}\ \emph {et~al.}(2025)\citenamefont
  {{Darius Shi}}, \citenamefont {{Zhang}},\ and\ \citenamefont
  {{Senthil}}}]{Shi2025_dopeMR}%
  \BibitemOpen
  \bibfield  {author} {\bibinfo {author} {\bibfnamefont {Z.}~\bibnamefont
  {{Darius Shi}}}, \bibinfo {author} {\bibfnamefont {C.}~\bibnamefont
  {{Zhang}}},\ and\ \bibinfo {author} {\bibfnamefont {T.}~\bibnamefont
  {{Senthil}}},\ }\bibfield  {title} {\bibinfo {title} {{Doping lattice
  non-abelian quantum {H}all states}},\ }\href
  {https://doi.org/10.48550/arXiv.2505.02893} {\bibfield  {journal} {\bibinfo
  {journal} {arXiv e-prints}\ ,\ \bibinfo {eid} {arXiv:2505.02893}} (\bibinfo
  {year} {2025})}\BibitemShut {NoStop}%
\bibitem [{\citenamefont {Behnia}\ and\ \citenamefont
  {Aubin}(2016)}]{Behnia2016_thermalReview}%
  \BibitemOpen
  \bibfield  {author} {\bibinfo {author} {\bibfnamefont {K.}~\bibnamefont
  {Behnia}}\ and\ \bibinfo {author} {\bibfnamefont {H.}~\bibnamefont {Aubin}},\
  }\bibfield  {title} {\bibinfo {title} {Nernst effect in metals and
  superconductors: a review of concepts and experiments},\ }\href
  {https://doi.org/10.1088/0034-4885/79/4/046502} {\bibfield  {journal}
  {\bibinfo  {journal} {Reports on Progress in Physics}\ }\textbf {\bibinfo
  {volume} {79}},\ \bibinfo {pages} {046502} (\bibinfo {year}
  {2016})}\BibitemShut {NoStop}%
\bibitem [{\citenamefont {Silaev}\ \emph {et~al.}(2015)\citenamefont {Silaev},
  \citenamefont {Garaud},\ and\ \citenamefont
  {Babaev}}]{Silaev2015_thermoelectricTRSB}%
  \BibitemOpen
  \bibfield  {author} {\bibinfo {author} {\bibfnamefont {M.}~\bibnamefont
  {Silaev}}, \bibinfo {author} {\bibfnamefont {J.}~\bibnamefont {Garaud}},\
  and\ \bibinfo {author} {\bibfnamefont {E.}~\bibnamefont {Babaev}},\
  }\bibfield  {title} {\bibinfo {title} {Unconventional thermoelectric effect
  in superconductors that break time-reversal symmetry},\ }\href
  {https://doi.org/10.1103/PhysRevB.92.174510} {\bibfield  {journal} {\bibinfo
  {journal} {Phys. Rev. B}\ }\textbf {\bibinfo {volume} {92}},\ \bibinfo
  {pages} {174510} (\bibinfo {year} {2015})}\BibitemShut {NoStop}%
\bibitem [{\citenamefont {Van~Harlingen}\ \emph {et~al.}(1980)\citenamefont
  {Van~Harlingen}, \citenamefont {Heidel},\ and\ \citenamefont
  {Garland}}]{VanHarlingen1980_thermoelectricity_SC}%
  \BibitemOpen
  \bibfield  {author} {\bibinfo {author} {\bibfnamefont {D.~J.}\ \bibnamefont
  {Van~Harlingen}}, \bibinfo {author} {\bibfnamefont {D.~F.}\ \bibnamefont
  {Heidel}},\ and\ \bibinfo {author} {\bibfnamefont {J.~C.}\ \bibnamefont
  {Garland}},\ }\bibfield  {title} {\bibinfo {title} {Experimental study of
  thermoelectricity in superconducting indium},\ }\href
  {https://doi.org/10.1103/PhysRevB.21.1842} {\bibfield  {journal} {\bibinfo
  {journal} {Phys. Rev. B}\ }\textbf {\bibinfo {volume} {21}},\ \bibinfo
  {pages} {1842} (\bibinfo {year} {1980})}\BibitemShut {NoStop}%
\bibitem [{\citenamefont {Cooper}\ \emph {et~al.}(1997)\citenamefont {Cooper},
  \citenamefont {Halperin},\ and\ \citenamefont
  {Ruzin}}]{Cooper1997_thermoelectric}%
  \BibitemOpen
  \bibfield  {author} {\bibinfo {author} {\bibfnamefont {N.~R.}\ \bibnamefont
  {Cooper}}, \bibinfo {author} {\bibfnamefont {B.~I.}\ \bibnamefont
  {Halperin}},\ and\ \bibinfo {author} {\bibfnamefont {I.~M.}\ \bibnamefont
  {Ruzin}},\ }\bibfield  {title} {\bibinfo {title} {Thermoelectric response of
  an interacting two-dimensional electron gas in a quantizing magnetic field},\
  }\href {https://doi.org/10.1103/PhysRevB.55.2344} {\bibfield  {journal}
  {\bibinfo  {journal} {Phys. Rev. B}\ }\textbf {\bibinfo {volume} {55}},\
  \bibinfo {pages} {2344} (\bibinfo {year} {1997})}\BibitemShut {NoStop}%
\bibitem [{\citenamefont {Potter}\ \emph {et~al.}(2016)\citenamefont {Potter},
  \citenamefont {Serbyn},\ and\ \citenamefont
  {Vishwanath}}]{Potter2016_thermoelectricCFL}%
  \BibitemOpen
  \bibfield  {author} {\bibinfo {author} {\bibfnamefont {A.~C.}\ \bibnamefont
  {Potter}}, \bibinfo {author} {\bibfnamefont {M.}~\bibnamefont {Serbyn}},\
  and\ \bibinfo {author} {\bibfnamefont {A.}~\bibnamefont {Vishwanath}},\
  }\bibfield  {title} {\bibinfo {title} {Thermoelectric transport signatures of
  {D}irac composite fermions in the half-filled {L}andau level},\ }\href
  {https://doi.org/10.1103/PhysRevX.6.031026} {\bibfield  {journal} {\bibinfo
  {journal} {Phys. Rev. X}\ }\textbf {\bibinfo {volume} {6}},\ \bibinfo {pages}
  {031026} (\bibinfo {year} {2016})}\BibitemShut {NoStop}%
\bibitem [{\citenamefont {Wang}\ \emph {et~al.}(2017)\citenamefont {Wang},
  \citenamefont {Cooper}, \citenamefont {Halperin},\ and\ \citenamefont
  {Stern}}]{Wang2017_CFLthermal}%
  \BibitemOpen
  \bibfield  {author} {\bibinfo {author} {\bibfnamefont {C.}~\bibnamefont
  {Wang}}, \bibinfo {author} {\bibfnamefont {N.~R.}\ \bibnamefont {Cooper}},
  \bibinfo {author} {\bibfnamefont {B.~I.}\ \bibnamefont {Halperin}},\ and\
  \bibinfo {author} {\bibfnamefont {A.}~\bibnamefont {Stern}},\ }\bibfield
  {title} {\bibinfo {title} {Particle-hole symmetry in the
  {F}ermion-{C}hern-{S}imons and {D}irac descriptions of a half-filled {L}andau
  level},\ }\href {https://doi.org/10.1103/PhysRevX.7.031029} {\bibfield
  {journal} {\bibinfo  {journal} {Phys. Rev. X}\ }\textbf {\bibinfo {volume}
  {7}},\ \bibinfo {pages} {031029} (\bibinfo {year} {2017})}\BibitemShut
  {NoStop}%
\bibitem [{\citenamefont {{Santos}}\ \emph {et~al.}(2019)\citenamefont
  {{Santos}}, \citenamefont {{Wang}},\ and\ \citenamefont
  {{Fradkin}}}]{Santos2018_CFPDW}%
  \BibitemOpen
  \bibfield  {author} {\bibinfo {author} {\bibfnamefont {L.~H.}\ \bibnamefont
  {{Santos}}}, \bibinfo {author} {\bibfnamefont {Y.}~\bibnamefont {{Wang}}},\
  and\ \bibinfo {author} {\bibfnamefont {E.}~\bibnamefont {{Fradkin}}},\
  }\bibfield  {title} {\bibinfo {title} {{Pair-Density-Wave Order and Paired
  Fractional Quantum Hall Fluids}},\ }\href
  {https://doi.org/10.1103/PhysRevX.9.021047} {\bibfield  {journal} {\bibinfo
  {journal} {Physical Review X}\ }\textbf {\bibinfo {volume} {9}},\ \bibinfo
  {eid} {021047} (\bibinfo {year} {2019})}\BibitemShut {NoStop}%
\bibitem [{\citenamefont {{Nosov}}\ and\ \citenamefont
  {{Shi}}(2026)}]{Nosov2026_C3ACFL}%
  \BibitemOpen
  \bibfield  {author} {\bibinfo {author} {\bibfnamefont {P.~A.}\ \bibnamefont
  {{Nosov}}}\ and\ \bibinfo {author} {\bibfnamefont {Z.~D.}\ \bibnamefont
  {{Shi}}},\ }\bibfield  {title} {\bibinfo {title} {Anomalous composite {F}ermi
  liquid with broken inversion symmetry}} (\bibinfo {year} {2026}),\ \bibinfo
  {note} {to appear}\BibitemShut {NoStop}%
\bibitem [{\citenamefont {Kivelson}\ \emph {et~al.}(1997)\citenamefont
  {Kivelson}, \citenamefont {Lee}, \citenamefont {Krotov},\ and\ \citenamefont
  {Gan}}]{Kivelson1997_HallCF}%
  \BibitemOpen
  \bibfield  {author} {\bibinfo {author} {\bibfnamefont {S.~A.}\ \bibnamefont
  {Kivelson}}, \bibinfo {author} {\bibfnamefont {D.-H.}\ \bibnamefont {Lee}},
  \bibinfo {author} {\bibfnamefont {Y.}~\bibnamefont {Krotov}},\ and\ \bibinfo
  {author} {\bibfnamefont {J.}~\bibnamefont {Gan}},\ }\bibfield  {title}
  {\bibinfo {title} {Composite-fermion hall conductance at
  \ensuremath{\nu}=1/2},\ }\href {https://doi.org/10.1103/PhysRevB.55.15552}
  {\bibfield  {journal} {\bibinfo  {journal} {Phys. Rev. B}\ }\textbf {\bibinfo
  {volume} {55}},\ \bibinfo {pages} {15552} (\bibinfo {year}
  {1997})}\BibitemShut {NoStop}%
\bibitem [{\citenamefont {Kumar}\ \emph {et~al.}(2019)\citenamefont {Kumar},
  \citenamefont {Raghu},\ and\ \citenamefont {Mulligan}}]{Kumar2019_HallCF}%
  \BibitemOpen
  \bibfield  {author} {\bibinfo {author} {\bibfnamefont {P.}~\bibnamefont
  {Kumar}}, \bibinfo {author} {\bibfnamefont {S.}~\bibnamefont {Raghu}},\ and\
  \bibinfo {author} {\bibfnamefont {M.}~\bibnamefont {Mulligan}},\ }\bibfield
  {title} {\bibinfo {title} {Composite fermion {Hall} conductivity and the
  half-filled {Landau} level},\ }\href
  {https://doi.org/10.1103/PhysRevB.99.235114} {\bibfield  {journal} {\bibinfo
  {journal} {Phys. Rev. B}\ }\textbf {\bibinfo {volume} {99}},\ \bibinfo
  {pages} {235114} (\bibinfo {year} {2019})}\BibitemShut {NoStop}%
\bibitem [{\citenamefont {{Barkeshli}}\ \emph {et~al.}(2013)\citenamefont
  {{Barkeshli}}, \citenamefont {{Yao}},\ and\ \citenamefont
  {{Kivelson}}}]{Barkeshli2012_spinonBFS}%
  \BibitemOpen
  \bibfield  {author} {\bibinfo {author} {\bibfnamefont {M.}~\bibnamefont
  {{Barkeshli}}}, \bibinfo {author} {\bibfnamefont {H.}~\bibnamefont {{Yao}}},\
  and\ \bibinfo {author} {\bibfnamefont {S.~A.}\ \bibnamefont {{Kivelson}}},\
  }\bibfield  {title} {\bibinfo {title} {{Gapless spin liquids: Stability and
  possible experimental relevance}},\ }\href
  {https://doi.org/10.1103/PhysRevB.87.140402} {\bibfield  {journal} {\bibinfo
  {journal} {\prb}\ }\textbf {\bibinfo {volume} {87}},\ \bibinfo {eid} {140402}
  (\bibinfo {year} {2013})}\BibitemShut {NoStop}%
\bibitem [{\citenamefont {Oganesyan}\ \emph {et~al.}(2001)\citenamefont
  {Oganesyan}, \citenamefont {Kivelson},\ and\ \citenamefont
  {Fradkin}}]{Oganesyan2001_nematic}%
  \BibitemOpen
  \bibfield  {author} {\bibinfo {author} {\bibfnamefont {V.}~\bibnamefont
  {Oganesyan}}, \bibinfo {author} {\bibfnamefont {S.~A.}\ \bibnamefont
  {Kivelson}},\ and\ \bibinfo {author} {\bibfnamefont {E.}~\bibnamefont
  {Fradkin}},\ }\bibfield  {title} {\bibinfo {title} {Quantum theory of a
  nematic {F}ermi fluid},\ }\href {https://doi.org/10.1103/PhysRevB.64.195109}
  {\bibfield  {journal} {\bibinfo  {journal} {Phys. Rev. B}\ }\textbf {\bibinfo
  {volume} {64}},\ \bibinfo {pages} {195109} (\bibinfo {year}
  {2001})}\BibitemShut {NoStop}%
\bibitem [{\citenamefont {Anakru}\ \emph {et~al.}(2025)\citenamefont {Anakru},
  \citenamefont {Gattu}, \citenamefont {Balram}, \citenamefont {Wu},
  \citenamefont {Kumar}, \citenamefont {Bi},\ and\ \citenamefont
  {Jain}}]{Anakru2025_noLogCFL}%
  \BibitemOpen
  \bibfield  {author} {\bibinfo {author} {\bibfnamefont {A.}~\bibnamefont
  {Anakru}}, \bibinfo {author} {\bibfnamefont {M.}~\bibnamefont {Gattu}},
  \bibinfo {author} {\bibfnamefont {A.~C.}\ \bibnamefont {Balram}}, \bibinfo
  {author} {\bibfnamefont {X.-C.}\ \bibnamefont {Wu}}, \bibinfo {author}
  {\bibfnamefont {P.}~\bibnamefont {Kumar}}, \bibinfo {author} {\bibfnamefont
  {Z.}~\bibnamefont {Bi}},\ and\ \bibinfo {author} {\bibfnamefont
  {J.}~\bibnamefont {Jain}},\ }\bibfield  {title} {\bibinfo {title} {Exploring
  the nature of the emergent gauge field in composite-fermion metals: A
  large-scale microscopic study},\ }\href@noop {} {\bibfield  {journal}
  {\bibinfo  {journal} {arXiv preprint arXiv:2509.07151}\ } (\bibinfo {year}
  {2025})}\BibitemShut {NoStop}%
\bibitem [{\citenamefont {{Shi}}\ and\ \citenamefont
  {{Senthil}}(2025)}]{Shi2024_doping}%
  \BibitemOpen
  \bibfield  {author} {\bibinfo {author} {\bibfnamefont {Z.~D.}\ \bibnamefont
  {{Shi}}}\ and\ \bibinfo {author} {\bibfnamefont {T.}~\bibnamefont
  {{Senthil}}},\ }\bibfield  {title} {\bibinfo {title} {{Doping a Fractional
  Quantum Anomalous {H}all Insulator}},\ }\href
  {https://doi.org/10.1103/kcm5-hx56} {\bibfield  {journal} {\bibinfo
  {journal} {Physical Review X}\ }\textbf {\bibinfo {volume} {15}},\ \bibinfo
  {eid} {031069} (\bibinfo {year} {2025})}\BibitemShut {NoStop}%
\bibitem [{\citenamefont {{Kim}}\ \emph {et~al.}(2025)\citenamefont {{Kim}},
  \citenamefont {{Timmel}}, \citenamefont {{Ju}},\ and\ \citenamefont
  {{Wen}}}]{Kim2024_anyonSC}%
  \BibitemOpen
  \bibfield  {author} {\bibinfo {author} {\bibfnamefont {M.}~\bibnamefont
  {{Kim}}}, \bibinfo {author} {\bibfnamefont {A.}~\bibnamefont {{Timmel}}},
  \bibinfo {author} {\bibfnamefont {L.}~\bibnamefont {{Ju}}},\ and\ \bibinfo
  {author} {\bibfnamefont {X.-G.}\ \bibnamefont {{Wen}}},\ }\bibfield  {title}
  {\bibinfo {title} {{Topological chiral superconductivity beyond pairing in a
  {F}ermi liquid}},\ }\href {https://doi.org/10.1103/PhysRevB.111.014508}
  {\bibfield  {journal} {\bibinfo  {journal} {\prb}\ }\textbf {\bibinfo
  {volume} {111}},\ \bibinfo {eid} {014508} (\bibinfo {year}
  {2025})}\BibitemShut {NoStop}%
\bibitem [{\citenamefont {{Divic}}\ \emph {et~al.}(2025)\citenamefont
  {{Divic}}, \citenamefont {{Cr{\'e}pel}}, \citenamefont {{Soejima}},
  \citenamefont {{Song}}, \citenamefont {{Millis}}, \citenamefont {{Zaletel}},\
  and\ \citenamefont {{Vishwanath}}}]{Divic2024_HofHubb}%
  \BibitemOpen
  \bibfield  {author} {\bibinfo {author} {\bibfnamefont {S.}~\bibnamefont
  {{Divic}}}, \bibinfo {author} {\bibfnamefont {V.}~\bibnamefont
  {{Cr{\'e}pel}}}, \bibinfo {author} {\bibfnamefont {T.}~\bibnamefont
  {{Soejima}}}, \bibinfo {author} {\bibfnamefont {X.-Y.}\ \bibnamefont
  {{Song}}}, \bibinfo {author} {\bibfnamefont {A.~J.}\ \bibnamefont
  {{Millis}}}, \bibinfo {author} {\bibfnamefont {M.~P.}\ \bibnamefont
  {{Zaletel}}},\ and\ \bibinfo {author} {\bibfnamefont {A.}~\bibnamefont
  {{Vishwanath}}},\ }\bibfield  {title} {\bibinfo {title} {{Anyon
  superconductivity from topological criticality in a {Hofstadter-Hubbard}
  model}},\ }\href {https://doi.org/10.1073/pnas.2426680122} {\bibfield
  {journal} {\bibinfo  {journal} {Proceedings of the National Academy of
  Science}\ }\textbf {\bibinfo {volume} {122}},\ \bibinfo {eid} {e2426680122}
  (\bibinfo {year} {2025})}\BibitemShut {NoStop}%
\bibitem [{\citenamefont {{Kuhlenkamp}}\ \emph {et~al.}(2025)\citenamefont
  {{Kuhlenkamp}}, \citenamefont {{Divic}}, \citenamefont {{Zaletel}},
  \citenamefont {{Soejima}},\ and\ \citenamefont
  {{Vishwanath}}}]{Kuhlenkamp2025_HofHubb}%
  \BibitemOpen
  \bibfield  {author} {\bibinfo {author} {\bibfnamefont {C.}~\bibnamefont
  {{Kuhlenkamp}}}, \bibinfo {author} {\bibfnamefont {S.}~\bibnamefont
  {{Divic}}}, \bibinfo {author} {\bibfnamefont {M.~P.}\ \bibnamefont
  {{Zaletel}}}, \bibinfo {author} {\bibfnamefont {T.}~\bibnamefont
  {{Soejima}}},\ and\ \bibinfo {author} {\bibfnamefont {A.}~\bibnamefont
  {{Vishwanath}}},\ }\bibfield  {title} {\bibinfo {title} {{Robust
  superconductivity upon doping chiral spin liquid and {C}hern insulators in a
  {Hubbard-Hofstadter} model}},\ }\href
  {https://doi.org/10.48550/arXiv.2509.02675} {\bibfield  {journal} {\bibinfo
  {journal} {arXiv e-prints}\ ,\ \bibinfo {eid} {arXiv:2509.02675}} (\bibinfo
  {year} {2025})}\BibitemShut {NoStop}%
\bibitem [{\citenamefont {{Nosov}}\ \emph {et~al.}(2025)\citenamefont
  {{Nosov}}, \citenamefont {{Han}},\ and\ \citenamefont
  {{Khalaf}}}]{Nosov2025_plateau}%
  \BibitemOpen
  \bibfield  {author} {\bibinfo {author} {\bibfnamefont {P.~A.}\ \bibnamefont
  {{Nosov}}}, \bibinfo {author} {\bibfnamefont {Z.}~\bibnamefont {{Han}}},\
  and\ \bibinfo {author} {\bibfnamefont {E.}~\bibnamefont {{Khalaf}}},\
  }\bibfield  {title} {\bibinfo {title} {{Anyon superconductivity and plateau
  transitions in doped fractional quantum anomalous {Hall} insulators}},\
  }\href {https://doi.org/10.48550/arXiv.2506.02108} {\bibfield  {journal}
  {\bibinfo  {journal} {arXiv e-prints}\ ,\ \bibinfo {eid} {arXiv:2506.02108}}
  (\bibinfo {year} {2025})}\BibitemShut {NoStop}%
\bibitem [{\citenamefont {{Darius Shi}}\ and\ \citenamefont
  {{Senthil}}(2025)}]{Shi2025_anyon_delocalization}%
  \BibitemOpen
  \bibfield  {author} {\bibinfo {author} {\bibfnamefont {Z.}~\bibnamefont
  {{Darius Shi}}}\ and\ \bibinfo {author} {\bibfnamefont {T.}~\bibnamefont
  {{Senthil}}},\ }\bibfield  {title} {\bibinfo {title} {{Anyon delocalization
  transitions out of a disordered {FQAH} insulator}},\ }\href
  {https://doi.org/10.48550/arXiv.2506.02128} {\bibfield  {journal} {\bibinfo
  {journal} {arXiv e-prints}\ ,\ \bibinfo {eid} {arXiv:2506.02128}} (\bibinfo
  {year} {2025})}\BibitemShut {NoStop}%
\bibitem [{\citenamefont {{Pichler}}\ \emph {et~al.}(2025)\citenamefont
  {{Pichler}}, \citenamefont {{Kuhlenkamp}}, \citenamefont {{Knap}},\ and\
  \citenamefont {{Vishwanath}}}]{Pichler2025_anyonSC}%
  \BibitemOpen
  \bibfield  {author} {\bibinfo {author} {\bibfnamefont {F.}~\bibnamefont
  {{Pichler}}}, \bibinfo {author} {\bibfnamefont {C.}~\bibnamefont
  {{Kuhlenkamp}}}, \bibinfo {author} {\bibfnamefont {M.}~\bibnamefont
  {{Knap}}},\ and\ \bibinfo {author} {\bibfnamefont {A.}~\bibnamefont
  {{Vishwanath}}},\ }\bibfield  {title} {\bibinfo {title} {{Microscopic
  Mechanism of Anyon Superconductivity Emerging from Fractional {C}hern
  Insulators}},\ }\href {https://doi.org/10.48550/arXiv.2506.08000} {\bibfield
  {journal} {\bibinfo  {journal} {arXiv e-prints}\ ,\ \bibinfo {eid}
  {arXiv:2506.08000}} (\bibinfo {year} {2025})}\BibitemShut {NoStop}%
\bibitem [{\citenamefont {Sharapov}\ \emph {et~al.}(2001)\citenamefont
  {Sharapov}, \citenamefont {Beck},\ and\ \citenamefont
  {Loktev}}]{Sharapov2001_d-wave}%
  \BibitemOpen
  \bibfield  {author} {\bibinfo {author} {\bibfnamefont {S.~G.}\ \bibnamefont
  {Sharapov}}, \bibinfo {author} {\bibfnamefont {H.}~\bibnamefont {Beck}},\
  and\ \bibinfo {author} {\bibfnamefont {V.~M.}\ \bibnamefont {Loktev}},\
  }\bibfield  {title} {\bibinfo {title} {Finite-temperature time-dependent
  effective theory for the phase field in two-dimensional d-wave neutral
  superconductors},\ }\href {https://doi.org/10.1103/PhysRevB.64.134519}
  {\bibfield  {journal} {\bibinfo  {journal} {Phys. Rev. B}\ }\textbf {\bibinfo
  {volume} {64}},\ \bibinfo {pages} {134519} (\bibinfo {year}
  {2001})}\BibitemShut {NoStop}%
\bibitem [{\citenamefont {Benfatto}\ \emph {et~al.}(2004)\citenamefont
  {Benfatto}, \citenamefont {Toschi},\ and\ \citenamefont
  {Caprara}}]{Benfatto2004_phase}%
  \BibitemOpen
  \bibfield  {author} {\bibinfo {author} {\bibfnamefont {L.}~\bibnamefont
  {Benfatto}}, \bibinfo {author} {\bibfnamefont {A.}~\bibnamefont {Toschi}},\
  and\ \bibinfo {author} {\bibfnamefont {S.}~\bibnamefont {Caprara}},\
  }\bibfield  {title} {\bibinfo {title} {Low-energy phase-only action in a
  superconductor: A comparison with the $\mathrm{XY}$ model},\ }\href
  {https://doi.org/10.1103/PhysRevB.69.184510} {\bibfield  {journal} {\bibinfo
  {journal} {Phys. Rev. B}\ }\textbf {\bibinfo {volume} {69}},\ \bibinfo
  {pages} {184510} (\bibinfo {year} {2004})}\BibitemShut {NoStop}%
\bibitem [{\citenamefont {Paramekanti}\ \emph {et~al.}(2000)\citenamefont
  {Paramekanti}, \citenamefont {Randeria}, \citenamefont {Ramakrishnan},\ and\
  \citenamefont {Mandal}}]{Paramekanti2000_d-wave}%
  \BibitemOpen
  \bibfield  {author} {\bibinfo {author} {\bibfnamefont {A.}~\bibnamefont
  {Paramekanti}}, \bibinfo {author} {\bibfnamefont {M.}~\bibnamefont
  {Randeria}}, \bibinfo {author} {\bibfnamefont {T.~V.}\ \bibnamefont
  {Ramakrishnan}},\ and\ \bibinfo {author} {\bibfnamefont {S.~S.}\ \bibnamefont
  {Mandal}},\ }\bibfield  {title} {\bibinfo {title} {Effective actions and
  phase fluctuations in d-wave superconductors},\ }\href
  {https://doi.org/10.1103/PhysRevB.62.6786} {\bibfield  {journal} {\bibinfo
  {journal} {Phys. Rev. B}\ }\textbf {\bibinfo {volume} {62}},\ \bibinfo
  {pages} {6786} (\bibinfo {year} {2000})}\BibitemShut {NoStop}%
\bibitem [{\citenamefont {Diamantini}\ \emph {et~al.}(1993)\citenamefont
  {Diamantini}, \citenamefont {Sodano},\ and\ \citenamefont
  {Trugenberger}}]{Diamantini1993_CS_monopole}%
  \BibitemOpen
  \bibfield  {author} {\bibinfo {author} {\bibfnamefont {M.~C.}\ \bibnamefont
  {Diamantini}}, \bibinfo {author} {\bibfnamefont {P.}~\bibnamefont {Sodano}},\
  and\ \bibinfo {author} {\bibfnamefont {C.~A.}\ \bibnamefont {Trugenberger}},\
  }\bibfield  {title} {\bibinfo {title} {Topological excitations in compact
  {M}axwell-{C}hern-{S}imons theory},\ }\href
  {https://doi.org/10.1103/PhysRevLett.71.1969} {\bibfield  {journal} {\bibinfo
   {journal} {Phys. Rev. Lett.}\ }\textbf {\bibinfo {volume} {71}},\ \bibinfo
  {pages} {1969} (\bibinfo {year} {1993})}\BibitemShut {NoStop}%
\bibitem [{\citenamefont {Einhorn}\ and\ \citenamefont
  {Savit}(1978)}]{Einhorn1978}%
  \BibitemOpen
  \bibfield  {author} {\bibinfo {author} {\bibfnamefont {M.~B.}\ \bibnamefont
  {Einhorn}}\ and\ \bibinfo {author} {\bibfnamefont {R.}~\bibnamefont
  {Savit}},\ }\bibfield  {title} {\bibinfo {title} {Topological excitations in
  the abelian {H}iggs model},\ }\href
  {https://doi.org/10.1103/PhysRevD.17.2583} {\bibfield  {journal} {\bibinfo
  {journal} {Phys. Rev. D}\ }\textbf {\bibinfo {volume} {17}},\ \bibinfo
  {pages} {2583} (\bibinfo {year} {1978})}\BibitemShut {NoStop}%
\bibitem [{\citenamefont {{Geraedts}}\ \emph {et~al.}(2016)\citenamefont
  {{Geraedts}}, \citenamefont {{Zaletel}}, \citenamefont {{Mong}},
  \citenamefont {{Metlitski}}, \citenamefont {{Vishwanath}},\ and\
  \citenamefont {{Motrunich}}}]{Geraedts2016_DiracCFL}%
  \BibitemOpen
  \bibfield  {author} {\bibinfo {author} {\bibfnamefont {S.~D.}\ \bibnamefont
  {{Geraedts}}}, \bibinfo {author} {\bibfnamefont {M.~P.}\ \bibnamefont
  {{Zaletel}}}, \bibinfo {author} {\bibfnamefont {R.~S.~K.}\ \bibnamefont
  {{Mong}}}, \bibinfo {author} {\bibfnamefont {M.~A.}\ \bibnamefont
  {{Metlitski}}}, \bibinfo {author} {\bibfnamefont {A.}~\bibnamefont
  {{Vishwanath}}},\ and\ \bibinfo {author} {\bibfnamefont {O.~I.}\ \bibnamefont
  {{Motrunich}}},\ }\bibfield  {title} {\bibinfo {title} {{The half-filled
  {L}andau level: The case for {D}irac composite fermions}},\ }\href
  {https://doi.org/10.1126/science.aad4302} {\bibfield  {journal} {\bibinfo
  {journal} {Science}\ }\textbf {\bibinfo {volume} {352}},\ \bibinfo {pages}
  {197} (\bibinfo {year} {2016})}\BibitemShut {NoStop}%
\bibitem [{\citenamefont {{Gaiotto}}\ \emph {et~al.}(2015)\citenamefont
  {{Gaiotto}}, \citenamefont {{Kapustin}}, \citenamefont {{Seiberg}},\ and\
  \citenamefont {{Willett}}}]{Gaiotto2015_gensym}%
  \BibitemOpen
  \bibfield  {author} {\bibinfo {author} {\bibfnamefont {D.}~\bibnamefont
  {{Gaiotto}}}, \bibinfo {author} {\bibfnamefont {A.}~\bibnamefont
  {{Kapustin}}}, \bibinfo {author} {\bibfnamefont {N.}~\bibnamefont
  {{Seiberg}}},\ and\ \bibinfo {author} {\bibfnamefont {B.}~\bibnamefont
  {{Willett}}},\ }\bibfield  {title} {\bibinfo {title} {{Generalized global
  symmetries}},\ }\href {https://doi.org/10.1007/JHEP02(2015)172} {\bibfield
  {journal} {\bibinfo  {journal} {Journal of High Energy Physics}\ }\textbf
  {\bibinfo {volume} {2015}},\ \bibinfo {eid} {172} (\bibinfo {year}
  {2015})}\BibitemShut {NoStop}%
\bibitem [{\citenamefont {{Kitaev}}(2006)}]{Kitaev2005_anyons}%
  \BibitemOpen
  \bibfield  {author} {\bibinfo {author} {\bibfnamefont {A.}~\bibnamefont
  {{Kitaev}}},\ }\bibfield  {title} {\bibinfo {title} {{Anyons in an exactly
  solved model and beyond}},\ }\href
  {https://doi.org/10.1016/j.aop.2005.10.005} {\bibfield  {journal} {\bibinfo
  {journal} {Annals of Physics}\ }\textbf {\bibinfo {volume} {321}},\ \bibinfo
  {pages} {2} (\bibinfo {year} {2006})}\BibitemShut {NoStop}%
\bibitem [{\citenamefont {{Lu}}\ and\ \citenamefont
  {{Vishwanath}}(2012)}]{Lu2012_bosoninvertible}%
  \BibitemOpen
  \bibfield  {author} {\bibinfo {author} {\bibfnamefont {Y.-M.}\ \bibnamefont
  {{Lu}}}\ and\ \bibinfo {author} {\bibfnamefont {A.}~\bibnamefont
  {{Vishwanath}}},\ }\bibfield  {title} {\bibinfo {title} {{Theory and
  classification of interacting integer topological phases in two dimensions: A
  {C}hern-{S}imons approach}},\ }\href
  {https://doi.org/10.1103/PhysRevB.86.125119} {\bibfield  {journal} {\bibinfo
  {journal} {\prb}\ }\textbf {\bibinfo {volume} {86}},\ \bibinfo {eid} {125119}
  (\bibinfo {year} {2012})}\BibitemShut {NoStop}%
\bibitem [{Note6()}]{Note6}%
  \BibitemOpen
  \bibinfo {note} {This family of states has previously appeared in Refs.~\cite
  {Yutushui2024_daughter,Zheltonozhskii2024_daughter} as tools for constructing
  daughter states of gapped phases at $\nu = 1/2$. However, these earlier works
  do not give a physical interpretation of these states for general $p, n$ as
  they do not naturally arise if the parent state is fully gapped.}\BibitemShut
  {Stop}%
\end{thebibliography}%

\onecolumngrid
\appendix

\newpage 

\section{Landau-Ginzburg theory for a Bogoliubov Fermi surface}\label{app:LGtheory}

The overarching goal of this section is to derive a Ginzburg-Landau theory for an electronic superconductor with a Bogoliubov Fermi surface (BFS) in 2+1D, treating the electromagnetic vector potential as a background gauge field $A$. These results can be directly transferred to the CBFL by replacing $A$ with the emergent gauge field $a$ and adding Chern-Simons terms for $a$. 

Our basic strategy will be as follows. In Section~\ref{subapp:Derivation_LG}, We begin with a microscopic lattice model of electrons (denoted $f$) with translation and $C_3$ rotational symmetry. Using a Hubbard Stratanovich transformation, we decouple the four-fermion interactions in the pairing channel and assume that the system enters a ground state in which the pairing order parameter $\Delta$ develops an expectation value. Then we consider phase fluctuations of the superconductor parametrized by $\Delta \sim \Delta_0 e^{i\phi}$ and derive an effective action for $\phi$ by integrating out the original fermions. The final action $S_{\rm eff}[\phi, 2A]$ (including the coupling between the gauge field $A$ and the phase field $\phi$) is summarized in \eqref{eq:IRaction_app} and takes the form
\begin{equation}
    S_{\rm eff}[\phi, 2A] = i \frac{\bar \rho_f}{2} \int_{\bs{r}, \tau} (\partial_{\tau}\phi - 2A_0) + \frac{1}{8} \int d^3 x (\partial_{\mu} \phi - 2A_{\mu}) \Pi^{\mu\nu}_{\rm BdG} (\partial_{\nu} \phi - 2A_{\nu}) \,, 
\end{equation}
where the coefficients $\Pi^{\mu\nu}_{\rm BdG}$ are given by correlation functions of fermion bilinears in the mean-field superconducting ground state. After integrating out the phase field $\phi$, we obtain the gauge-invariant polarization function $\Pi^{\mu\nu}_{\rm BFL}$ which enters the gauge field action as
\begin{equation}
    S_{\rm eff}[A] = \frac{1}{2} \int \frac{d^3 q}{(2\pi)^3} A_{\mu}^*(q) \Pi^{\mu\nu}_{\rm BFL}(q) A_{\nu}(q) \,, \quad \Pi^{\mu\nu}_{\rm BFL}(q) = \Pi^{\mu\nu}_{\rm BdG}(q) - \frac{\Pi^{\mu\nu_1}_{\rm BdG}(q) q_{\nu_1} q_{\mu_1} \Pi^{\mu_1\nu}_{\rm BdG}(q)}{q_{\mu_2} \Pi^{\mu_2\nu_2}_{\rm BdG}(q) q_{\nu_2}} \,.  
\end{equation}
These results are summarized in \eqref{eq:BFS_response_function}. 

The rest of the appendix will be dedicated to the computation of $\Pi^{\mu\nu}_{\rm BdG}$ and $\Pi^{\mu\nu}_{\rm BFL}$. In Section~\ref{subapp:BFS_correlator_prelim}, we set up some basic notations that allow us to write down the BdG Hamiltonian for any gap function $\Delta(\bs{k})$. In Sec.~\ref{subapp:gapequation}, we introduce a toy model with a specific attractive interaction in the $p$-wave channel and solve the self-consistent gap equation to determine $\Delta(\bs{k})$. Expanding around the mean-field ground state, we can identify the density operator $\rho$ and the current operator $J^i$ in the Bogoliubov basis and compute their correlation functions (which in turn determine $\Pi^{\mu\nu}_{\rm BdG}$). Secs.~\ref{subapp:BdG_rhorho},~\ref{subapp:BdG_JJ},~\ref{subapp:BdG_rhoJ} treat the density-density, current-current, and density-current correlation functions in order. The key results are summarized in Eqs.~\eqref{eq:Grhorhofull},~\eqref{eq:GJJfull},~\eqref{eq:GrhoJ_full}. 

While the phase-only effective action $S_{\rm eff}[\phi, 2A]$ allows us to calculate all gauge-invariant correlation functions in the system, it is useful to manifestly keep track of the gapless Bogoliubov excitations for other purposes (e.g. deducing the global topological properties of the CBFL phase). In Section~\ref{subapp:LG_withBFS}, we describe this procedure within the Wilsonian renormalization group framework and arrive at an alternative effective action \eqref{eq:LG_withBFS_app} which plays a prominent role in the main text. 

\subsection{General derivation of the quantum Ginzburg-Landau action for a superconductor of spinless fermions}\label{subapp:Derivation_LG}

Our starting point is a model of electrons (denoted $f$) with a $C_3$-symmetric dispersion $\epsilon_f(\bs{k})$. Including a general four-fermion interaction and setting the external gauge field $A$ to zero, we can write the action as
\begin{equation}\label{eq:UVaction_app}
    S = \int_{\tau, \bs{r}} \bar f(\bs{r},\tau) [\partial_{\tau} + \xi_{-i \nabla}] f(\bs{r},\tau) - \frac{1}{4}\int_{\tau, \bs{r}, \bs{r}'} V(\bs{r} - \bs{r}') \bar f(\bs{r}, \tau) \bar f(\bs{r}', \tau) f(\bs{r}',\tau) f(\bs{r},\tau) \,,
\end{equation} 
where $\xi_{\bs{k}} = \epsilon_f(\bs{k}) - \mu$ is the dispersion relative to the chemical potential, $V(\bs{r} - \bs{r}')$ is an unspecified two-body interaction that decays at large $|\bs{r} - \bs{r}'|$, and $\int_{\tau} \equiv \int d \tau, \int_{\bs{r}} \equiv \int d^2 \bs{r}$. 

Now let us introduce a bilocal Hubbard-Stratanovich field $\Delta(\bs{r}, \bs{r}', \tau)$ to decouple the four-fermion interaction. The full action after decoupling takes the form
\begin{equation}
    \begin{aligned}
    S[f, \Delta] &= \int_{\tau, \bs{r}} \bar f(\bs{r},\tau) [\partial_{\tau} + \,\xi_{-i \nabla}] f(\bs{r},\tau) + \int_{\tau, \bs{r}, \bs{r}'} \frac{|\Delta(\bs{r}, \bs{r}', \tau)|^2}{V(\bs{r} - \bs{r}')} \\
    &\hspace{0.3cm} + \frac{1}{2} \int_{\tau, \bs{r}, \bs{r}'} \, \Delta(\bs{r}, \bs{r}', \tau) \, \bar f(\bs{r}, \tau) \bar f(\bs{r}', \tau) + \frac{1}{2} \int_{\tau, \bs{r}, \bs{r}'} \, \Delta^*(\bs{r}, \bs{r}', \tau) f(\bs{r}', \tau) \bar f(\bs{r}, \tau) \,. 
    \end{aligned} 
\end{equation}
We now tune the interaction $V$ to induce a superconducting ground state with some gap function $\bar \Delta$. In this work, our interest is not in precisely determining the form of $\bar \Delta$ for a specific choice of $V$. Rather, we want to qualitatively understand universal effects of the BFS on the superconducting ground state. Therefore, we will simply take $\bar \Delta$ as given and proceed to derive the effective Ginzburg-Landau theory by expanding $\Delta$ around its mean-field value. 

In general, the bilinear gap function $\Delta$ can be expanded as an infinite series
\begin{equation}
    \Delta(\bs{r}, \bs{r}', \tau) = \sum_{n=0}^{\infty} \Delta_n(\bs{R}, \tau) \cdot g_n(\bs{s}) \,, \quad \bs{R} = \frac{\bs{r} + \bs{r}'}{2} \,, \quad \bs{s} = \bs{r} - \bs{r}' \,,
\end{equation}
where $\bs{R}$ is the center-of-mass coordinate of the Cooper pair, $\bs{s}$ is the relative coordinate, and $g_n(\bs{s})$ is a complete orthonormal basis for regular functions of the relative coordinate $\bs{s}$. We refer to $g_n(\bs{s})$ as the $n$-th Cooper pair wavefunction. The solution of the gap equation selects a particular Cooper pair wavefunction that we can choose to be $g_0$ without loss of generality. When expanding around the ground state, amplitude fluctuations of $\Delta_n$ and fluctuations to orthogonal pairing channels $n > 0$ are both gapped. Therefore, in the low energy limit, it suffices to project out all modes with $n > 0$ and focus on the phase fluctuations of $\Delta_0$ parametrized as
\begin{equation}
    \Delta(\bs{r}, \bs{r}', \tau) \approx \bar \Delta(\bs{s}) \, e^{i \phi(\bs{R},\tau)} \,, \quad \bar \Delta(\bs{s}) = \Delta_0 \, g_0(\bs{s}) \,. 
\end{equation}
We will refer to the positive constant $\Delta_0$ as the pairing amplitude and $g_0$ as the pairing wavefunction. Plugging this ansatz into $S[f,\Delta]$ and dropping constant terms independent of the dynamical fields, we find a low energy effective action
\begin{equation}
    S_{\rm eff}[f,\Delta] = \int_{\tau,\bs{r}} \, \bar f(\bs{r},\tau) [\partial_{\tau} + \xi_{-i \nabla}] f(\bs{r},\tau) + \frac{\Delta_0}{2} \int_{\tau, \bs{s}, \bs{R}}  \, g_0(\bs{s}) e^{i \phi(\bs{R}, \tau)} \bar f(\bs{R} + \bs{s}/2, \tau) \bar f(\bs{R} - \bs{s}/2, \tau) + \mathrm{h.c.} \,. 
\end{equation}
As usual, a uniform phase $\phi(\bs{R}, \tau) = \phi_0$ can be absorbed by multiplying $f$ by $e^{-i\phi_0/2}$. Therefore, without loss of generality, we can expand $\phi$ near zero. Now consider a general $e^{i\phi(\bs{R}, \tau)}$ with $\phi \ll 1$. Via a local redefinition of fermion fields 
\begin{equation}
    f(\bs{r}, \tau) \rightarrow e^{i \phi(\bs{r}, \tau)/2} f(\bs{r}, \tau) \,,
\end{equation}
we can transform the effective action to
\begin{equation}
    \begin{aligned}
        S_{\rm eff}[f,\Delta] &= \int_{\tau,\bs{r}} \, \bar f(\bs{r},\tau) e^{-i \phi(\bs{r}, \tau)/2}  [\partial_{\tau} + \xi_{-i \nabla}] \bigg\{e^{i \phi(\bs{r}, \tau)/2}  f(\bs{r},\tau)\bigg\} \\
        &+ \frac{1}{2} \int_{\tau, \bs{s}, \bs{R}} \, \Delta_0 \, g_0(\bs{s}) \, \bar f(\bs{R} + \bs{s}/2, \tau) \bar f(\bs{R} - \bs{s}/2, \tau) \, e^{i \phi(\bs{R}, \tau) - i \phi(\bs{R} + \bs{s}/2, \tau)/2 - i\phi(\bs{R} - \bs{s}/2, \tau)/2} + \mathrm{h.c.} \,. 
    \end{aligned}
\end{equation}
The first term can be expanded in powers of $\partial_{\mu} \phi$. Naively, to leading order in a gradient expansion, we should only keep terms linear in $\partial_{\mu} \phi$. Integrating out the fermions would then generate an effective action which is quadratic in $\partial_{\mu} \phi$ as expected. However, this procedure misses an important ``diamagnetic term" which is quadratic in $\partial_{\mu} \phi$
\begin{equation}
    \textrm{Diamagnetic term} = \frac{1}{8} \int_{\tau, \bs{r}} \bar f [\nabla_i \nabla_j \xi_{-i\nabla}] f \nabla_i \phi \nabla_j \phi \,. 
\end{equation}
The fermion bilinear appearing in the diamagnetic term has a nonzero ground state expectation value 
\begin{equation}
    K^{ij}_{\rm diam} = \int_{\bs{k}} \ev{f^{\dagger}_{\bs{k}} \frac{\partial^2 \xi(\bs{k})}{\partial k_i \partial k_j} f_{\bs{k}}}_{\rm gs} \,.
\end{equation}
As a result, an additional term in the effective action for $\phi$ can be obtained by replacing $\bar f [\nabla_i \nabla_j \xi_{-i\nabla}] f$ with its expectation value $K^{ij}_{\rm diam}$ inside the diamagnetic term.

As for the second term, we can expand $\phi(\bs{R} \pm \bs{s}/2, \tau)$ around $\bs{s} = 0$ and see that the three phase factors almost cancel each other up to a term proportional to the second derivative $\nabla_{R_i} \nabla_{R_j} \phi(\bs{R}, \tau)$. If we include this coupling and integrate out the fermions, then the gradient expansion for $\phi$ contains a term which is proportional to $\phi \partial^3 \phi$. This term is subleading relative to the dominant terms with two gradients and will be neglected from now on.

Putting everything together, the effective action to leading order in the gradient expansion can be written as
\begin{equation}\label{eq:IRaction_app_withfermion}
    \begin{aligned}
    S_{\rm eff}[f, \phi] &\approx \int_{\tau,\bs{r}} \, \bar f(\bs{r},\tau) [\partial_{\tau} + i \partial_{\tau} \phi/2 + \xi_{-i \nabla + \nabla \phi/2}] f(\bs{r},\tau) + \frac{\Delta_0}{2} \int_{\tau, \bs{s}, \bs{R}}  \, g_0(\bs{s}) \, \bar f(\bs{R} + \bs{s}/2, \tau) \bar f(\bs{R} - \bs{s}/2, \tau) + \mathrm{h.c.} \\
    &\approx \int_{\tau,\bs{r}}\, \bar f [\partial_{\tau} + \xi_{-i \nabla}] f + i \int_{\tau, \bs{r}} \, \rho_f \, \partial_{\tau} \phi/2  + \int_{\tau,\bs{r}}\, \bs{J}_f \cdot \nabla \phi/2 + \frac{1}{8} \int_{\tau, \bs{r}} \, \bar f \,[\nabla_i \nabla_j \xi_{-i\nabla} f] \, \nabla_i \phi \nabla_j \phi  \\
    &\hspace{0.5cm}+ \frac{\Delta_0}{2} \, \int_{\tau, \bs{s}, \bs{R}} \, g_0(\bs{s}) \, \bar f(\bs{R} + \bs{s}/2, \tau) \bar f(\bs{R} - \bs{s}/2, \tau) + \mathrm{h.c.} \,, 
    \end{aligned}
\end{equation}
where we omitted the spacetime arguments of various fields and defined the density $\rho_f$ and paramagnetic current $J_f$
\begin{equation}
    \rho_f = \bar f f \,, \quad \bs{J}_f = \bar f \nabla \xi_{-i\nabla} f \quad \rightarrow \quad \bs{J}_f(\bs{q}, \Omega) = \int_{\bs{k}, \omega} \, \bs{v}_{\bs{k}} \, \bar f(\bs{k} + \bs{q}/2, \omega + \Omega/2) \, f(\bs{k} - \bs{q}/2, \omega - \Omega/2) \,,
\end{equation}
using the integration shorthand $\int_{\bs{k}, \omega} \equiv \int \frac{d^2 \bs{k} d \omega}{(2\pi)^3}$. Further integrating out the fermion fields $\bar f, f$ and keeping terms up to quadratic order in $\phi, \partial_{\tau} \phi, \nabla \phi$, we arrive at a compact effective action for $\phi$
\begin{equation}\label{eq:IRaction_app}
    S_{\rm eff}[\phi] \approx i \frac{\bar \rho_f}{2} \int_{\tau,\bs{r}} \partial_{\tau} \phi + \frac{1}{8} \int d^3 x \partial_{\mu} \phi \Pi^{\mu\nu}_{\rm BdG} \partial_{\nu} \phi \,,
\end{equation}
where the BdG response function $\Pi^{\mu\nu}_{\rm BdG}$ is defined as
a $3$x$3$ matrix of correlators evaluated in the mean-field SC state described by a corresponding BdG Hamiltonian 
\begin{equation}
    \Pi_{\rm BdG}^{\mu\nu}(\bm{q},\Omega) =\begin{pmatrix}
       G_{\rho_f \rho_f}(\bs{q}, \Omega)  & -i G_{\rho_f J^j_f}(\bs{q}, \Omega)\\
        -i G_{J^i_f\rho_f }(\bs{q}, \Omega) & K_{\rm diam}^{ij} - G_{J^i_f J^j_f}(\bs{q}, \Omega)
\end{pmatrix}\;.\label{eq:Pi_MF_munu}
\end{equation}

We recognize $\chi = \Pi^{00}_{\rm BdG}(q \rightarrow 0, \Omega = 0)/4$ as the generalized compressibility and $\rho^s = \Pi^{ii}_{\rm BdG}(q \rightarrow 0, \Omega = 0)/4$ as the superfluid stiffness. In a gapped superconductor, the coefficients $\chi$ and $\rho_s$ are positive and $\Pi^{\mu\nu}_{\rm BdG}$ admits analytic expansions in powers of $\bs{q}, \Omega$. In the presence of a gapless Bogoliubov Fermi surface, these coefficients generally become non-analytic. This non-analyticity translates to a non-local action in spacetime coordinates. 

The mixed current-density response function $\Pi^{0 i}_{\rm BdG}$ is more exotic. In a system with inversion symmetry, $\Pi^{0 i}_{\rm BdG}$ vanishes in the static limit and can be dropped to leading order in the gradient expansion \cite{Sharapov2001_d-wave,Benfatto2004_phase,Paramekanti2000_d-wave}.  However, since we are interested in systems with broken inversion symmetry, this constraint no longer applies and $\Pi^{0 i}_{\rm BdG}$ must be computed on a case by case basis. 

From Eq.~\eqref{eq:IRaction_app} we can also infer the dispersion relation of the phase mode. As we will see in the next section, all correlation functions evaluated in the mean-field state with BFS contain non-analytic Landau-damping terms that depend on the ratio $\Omega/q$. Thus, after performing analytic continuation to real frequencies $i\Omega\rightarrow \omega+i0^+$ and using the following ansatz
\begin{equation}
    \omega_{\rm phase}(\bm{q})= c_{\hat{\bm{q}}}q\label{eq:phase_mode}
\end{equation}
we arrive at the equation for the direction-dependent ``velocity" $c_{\hat{\bm{q}}}$
\begin{equation}
c_{\hat{\bm{q}}}^2 \Pi^{00}_{\rm BdG}(\bm{q},c_{\hat{\bm{q}}}q)+2c_{\hat{\bm{q}}}\hat{q}_i\Pi^{0i}_{\rm BdG}(\bm{q},c_{\hat{\bm{q}}}q) - \hat{q}_i\hat{q}_j\Pi^{ij}_{\rm BdG}(\bm{q},c_{\hat{\bm{q}}}q)=0
\end{equation}
In a regular fully gapped superfluid with inversion symmetry, $\chi$ and $\rho_s$ can be approximated as constants, while $\Pi^{0i}_{\rm BdG}$ can be neglected because it vanishes as $q$ goes to zero. This leads to a conventional result $c_{\hat{\bm{q}}}= \pm \sqrt{\rho_s/\chi}$. With BFS, $c_{\hat{\bm{q}}}$ is generally complex, and its imaginary part is negative (as required by causality) and describes the broadening of the Goldstone mode due to its coupling to gapless Bogoliubov excitations. 

Finally, let us think through the coupling of the effective action to the external gauge field $A$. Recall that $\phi$ parametrizes phase fluctuations of the pair order parameter $\Delta_0(\bs{R}, \tau)$. As a result, under a $U(1)_A$ gauge transformation generated by $\alpha(\bs{R}, \tau)$, the phase $\phi$ transforms as $\phi \rightarrow \phi + 2 \alpha$ while the gauge field transforms as $A_{\mu} \rightarrow A_{\mu} + \partial_{\mu} \alpha$. To write down a gauge-invariant Lagrangian, it therefore suffices to replace $\partial_{\mu} \phi$ with $\partial_{\mu} \phi - 2 A_{\mu}$ in the effective action. The final gauge-invariant effective action thus takes the form
\begin{equation}\label{eq:LG_phaseonly_app}
    \begin{aligned}
    &S_{\rm eff}[\phi, A] \approx i \frac{\bar \rho_f}{2} \int_{\tau,\bs{r}} \, (\partial_{\tau} \phi - 2A_{0}) + \frac{1}{8} \int d^3 x \, (\partial_{\mu} \phi - 2A_{\mu}) \,\Pi^{\mu\nu}_{\rm BdG} \, (\partial_{\nu} \phi - 2A_{\nu}) \,. 
    \end{aligned}
\end{equation}
After integrating out the phase mode, we arrive at the following action for the external gauge field from which all gauge-invariant response functions can be obtained
\begin{equation}
\begin{aligned}
S_{\rm eff}[ A]&= \frac{1}{2}\int_{\bs{q}, \Omega}  A^*_{\mu}(\bs{q}, \Omega) \Pi_{\rm BFL}^{\mu\nu}(\bs{q}, \Omega) A_{\nu}(\bs{q}, \Omega)\;,\\
    \Pi_{\rm BFL}^{\mu\nu}(\bs{q}, \Omega)&=\Pi_{\rm  BdG}^{\mu\nu}(\bm{q},\Omega)- \frac{\Pi_{\rm BdG}^{\mu\nu_1}(\bm{q},\Omega)q_{\nu_1} q_{\mu_1}\Pi_{\rm BdG}^{\mu_1\nu}(\bm{q},\Omega)}{q_{\mu_2} \Pi_{\rm BdG}^{\mu_2\nu_2}(\bm{q},\Omega) q_{\nu_2}}\;,  
    \label{eq:BFS_response_function}
\end{aligned}
\end{equation}
where $\mu=0,1,2$, $q_0=\Omega$. Note that 
\begin{equation}
     \Pi_{\rm BFL}^{\mu\nu}(\bs{q}, \Omega)q_\nu = q_\mu\Pi_{\rm BFL}^{\mu\nu}(\bs{q}, \Omega)=0\label{eq:gauge_inv}
\end{equation}
as required by gauge invariance.

\subsection{Correlators in the mean-field SC state with broken inversion symmetry: preliminaries}\label{subapp:BFS_correlator_prelim}

Let us now calculate the coefficients entering the effective action \eqref{eq:IRaction_app}, which are related to density/current correlation function in the mean-field superconducting ground state. Following the notation of the previous section, we use $\xi_{\bs{k}} \equiv \epsilon_f(\bs{k}) - \mu$ to denote the fermion dispersion in the parent band. This dispersion can be decomposed into a rotationally invariant term $\varepsilon_{+,\bs{k}}$ and a warping term $\varepsilon_{-,\bs{k}}$ that only respects the $C_3$ symmetry:
\begin{equation}
    \xi_{\bs{k}} = \varepsilon_{+,\bs{k}} + \varepsilon_{-,\bs{k}} \,, \quad \varepsilon_{+,\bs{k}} = \frac{\xi_{\bs{k}}+\xi_{-\bs{k}}}{2} \,, \quad \varepsilon_{-,\bs{k}} = \frac{\xi_{\bs{k}}-\xi_{-\bs{k}}}{2} \,.
\end{equation}
It is clear that $\varepsilon_{+,\bs{k}}$ is inversion-even while $\varepsilon_{-,\bs{k}}$ is inversion-odd. We will assume that $\varepsilon_{-,\bs{k}}$ is weak so that we are close to the inversion-symmetric limit. As argued in the main text, this limit is reasonable in approximate models for twisted MoTe$_2$ as well as rhombohedral multilayer graphene, in which a $C_6$ lattice rotation symmetry is broken down to $C_3$ by trigonal warping. In the Nambu basis $\Psi_{\bs{k}} = (f_{\bs{k}}, f^{\dagger}_{-\bs{k}})^T $, the Bogoliubov-de Gennes (BdG) Hamiltonian with an arbitrary gap function $\Delta_{\bs{k}}$ takes the form
\begin{equation}
    \mathcal{H}_{\rm BdG} = \frac{1}{2} \sum_{\bs{k}} \Psi^{\dagger}_{\bs{k}} \hat{H}_{\rm BdG} \Psi_{\bs{k}} \,, \quad 
    \hat{H}_{\rm BdG} = \begin{pmatrix}
        \xi_{\bs{k}} & \Delta_{\bs{k}} \\ \Delta_{\bs{k}}^* & - \xi_{-\bs{k}}
    \end{pmatrix} = \varepsilon_{+,\bs{k}} \hat{\tau}_z + \varepsilon_{-,\bs{k}} \hat{\tau}_0 + \Re \Delta_{\bs{k}} \hat{\tau}_x - \Im \Delta_{\bs{k}} \hat{\tau}_y \,. 
\end{equation}
This Hamiltonian is diagonalized by a standard Bogoliubov transformation 
\begin{equation}
    \gamma_{+,\bs{k}} = u_{\bs{k}} f_{\bs{k}} + v_{\bs{k}} f_{-\bs{k}}^{\dagger} \,, \quad \gamma_{-,\bs{k}} = v_{\bs{k}}^* f_{\bs{k}} - u_k^* f^{\dagger}_{-\bs{k}} \,, \quad H = \frac{1}{2} \sum_{\bs{k}} \left[E_+(\bs{k}) \gamma^{\dagger}_{+,\bs{k}} \gamma_{+,\bs{k}} + E_-(\bs{k}) \gamma^{\dagger}_{-,\bs{k}} \gamma_{-,\bs{k}} \right] \,,
\end{equation}
where the energy dispersion of the two branches are given by
\begin{equation}
    E_{\pm}(\bs{k}) = \varepsilon_{-,\bs{k}} \pm E_0(\bs{k}) \,, \quad E_0(\bs{k}) = \sqrt{|\Delta_{\bs{k}}|^2 + \varepsilon_{+,\bs{k}}^2} \,,
\end{equation}
and the coherence factors $u_{\bs{k}}, v_{\bs{k}}$ are
\begin{equation}
    u_{\bs{k}} = \sqrt{\frac{1}{2} \left[1 + \frac{\varepsilon_{+,\bs{k}}}{E_0(\bs{k})}\right]} \,, \quad v_{\bs{k}} =  \frac{\Delta_{\bs{k}}}{|\Delta_{\bs{k}}|} \sqrt{\frac{1}{2} \left[1 - \frac{\varepsilon_{+,\bs{k}}}{E_0(\bs{k})}\right]} \,.
\end{equation}
Note that since $\tau_0$ commutes with $\tau_x, \tau_y, \tau_z$, the inclusion of $\varepsilon_{-, \bs{k}}$ changes the dispersion $E_{\pm}(\bs{k})$ but not the coherence factors $u_{\bs{k}}, v_{\bs{k}}$. 

Without inversion-breaking, we reproduce the gapped BdG bands of a conventional BCS superconductor. However, when inversion-symmetry is broken, Bogoliubov Fermi surfaces form at momenta $\bs{k}$ satisfying
\begin{equation}
    \varepsilon_{-,\bs{k}}^2 = |\Delta_{\bs{k}}|^2 + \varepsilon_{+,\bs{k}}^2 \quad \rightarrow \quad \xi_{\bs{k}} \xi_{-\bs{k}} + |\Delta_{\bs{k}}|^2 = 0 \,. 
\end{equation}
For a general inversion-asymmetric dispersion, nesting only occurs at discrete points. Precisely at these points, $\varepsilon_{-\bs{k}} = 0$ and the spectrum must be gapped by the pairing. Moreover, by the final condition, it is clear that the original Fermi surface located at $\xi_{\bs{k}} = 0$ is fully gapped. Therefore, the BFS can only exist away from the original Fermi surface in pockets that do not touch the nesting points. 

\begin{figure}
    \centering
    \includegraphics[width=0.5\linewidth]{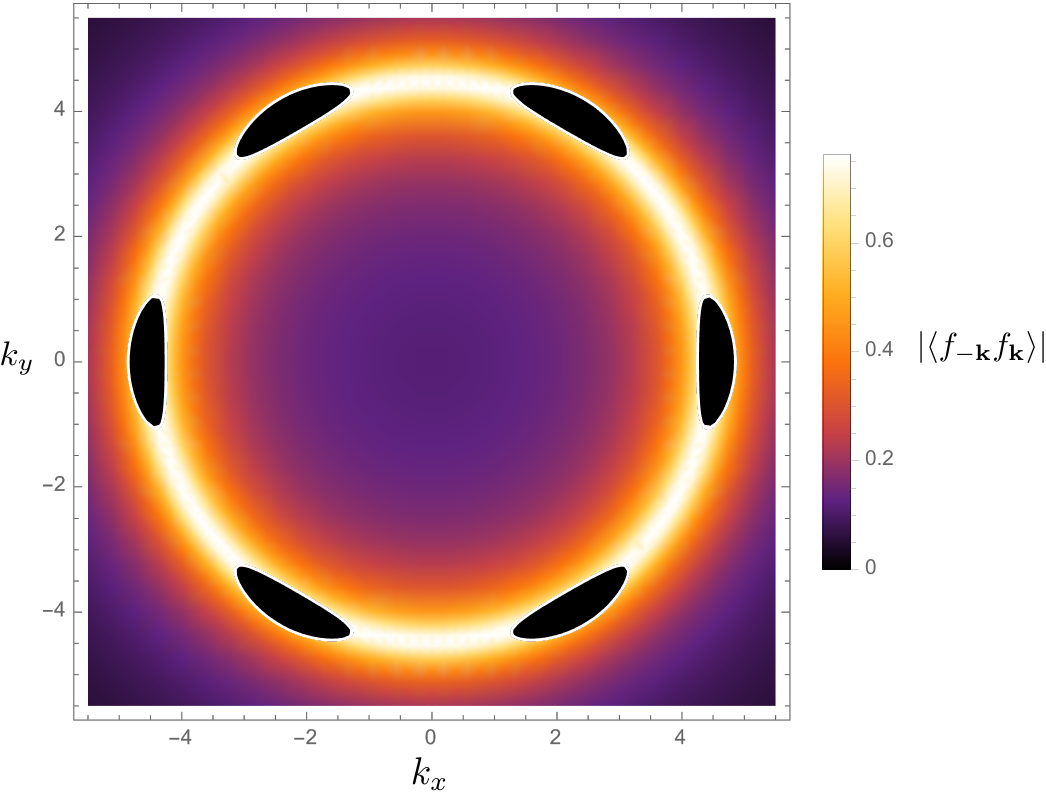}
    \caption{Illustration of the zero-temperature pairing amplitude $|\ev{f_{-\bs{k}} f_{\bs{k}}}|$ for $\Delta(\bs{k}) = 1.525 e^{i\theta_{\bs{k}}}, \;\lambda = 0.2,\; \mu = 10, \;m = 1$.  } 
    \label{fig:BFS_amplitude}
\end{figure}

Since the BdG spectrum has two branches, we can write the Green's function at fixed $\bs{k}$ in terms of projectors $P_{\pm}$ onto these bands:
\begin{equation}
    \hat{\mathcal{G}}(\bs{k}, \omega) = \begin{pmatrix}
        \langle f^{\dagger}_{\bs{k}, i\omega} f_{\bs{k}, i\omega}\rangle & \langle f_{-\bs{k}, -i\omega} f_{\bs{k}, i\omega}\rangle \\ \langle f^{\dagger}_{\bs{k},i\omega_n} f^{\dagger}_{-\bs{k}, -i\omega}\rangle  & \langle f_{-\bs{k}, -i\omega} f^{\dagger}_{-\bs{k}, -i\omega}\rangle 
    \end{pmatrix}= \frac{1}{i\omega \hat{\tau}_0- \hat{H}_{\rm BdG}} = \sum_{s = \pm} \frac{\hat{P}_{s}(\bs{k})}{i\omega - E_s(\bs{k})} \,, \quad \hat{P}_s(\bs{k}) = \frac{1}{2} \left[\hat{\tau}_0 + s\, \bs{d} \cdot \hat{\bs{\tau}} \right] \,, \label{eq:G_P_projector}
\end{equation}
where we have defined the normalized vector
\begin{equation}
    \bs{d} = \frac{1}{E_0(\bs{k})} \left(\Re \Delta_{\bs{k}}, - \Im \Delta_{\bs{k}}, \varepsilon_{+,\bs{k}} \right) \,. 
\end{equation}

It is straightforward to verify that $P_s(\bs{k})$ is a properly normalized projector
\begin{equation}
    \hat{P}_s(\bs{k})^2 = \frac{1}{4} \left[\hat{\tau}_0 + 2s \bs{d} \cdot \hat{\bs{\tau}} + d_a d_b \{\hat{\tau}_a, \hat{\tau}_b\}/2\right] = \frac{1}{4} \left[\hat{\tau}_0 + |\bs{d}|^2 \hat{\tau}_0 + 2s \bs{d} \cdot \hat{\bs{\tau}} \right] = \hat{P}_s(\bs{k}) \,. 
\end{equation}
In Euclidean time, the matrix Green's function in Eq.~\eqref{eq:G_P_projector} reads as
\begin{equation}
   \hat{\mathcal{G}}(\bm{k},\tau) \equiv \int \frac{d \omega}{2\pi} \hat{\mathcal{G}}(\bs{k}, \omega) e^{i\omega \tau} =- \begin{pmatrix}
        \langle f_{\bm k}(\tau)f_{\bm{k}}^\dagger\rangle &   \langle  f_{\bm k}(\tau)f_{-\bm{k}}\rangle\\
         \langle f^\dagger_{-\bm k}(\tau)f^\dagger_{\bm{k}}\rangle &   \langle f_{-\bm k}^\dagger(\tau)f_{-\bm{k}}\rangle
    \end{pmatrix}\;.
\end{equation}
From this Green's function, we see that the pair amplitude is given by 
\begin{equation}
    \ev{f_{-\bs{k}} f_{\bs{k}}} =\hat{\mathcal{G}}_{12}(\bs{k}, \tau=0^-)= \int \frac{d\omega}{2\pi} \hat{\mathcal{G}}_{12}(\bs{k}, \omega) e^{i \omega 0^+} = \frac{\Delta_{\bs{k}}}{2 E_0(\bs{k})}  \Big(f(E_{+}(\bs{k}))-f(E_{-}(\bs{k}))\Big)\label{eq:ckc-k}\;,
\end{equation}
where $f(E)$ is the Fermi-Dirac distribution function. Therefore, the electron pairing amplitude vanishes whenever $E_+$ and $E_-$ have the same sign. These are precisely the Bogoliubov Fermi pockets in momentum space. The Bogoliubov Fermi surface is located where one of $E_+, E_-$ changes sign. A visualization of the pairing amplitude is shown in Fig.~\ref{fig:BFS_amplitude}.

Next, we note the following identities involving the Nambu spinors
\begin{equation}
    -\langle \Psi_{\bm{k},\alpha}(\tau) \Psi^\dagger_{\bm{k},\beta} \rangle = [\hat{\mathcal{G}}(\bm{k},\tau)]_{\alpha,\beta}\;, \quad  -\langle \Psi_{-\bm{k},\alpha}^\dagger(\tau) \Psi^\dagger_{\bm{k},\beta} \rangle = [\hat{\tau}_x\hat{\mathcal{G}}(\bm{k},\tau)]_{\alpha,\beta}, \;\quad -\langle \Psi_{\bm{k},\alpha}(\tau) \Psi_{-\bm{k},\beta} \rangle =[\hat{\mathcal{G}}(\bm{k},\tau)  \hat{\tau}_x]_{\alpha,\beta}\label{eq:Psi_Psi}
\end{equation}
We are now in a position to compute the correlation functions. The scalar and vector potentials couple through the BdG Hamiltonian as
\begin{equation}
   \mathcal{H}_{\rm BdG} (A) = \frac{1}{2} \sum_{\bs{k}} \Psi^{\dagger}_{\bs{k}} \begin{pmatrix}
        \xi_{\bs{k}-\bm{A}}-iA_0 & \Delta_{\bs{k}} \\ \Delta_{\bs{k}}^* & - \xi_{-\bs{k} -\bm{A}}+iA_0
    \end{pmatrix} \Psi_{\bs{k}} \,. 
\end{equation}
Cf. our convention in Eq.~\eqref{eq:IRaction_app_withfermion}. From this coupling, we can infer the density and current operators in Nambu basis as
\begin{equation}
\begin{aligned}
   \rho(\bm{q})&=\partial_{A_0} \mathcal{H}_{\rm BdG} (A)|_{A=0}=\frac{1}{2}\sum\limits_{\bm{k}}  \Psi^\dagger_{\bm{k}+\bm{q}/2} \hat{\tau}_z\Psi_{\bm{k}-\bm{q}/2}\;,
   \\
   \bm{J}(\bm{q})&=-\partial_{\bm{A}} \mathcal{H}_{\rm BdG} (A)|_{A=0}=\frac{1}{2}\sum\limits_{\bm{k}}  \Psi^\dagger_{\bm{k}+\bm{q}/2}  \hat{\tau}_z\left[\nabla_{\bm{k}}\varepsilon_{+,\bm{k}}\hat{\tau}_z+\nabla_{\bm{k}}\varepsilon_{-,\bm{k}}\hat{\tau}_0\right]\Psi_{\bm{k}-\bm{q}/2}\;.
   \end{aligned}
\end{equation}
Thus the density-density correlator is
\begin{equation}
       \langle \rho(\bm{q},\tau) \rho(-\bm{q})\rangle =\frac{1}{4}\sum\limits_{\bm{k},\bm{k}'}  \langle \Psi^\dagger_{\bm{k}+\bm{q}/2}(\tau)\hat{\tau}_z\Psi_{\bm{k}-\bm{q}/2}(\tau) \Psi^\dagger_{\bm{k}'-\bm{q}/2} \hat{\tau}_z\Psi_{\bm{k}'+\bm{q}/2}\rangle\;. \label{eq:rho_rho_nambu}
\end{equation}
Crucially, when we perform Wick's contractions, we should consider both $\Psi \Psi^\dagger$ and $\Psi \Psi$, etc., pairings. Using Eq.~\eqref{eq:Psi_Psi}, we find that the result is doubled
\begin{equation}
     G_{\rho\rho}(\bm{q},\Omega)=\langle \rho(\bm{q},\Omega) \rho(-\bm{q},-\Omega)\rangle  = - \frac{1}{2}\int \frac{d^2 \bs{k} d \omega}{(2\pi)^3} \Tr \left[ \hat{\mathcal{G}}(\bs{k} + \bs{q}/2, i\omega + i\Omega/2) \tau_z \hat{\mathcal{G}}(\bs{k} - \bs{q}/2, i\omega - i\Omega/2)\tau_z\right]\;. 
\end{equation}
Similarly, the current-current correlator is given by
\begin{equation}
\begin{aligned}
   G_{J^i J^j}(\bm{q},\Omega)=  \langle J^i(\bm{q},\Omega) J^j(-\bm{q},-\Omega)\rangle  = - \frac{1}{2}\int \frac{d^2 \bs{k} d \omega}{(2\pi)^3} &\Tr \left[ \hat{\mathcal{G}}(\bs{k} + \bs{q}/2, i\omega + i\Omega/2) \left(\nabla_{k_i}\varepsilon_{+,\bm{k}}\hat{\tau}_0+\nabla_{k_i}\varepsilon_{-,\bm{k}}\hat{\tau}_z\right)\right. \\
&\times\left. \hat{\mathcal{G}}(\bs{k} - \bs{q}/2, i\omega - i\Omega/2)\left(\nabla_{k_j}\varepsilon_{+,\bm{k}}\hat{\tau}_0+\nabla_{k_j}\varepsilon_{-,\bm{k}}\hat{\tau}_z\right)\right]\;.
     \end{aligned}
\end{equation}
Finally, the current-density correlator reads
\begin{equation}
     \langle J^i(\bm{q},\Omega) \rho(-\bm{q},-\Omega)\rangle  = - \frac{1}{2}\int \frac{d^2 \bs{k} d \omega}{(2\pi)^3} \Tr \left[ \hat{\mathcal{G}}(\bs{k} + \bs{q}/2, i\omega + i\Omega/2) \left(\nabla_{k_i}\varepsilon_{+,\bm{k}}\hat{\tau}_0+\nabla_{k_i}\varepsilon_{-,\bm{k}}\hat{\tau}_z\right)\hat{\mathcal{G}}(\bs{k} - \bs{q}/2, i\omega - i\Omega/2)\hat{\tau}_z\right],
\end{equation}
and the diamagnetic part reads as
\begin{equation}
    K^{ij}_{\rm diam} = \int \frac{d^2\bm{k} d\omega}{(2\pi)^3} \frac{\partial^2 \xi(\bs{k})}{\partial k_i \partial k_j}  \hat{\mathcal{G}}_{11}(\bs{k}, \omega) e^{i \omega 0^+}=\frac{1}{2}\int \frac{d^2\bm{k} }{(2\pi)^2} \frac{\partial^2 \xi(\bs{k})}{\partial k_i \partial k_j}\sum\limits_{s=\pm} \left(1+\frac{s \varepsilon_{+,\bm{k}}}{E_{0}(\bm{k})}\right)f(E_s(\bm{k}))\equiv k_{\rm diam} \delta_{ij}\;.\label{eq:diamag_BdG}
\end{equation}
Here $k_{\rm diam}\equiv K^{11}_{\rm diam}\equiv K^{22}_{\rm diam} $ since $K^{ij}_{\rm diam} $ is proportional to the identity tensor (the latter follows from the $C_3$ rotational symmetry).

Our next goal is to understand the general structure of these correlators in the presence of the BFS.

\subsection{Self-consistent calculation of the order parameter}\label{subapp:gapequation}

Before discussing the calculation of the correlation functions, we briefly comment on the solution of the self-consistent gap equation $\Delta_{\bm{k}} = -\int \frac{d^2\bm{k}'}{(2\pi)^2}V_{\bm{k},\bm{k}'} \ev{f_{-\bs{k}'} f_{\bs{k}'}}$, where $V_{\bm{k},\bm{k}'}$ is the effective attraction in the Cooper channel, and the pair function is given in Eq.~\eqref{eq:ckc-k}. Thus, we obtain the following equation 
\begin{equation}\label{eq:self-consistent_eq}
    \Delta_{\bm{k}}= \int \frac{d^2\bm{k}'}{(2\pi)^2} V_{\bm{k},\bm{k}'} \frac{\Delta_{\bm{k}'}}{2E_0(\bm{k}')}  \Big\{f(E_{-}(\bm{k}'))-f(E_{+}(\bm{k}'))\Big\}\;.
\end{equation}
We note that in the absence of the BFS, the factor in brackets reduces to unity when the zero temperature limit is taken since $E_-\leq 0$ and $E_+\geq 0$. In contrast, the BFS introduces a non-trivial structure in momentum space illustrated in Fig.~\ref{fig:BFS_amplitude}.

The emergence and stability of the Bogoliubov Fermi surface depend on specific model details. Although here we are mainly interested in the observable consequences of BFS assuming it is formed, it is still instructive to provide a minimal example demonstrating its appearance within the simplest possible setup. As a concrete example, we consider the dispersion relation $ \xi_{\bm{k}}= \epsilon_f(\bm{k})-\mu$ given by
\begin{equation}\label{eq:xi_C3}
     \xi_{\bm{k}} = \frac{k^2}{2m}\left(1+\lambda \cos 3\theta_{\bm{k}}\right)-\mu\;,
\end{equation}
where $\mu$ is the chemical potential, $\cos3\theta_{\bm{k}}= (k_x^3-3k_y^2 k_x)/k^3$, and we assume $|\lambda|<1$ for stability. We also choose the pairing interaction $\lambda_{\bm{k},\bm{k}'}$ of the form $\lambda_{\bm{k},\bm{k}'}=V_0 e^{i\theta_{\bm{k}}}e^{-i\theta_{\bm{k}'}}$.  Since this attraction is in the chiral $l=1$ channel, we seek a solution in the form
$\Delta_{\bm{k}}=\Delta(0) e^{i\theta_{\bm{k}}}$ and $|\Delta_{\bm{k}}|^2=\Delta(0)^2 $. In this case, Eq.~\eqref{eq:self-consistent_eq} is simplified to a transcendental equation determining $\Delta(0)$. We also allow for solutions with a finite center-of-mass momentum $\bm{q}$. In this case, $\Delta(0)$ is replaced with $\Delta({\bm{q}})$, the dispersion $\xi_{\bm{k}}$ is replaced by $ \xi_{\bm{k}+\bm{q}/2}$, and $\xi_{-\bm{k}}$ is replaced by $ \xi_{-\bm{k}+\bm{q}/2}$. The associated Bogoliubov energies are denoted as $E_{\pm,\bm{q}}(\bm{k}')$ and $E_{0,\bm{q}}(\bm{k}')$. The quadratic part of the linearized Ginzburg-Landau action $\sim K(\bm{q})|\Delta(\bm{q})|$ at $T=T_c$ is given by
\begin{equation}\label{eq:GL_Tc}
  K(\bm{q})=  \frac{1}{V_0} -\int \frac{d^2\bm{k}'}{(2\pi)^2}\frac{1}{2E_{0,\bm{q}}(\bm{k}')}  \Big\{f(E_{-,\bm{q}}(\bm{k}'))-f(E_{+,\bm{q}}(\bm{k}'))\Big\}_{\Delta=0}\;,
\end{equation}
where the second term is just the normal state pairing susceptibility at external momenta $\bm{q}$.

The detailed phase diagram depends on the values of parameters $\lambda$ and $V_0$, and solutions with both $q=0$ and finite $q$ are possible. To obtain the solution with BFS shown in Fig.~\ref{fig:fig1} in the main text, we used $\mu=10$, $m=1$, $\lambda=0.2$, and $1/V_{0}=0.73$. We also implemented a UV momentum cutoff $k_{\bm UV}=10 k_F$, where $k_F=\sqrt{2m\mu}$. For these parameters, we find that the first solution of the linearized equation appears at $T=T_c\approx 0.342$ for $q=0$. The associated quadratic part of the Ginzburg-Landau action $K(\bm{q})$ at $T_c$ exhibits $C_3$ rotational symmetry, as shown in Fig.~\ref{fig:fig1}(b). We then followed the evolution of the corresponding order parameter $\Delta(0)$ below $T_c$, and found its magnitude at $T= 0$ to be $\Delta(0) \approx 1.525$. We note that the standard BCS relation between $T_c$ and $\Delta(0)|_{T=0}$ is not expected to hold in the presence of the BFS. 

\subsection{Density-density correlation functions}\label{subapp:BdG_rhorho}

We begin with the zero-temperature density-density correlation function
\begin{equation}
    \begin{aligned}
    &G_{\rho\rho}(\bs{q}, \Omega) \equiv \ev{\rho(\bs{q}, \Omega) \rho(-\bs{q}, -\Omega)}_c = - \frac{1}{2}\int \frac{d^2 \bs{k} d \omega}{(2\pi)^3} \Tr \left[\hat{\mathcal{G}}(\bs{k} + \bs{q}/2, i\omega + i\Omega/2) \hat{\tau}_z  \hat{\mathcal{G}}(\bs{k} - \bs{q}/2, i\omega - i\Omega/2)\hat{\tau}_z \right] \\
    &= - \frac{1}{2} \int \frac{d^2 \bs{k} d \omega}{(2\pi)^3} \sum_{s_1, s_2} \frac{1}{i(\omega + \Omega/2) - E_{s_1}(\bs{k}+\bs{q}/2)} \frac{1}{i(\omega - \Omega/2) - E_{s_2}(\bs{k}-\bs{q}/2)} \Tr \left[ \hat{P}_{s_1}(\bs{k}+\bs{q}/2) \hat{\tau}_z \hat{P}_{s_2}(\bs{k} - \bs{q}/2)\hat{\tau}_z\right]  \,.
    \end{aligned}
\end{equation}
The integral over $\omega$ can be readily done and expressed in terms of the Fermi function $f(E)$
\begin{equation}
    - \int \frac{d \omega}{2\pi} \frac{1}{i(\omega + \Omega/2) - E_{s_1}(\bs{k}+\bs{q}/2)} \frac{1}{i(\omega - \Omega/2) - E_{s_2}(\bs{k}-\bs{q}/2)} = \frac{f\left[E_{s_1}(\bs{k}+\bs{q}/2)\right] - f\left[E_{s_2}(\bs{k}-\bs{q}/2)\right]}{i\Omega - E_{s_1}(\bs{k}+\bs{q}/2) + E_{s_2}(\bs{k}-\bs{q}/2)}  \,.
\end{equation}
The trace can also be evaluated for general pairing vector $\bs{d}$
\begin{equation}
    \begin{aligned}
    &\Tr \left[\hat{\tau}_z \hat{P}_{s_1}(\bs{k}+\bs{q}/2) \hat{\tau}_z \hat{P}_{s_2}(\bs{k} - \bs{q}/2)\right] \\
    &= \frac{1}{4} \Tr \left[\hat{\tau}_z (\hat{\tau}_0 + s_1 d^a(\bs{k}+\bs{q}/2) \hat{\tau}_a)\hat{\tau}_z (\hat{\tau}_0 + s_2 d^b(\bs{k}-\bs{q}/2) \hat{\tau}_b)\right] \\
    &= \frac{1}{4} \left[2 + s_1 s_2 d^a(\bs{k}+\bs{q}/2) d^b(\bs{k}-\bs{q}/2) \Tr \hat{\tau}_z \hat{\tau}_a \hat{\tau}_z \hat{\tau}_b\right] \\
    &= \frac{E_0(\bs{k}+\bs{q}/2) E_0(\bs{k}-\bs{q}/2) + s_1 s_2 (\varepsilon_{+,\bs{k}+\bs{q}/2}\varepsilon_{+,\bs{k}-\bs{q}/2} - \Re \Delta_{\bs{k}+\bs{q}/2} \Re \Delta_{\bs{k}-\bs{q}/2} - \Im \Delta_{\bs{k}+\bs{q}/2} \Im \Delta_{\bs{k}+\bs{q}/2})}{2 E_0(\bs{k}+\bs{q}/2) E_0(\bs{k}-\bs{q}/2)} \,.
    \end{aligned}
\end{equation}
Up to corrections that are quadratically suppressed in $q$, we can approximate this trace by its value at $q = 0$, which is significantly simpler
\begin{equation}
    \Tr \left[\hat{\tau}_z \hat{P}_{s_1}(\bs{k}) \hat{\tau}_z \hat{P}_{s_2}(\bs{k})\right] = \frac{E_0(\bs{k})^2 + s_1 s_2 (\varepsilon_{+,\bs{k}}^2 - |\Delta_{\bs{k}}|^2)}{2 E_0(\bs{k})^2} = \frac{(1 + s_1 s_2) \varepsilon_{+,\bs{k}}^2}{2 E_0(\bs{k})^2} + \frac{(1 - s_1 s_2) |\Delta_{\bs{k}}|^2}{2 E_0(\bs{k})^2}  \,.
\end{equation}
Therefore, we end up with a general small-$|\bs{q}|$ approximation for the density correlation function 
\begin{equation}
    \begin{aligned}
       G_{\rho\rho}(\bs{q}, \Omega)
        &\approx \sum_{s_1, s_2 = \pm } \frac{1}{2}\int \frac{d^2 \bs{k}}{(2\pi)^2} \frac{f\left[E_{s_1}(\bs{k}+\bs{q}/2)\right] - f\left[E_{s_2}(\bs{k}-\bs{q}/2)\right]}{i\Omega - E_{s_1}(\bs{k}+\bs{q}/2) + E_{s_2}(\bs{k}-\bs{q}/2)} \left[\frac{(1 + s_1 s_2) \varepsilon_{+,\bs{k}}^2}{2 E_0(\bs{k})^2} + \frac{(1 - s_1 s_2) |\Delta_{\bs{k}}|^2}{2 E_0(\bs{k})^2} \right] \;.
    \end{aligned}
\end{equation}
There are two qualitatively distinct contributions to the above correlator. When $s_1 = s_2$, we have scattering within the same BFS pocket (such processes are typically absent in $T=0$ SCs with inversion symmetry because $E_{s}(\bm{k})$ is sign-definite in that case, and the distribution functions cancel). At small $\bs{q}$ and $\Omega$, since $E_{s}(\bs{k} + \bs{q}/2) \approx E_{s}(\bs{k} - \bs{q}/2)$, we get singular contributions from the particle-hole continuum within a band. In contrast, when $s_1 \neq s_2$, a singular denominator requires $\bs{k}+\bs{q}/2$ to be on the mass shell of the $s_1$-band and $\bs{k}-\bs{q}/2$ to be on the mass shell of the $s_2$-band. This is impossible unless the momentum $\bs{q}$ gets sufficiently large to induce interband scattering. Therefore, the infrared singular contributions to the correlator arise entirely from the intraband contribution, while the interband contribution gives a regular analytic part. Let us first determine the singular contributions from intraband scattering
\begin{equation}
    G^{\rm intra}_{\rho\rho}(\bs{q}, \Omega) \approx \sum_{s = \pm} \frac{1}{2}\int \frac{d^2 \bs{k}}{(2\pi)^2} \frac{f\left[E_{s}(\bs{k}+\bs{q}/2)\right] - f\left[E_{s}(\bs{k}-\bs{q}/2)\right]}{i\Omega - E_{s}(\bs{k}+\bs{q}/2) + E_{s}(\bs{k}-\bs{q}/2)} \frac{\varepsilon_{+,\bs{k}}^2}{E_0(\bs{k})^2} \,.
\end{equation}
This formula admits a simple interpretation. We can define an effective charge on the BFS
\begin{equation}
    e_*(\bs{k}) = \frac{\varepsilon_{+,\bs{k}}}{E_0(\bs{k})} \,. 
\end{equation}
Then the density correlation is equivalent to the sum of two density correlation functions of Fermi gases with dispersion $E_+(\bs{k})$ and $E_-(\bs{k})$, weighted by the effective charge $e_*(\bs{k})$. Generally, the effective charge $|e_*(\bs{k})|$ vanishes close to the nested points and approaches 1 farther away from the nested points. Therefore, the contribution from the BFS to the density correlation function is reduced relative to a regular Fermi gas with the same quasiparticle dispersion. Nevertheless, the qualitative structure will be the same: there will be a nonzero compressibility as well as Landau damping. 

Assuming small external momenta and frequency but fixed ratio $\Omega/q$, we can approximate
$   f\left[E_{s}(\bs{k}+\bs{q}/2)\right] - f\left[E_{s}(\bs{k}-\bs{q}/2)\right]\approx - (\bm{v}_s\cdot \bm{q})\delta(E_{s}(\bs{k}))$, where $\bm{v}_s=\nabla_{\bm{k}}E_{s}(\bs{k})$ is the group velocity of the Bogoliubov quasiparticles. As a result, we arrive at the following simple expression
\begin{equation}
     G^{\rm intra}_{\rho \rho}(\bm{q},\Omega) \approx \chi_0 -\frac{i\Omega}{ q}  \Gamma_{\hat{\bm{q}}}\left(\frac{\Omega}{q}\right) \;.\label{eq:Pi_rho_rho_general}
\end{equation}
The weighted density of states at the Fermi level of Bogoliubov quasiparticles $\chi_0$, and $\Gamma_{\hat{\bm{q}}}\left(q/\Omega\right)$ are given by
\begin{equation}
     \chi_0 = \frac{1}{2}\sum_{s = \pm} \int\frac{d^2\bm{k} }{(2\pi)^2} |  e_*(\bs{k})|^2\delta(E_s(\bm{k}))\;,\quad    \Gamma_{\hat{\bm{q}}}\left(\Omega/q\right) = \frac{1}{2}\sum_{s = \pm} \int\frac{d^2\bm{k} }{(2\pi)^2} \frac{ |  e_*(\bs{k})|^2\delta(E_s(\bm{k})) }{i\Omega/q-\bm{v}_s\cdot \hat{\bm{q}} } \;.\label{eq:def-Gamma}
\end{equation}
Finally, let us include the regular interband contributions, which are analytic at small $\bs{q}, \Omega$
\begin{equation}
      (\text{regular interband contributions})
        \approx \frac{1}{2}\int \frac{d^2 \bs{k}}{(2\pi)^2} \Big(f\left[E_{-}(\bs{k})\right] - f\left[E_{+}(\bs{k})\right]\Big)\frac{|\Delta_{\bs{k}}|^2}{E_0(\bs{k})^3} +\mathcal{O}(\Omega^2, \; q^2)  \;.\label{eq:interband}
\end{equation}
The full density correlator therefore takes the form 
\begin{equation}\label{eq:Grhorhofull}
    G_{\rho\rho}(\bs{q},\Omega) = \tilde \chi_0 - \frac{i\Omega}{q} \Gamma_{\hat{\bs{q}}}\left(\frac{\Omega}{q}\right) + \mathcal{O}(\Omega^2, q^2) \,,
\end{equation}
where $\tilde \chi_0$ is the sum of $\chi_0$ and the leading term in the regular interband contribution. Further assuming the limit $|\Omega|\ll q$, we find Eq.~\eqref{eq:density_f_main} from the main text, where we identified $\Gamma_{\hat{\bm{q}}}\equiv\Gamma_{\hat{\bm{q}}}\left(0^+\right)$.

Let us now illustrate this general structure by calculating the form of the Landau damping assuming a dispersion relation in Eq.~\eqref{eq:xi_C3}. In this case, we find
\begin{equation}
    \varepsilon_{+,\bs{k}} = \frac{k^2}{2m}-\mu\,, \quad \quad \varepsilon_{-,\bs{k}} =\frac{\lambda k^2}{2m} \cos 3\theta_{\bm{k}},\quad\quad E_{\pm}(\bs{k}) = \frac{\lambda k^2}{2m} \cos 3\theta_{\bm{k}} \pm \sqrt{\Delta^2 + \left(\frac{k^2}{2m}-\mu\right)^2}\;.
\end{equation}
Here $\Delta\equiv |\Delta_{\bm{k}}|$ which we assume to be momentum-independent. It is convenient to first integrate= over the radial component of $\bm{k}$, with fixed $\theta_{\bm{k}}$. To this end, we note that
\begin{equation}
     \delta(E_s(\bm{k})) =  \frac{\delta(k-k_{F,s}(\theta_{\bm{k}})) }{|\partial_k E_s(k,\theta_{\bm{k}})|}\;,
\end{equation}
where $\partial_k$ denotes the radial derivative with respect to the absolute value of $k$
\begin{equation}
    \partial_k E_s = \frac{k}{m}\left(\lambda \cos 3\theta_{\bm{k}} +\frac{s(k^2/2m\mu-1)}{\sqrt{\bar{\Delta}^2+(k^2/2m\mu-1)^2}}\right)\;.
\end{equation}
where we defined a dimensionless gap $\bar{\Delta}=\Delta/\mu$.

The Fermi momenta $k_{F,s}(\theta_{\bm{k}})$ are positive, real solutions to the equation $E_{s}(k_F,\theta_{\bm{k}})=0$ with fixed angle $\theta_{\bm{k}}$. Taking a square of this equation, we find
\begin{equation}
k_{F,\pm}(\theta_{\bm{k}})\overset{?}{=}\sqrt{\frac{2m\mu}{1-\lambda^2\cos^2(3\theta_{\bm{k}})}} \;\sqrt{1\pm \sqrt{\lambda^2\cos^2(3\theta_{\bm{k}})-(1-\lambda^2\cos^2(3\theta_{\bm{k}}))\bar{\Delta}^2}}\;,\label{eq:k_F}
\end{equation}
independent of $s$ (we dismissed negative solutions). The question mark indicates that $k_{F,l}(\theta_{\bm{k}})$ is not always a true solution to $E_{s}(k,\theta_{\bm{k}})=0$ with a particular $s$. In order for it to be a proper solution, several consistency conditions must be met. One obvious condition follows from the fact that the inner square root in Eq.~\eqref{eq:k_F} is real
\begin{equation}\label{eq:cond_1}
    \lambda^2 \cos^2 3\theta_{\bm{k}} \geq (1-\lambda^2 \cos ^23\theta_{\bm{k}} )\bar{\Delta}^2 \;.
\end{equation}
The outer square root is always real. Moreover, we find that both solutions $k_{F,\pm}(\theta_{\bm{k}})$ with $l=\pm$ are valid solution to $E_{+}(k,\theta_{\bm{k}})=0$ provided that 
\begin{equation}\label{eq:cond2}
\lambda \cos 3\theta_{\bm{k}} \left(1-\sqrt{\lambda^2 \cos^2 3\theta_{\bm{k}} -(1-\lambda^2 \cos ^23\theta_{\bm{k}} )\bar{\Delta}^2 }\right) \leq 0\;.
\end{equation}

Similarly, both solutions $k_{F,\pm}(\theta_{\bm{k}})$ are valid solution to $E_{-}(k,\theta_{\bm{k}})=0$ provided that 
\begin{equation}\label{eq:cond3}
\lambda \cos 3\theta_{\bm{k}} \left(1-\sqrt{\lambda^2 \cos^2 3\theta_{\bm{k}} -(1-\lambda^2 \cos ^23\theta_{\bm{k}} )\bar{\Delta}^2}\right) \geq 0\;.
\end{equation}

\begin{figure*}[t!]
    \centering
\includegraphics[width=0.4\linewidth]{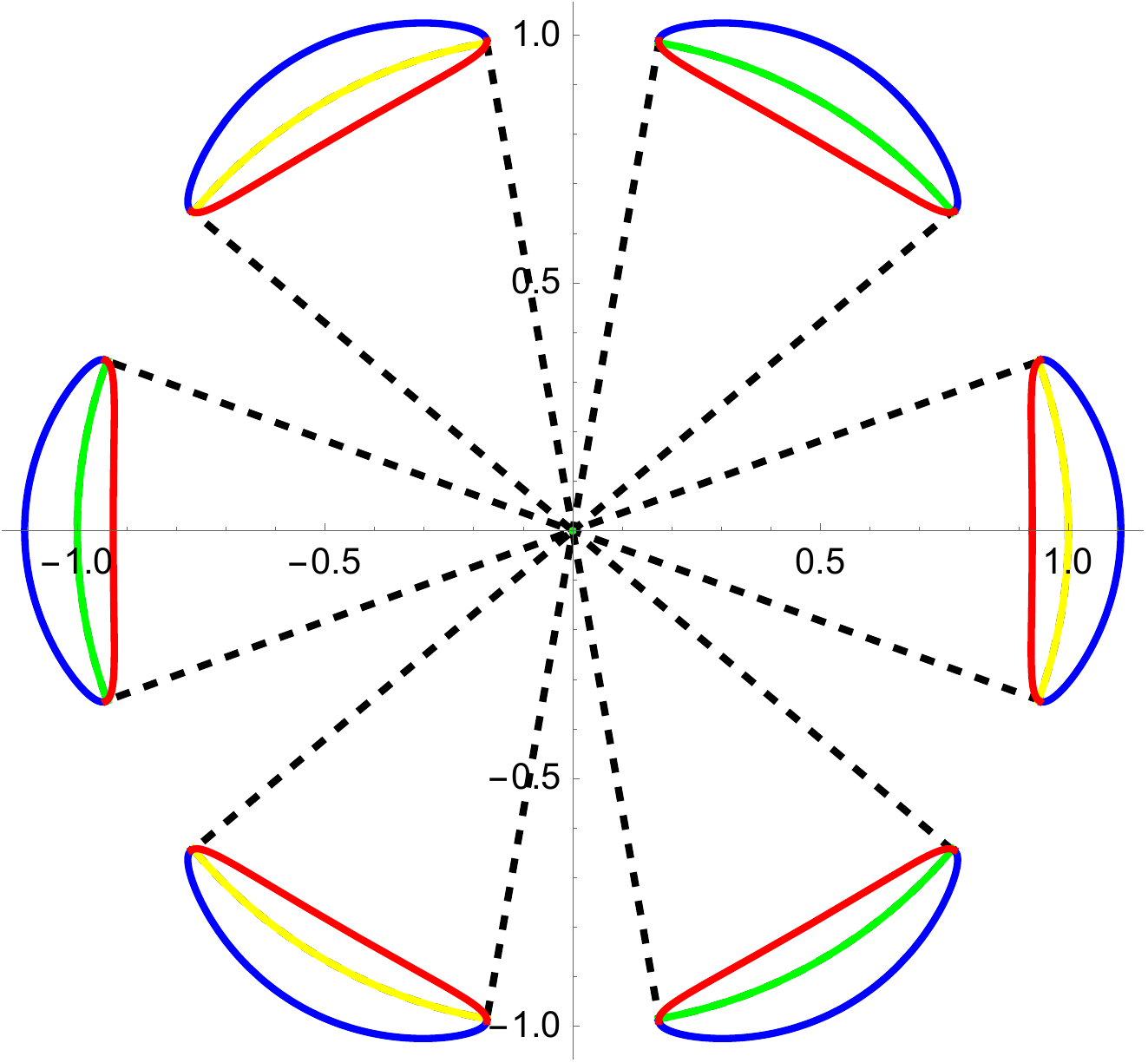}
    \caption{Illustration of various angular-dependent quantities parameterizing the BFS. Sectors indicated by the black dashed line are determined from the condition in Eq.~\eqref{eq:cond_1}. Blue (red) solid lines show the value of $k_{F,+}(\theta)/\sqrt{2m\mu}$ ($k_{F,-}(\theta)/\sqrt{2m\mu}$). From this, we can see that $k_{F,+}(\theta)$ parameterizes the outer segment of the BFS pockets, whereas $k_{F,-}(\theta)$ parameterizes the inner segments. Yellow lines indicate the region allowed by the inequality in Eq.~\eqref{eq:cond3}, whereas the green lines show the region allowed by Eq.~\eqref{eq:cond2}.
    Parameters used: $\bar{\Delta}=0.1$, $\lambda=0.2$, $\sqrt{2m\mu}=1$. } \label{fig:fig2}
\end{figure*}
We demonstrate these inequalities in Fig.~\ref{fig:fig2}. As a result, the static part $\chi_0$ is given by
\begin{equation}
\begin{aligned}
        \chi_0 &= \sum_{s,l= \pm} \frac{1}{2}\int\frac{d\theta_{\bm{k}} }{(2\pi)^2}   \frac{k_{F,l}\;|  e_*(k_{F,l})|^2}{|\partial_k E_s(k_{F,l})|} \theta\left(s\lambda \cos 3\theta_{\bm{k}} \left(\sqrt{\lambda^2 \cos^2 3\theta_{\bm{k}} -(1-\lambda^2 \cos ^23\theta_{\bm{k}} )\bar{\Delta}^2} -1\right)  \right) \\
    &\times\theta\left(\lambda^2\cos^2(3\theta_{\bm{k}})-(1-\lambda^2\cos^2(3\theta_{\bm{k}}))\bar{\Delta}^2\right)\;.
    \end{aligned}
\end{equation}
Here $k_{F,l}\equiv k_{F,l}(\theta_{\bm{k}})$ is given in Eq.~\eqref{eq:k_F}. The prefactor of the Landau damping term, $ \Gamma_{\hat{q}}\left(\alpha\right)$, is given by
\begin{equation}
\begin{aligned}
 \Gamma_{\hat{q}}\left(\alpha\right) &=\sum_{s,l =\pm} \frac{1}{2}\int\frac{d\theta_{\bm{k}} }{(2\pi)^2} \frac{1}{(i\alpha-\bm{v}_s(k_{F,l})\cdot \hat{\bm{q}} )}  \frac{k_{F,l}\;|  e_*(k_{F,l})|^2}{|\partial_k E_s(k_{F,l})|} \theta\left(s \lambda \cos 3\theta_{\bm{k}} \left(\sqrt{\lambda^2 \cos^2 3\theta_{\bm{k}} -(1-\lambda^2 \cos ^23\theta_{\bm{k}} )\bar{\Delta}^2} -1\right)  \right) \\
    &\times\theta\left(\lambda^2\cos^2(3\theta_{\bm{k}})-(1-\lambda^2\cos^2(3\theta_{\bm{k}}))\bar{\Delta}^2\right)\;.\label{eq:Gamma_explicit}
    \end{aligned}
\end{equation}
The velocity $\bm{v}_s(k_{F,l})$ denotes $\nabla_{\bm{k}}E_{s}(\bs{k})$ evaluated at $(k,\theta_{\bm{k}})\rightarrow (k_{F,l}(\theta_{\bm{k}}),\; \theta_{\bm{k}})$. Explicitly,
\begin{equation}
    \begin{aligned}
     v_{x,s}&= \frac{k }{m} \left(\frac{s \cos \theta_{\bm{k}} \left(k^2/2 \mu  m-1\right)}{\sqrt{\left(k^2/2 \mu  m-1\right)^2+\bar{\Delta}^2 }}-\frac{\lambda }{4} (\cos 4\theta_{\bm{k}}-5 \cos 2\theta_{\bm{k}})\right)\;,\\
     v_{y,s}&= \frac{k  }{ m}\left(\frac{s \sin \theta_{\bm{k}} \left(k^2/2 \mu  m-1\right)}{\sqrt{\left(k^2/2 \mu  m-1\right)^2+\bar{\Delta}^2 }}-\frac{\lambda }{2} \sin \theta_{\bm{k}} (6 \cos \theta_{\bm{k}} +\cos 3 \theta_{\bm{k}} )\right)\;.
    \end{aligned}
\end{equation}
Putting everything together, we find
\begin{equation}
    \begin{aligned}
\frac{k_{F,l}\;|  e_*(k_{F,l})|^2}{|\partial_k E_s(k_{F,l})|}  &= \frac{m}{\left|\lambda \cos 3\theta_{\bm{k}} +\frac{s(B_{l}(\theta_{\bm{k}})-1)}{\sqrt{\bar{\Delta}^2+(B_{l}(\theta_{\bm{k}})-1)^2}}\right|}\left(1-\frac{\bar{\Delta}^2}{\lambda^2 B_{l}^2(\theta_{\bm{k}}) \cos^23\theta_{\bm{k}}}\right),\\
    \bm{v}_s(k_{F,l})\cdot \hat{\bm{q}}&= \sqrt{\frac{2\mu}{m}}B^{1/2}_{l}(\theta_{\bm{k}}) \left(\frac{s \cos \theta_{\bm{k}} \left(B^{}_{l}(\theta_{\bm{k}})-1\right)}{\sqrt{\left(B^{}_{l}(\theta_{\bm{k}})-1\right)^2+\bar{\Delta}^2 }}-\frac{\lambda }{4} (\cos 4\theta_{\bm{k}}-5 \cos 2\theta_{\bm{k}})\right)\cos\phi_{\bm{q}}\\
    &+\sqrt{\frac{2\mu}{m}}B^{1/2}_{l}(\theta_{\bm{k}}) \left(\frac{s \sin \theta_{\bm{k}} \left(B^{}_{l}(\theta_{\bm{k}})-1\right)}{\sqrt{\left(B^{}_{l}(\theta_{\bm{k}})-1\right)^2+\bar{\Delta}^2 }}-\frac{\lambda }{2} \sin \theta_{\bm{k}} (6 \cos \theta_{\bm{k}} +\cos 3 \theta_{\bm{k}} )\right)\sin\phi_{\bm{q}}
    \end{aligned}
\end{equation}
Here we used a parameterization $\hat{\bm{q}}=(\cos\phi_{\bm{q}},\;\sin\phi_{\bm{q}})$, and defined 
\begin{equation}
B_{l}(\theta_{\bm{k}})\equiv \frac{k_{F,l}^2(\theta_{\bm{k}})}{2m\mu}= \frac{1}{1-\lambda^2\cos^2(3\theta_{\bm{k}})}\left(1+l \sqrt{\lambda^2\cos^2(3\theta_{\bm{k}})-(1-\lambda^2\cos^2(3\theta_{\bm{k}}))\bar{\Delta}^2}\right)\;.
\end{equation}
With all these explicit expressions at hand, one could in principle consider different expansions such as small $\lambda$ and/or $\bar{\Delta}$, which we leave for future work. Here we just numerically evaluate these integrals as a function of $\phi_{\bm{q}}$ where $\hat{\bm{q}}=(\cos\phi_{\bm{q}},\;\sin\phi_{\bm{q}})$, see Fig.~\ref{fig:fig_Gamma}. We note that $\operatorname{Im}\Gamma_{\hat{\bm{q}}}<0$ has a period $\pi/3$, whereas $\operatorname{Re}\Gamma_{\hat{\bm{q}}}$ is periodic with period $2\pi/3$ and vanishes at $\phi_{\bm{q}}=\pi/6$.

\begin{figure}[t!]
    \centering
\includegraphics[width=0.85\linewidth]{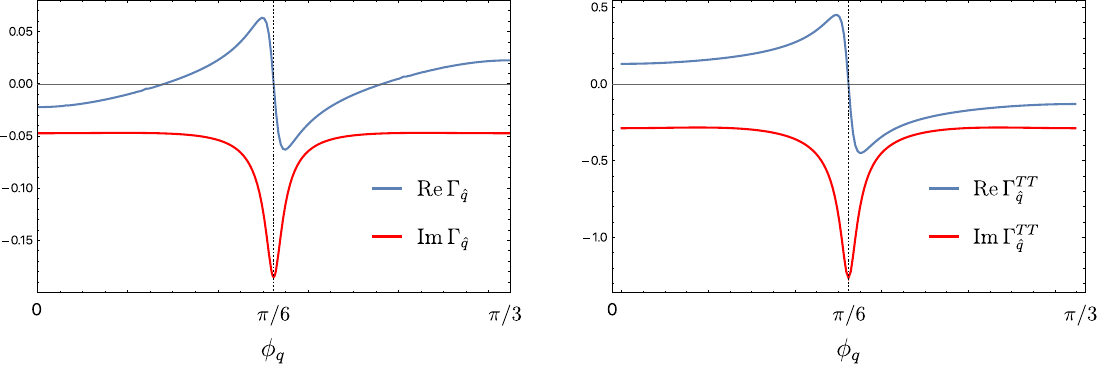}
    \caption{Left panel: angular dependence of $\Gamma_{\hat{\bm{q}}}\equiv\Gamma_{\hat{\bm{q}}}(\alpha\rightarrow0^+)$ in units of $m^{3/2}/\mu^{1/2}$, defined in Eq.~\eqref{eq:def-Gamma} (and also given more explicitly for our illustrative model in Eq.~\eqref{eq:Gamma_explicit}). Reft panel: angular dependence of $\Gamma^{TT}_{\hat{\bm{q}}}\equiv\Gamma_{\hat{\bm{q}}}^{TT}(\alpha\rightarrow0^+)$ in units of $\sqrt{m\mu}$, defined in Eq.~\eqref{eq:Gamma_IJ}. Parameters: $\lambda=0.2$, and $\Delta/\mu=0.1525$. }
    \label{fig:fig_Gamma}
\end{figure}

\subsection{Current-current correlation function}\label{subapp:BdG_JJ}
Next we calculate the current-current correlator. As before, we will focus on the London limit when $q$ is much smaller than both the inverse coherence length and the Fermi momentum of the BFS pockets. In this case, the matrix elements can be evaluated at $q=0$
\begin{equation}
\begin{aligned}
    &\Tr \left[ \hat{P}_s(\bm{k})\left(\nabla_{k_i}\varepsilon_{+,\bm{k}}\hat{\tau}_0+\nabla_{k_i}\varepsilon_{-,\bm{k}}\hat{\tau}_z\right)\hat{P}_{s'}(\bm{k})\left(\nabla_{k_j}\varepsilon_{+,\bm{k}}\hat{\tau}_0+\nabla_{k_j}\varepsilon_{-,\bm{k}}\hat{\tau}_z\right)\right]\\  
    &=\delta_{s',s} \left\{v_{i,s} v_{j,s}+\frac{|\Delta_{\bm{k}}|^2}{E_0^2(\bm{k})}\left(\nabla_{k_i}\varepsilon_{+,\bm{k}}\nabla_{k_j}\varepsilon_{+,\bm{k}}- \nabla_{k_i}\varepsilon_{-,\bm{k}}\nabla_{k_j}\varepsilon_{-,\bm{k}}\right)\right\}+\frac{\delta_{s',-s} |\Delta_{\bm{k}}|^2}{E_0^2(\bm{k})} \nabla_{k_i}\varepsilon_{-,\bm{k}}\nabla_{k_j}\varepsilon_{-,\bm{k}}
    \end{aligned}
\end{equation}
where $v_{i,s}$ is a velocity of Bogoliubov quasiparticles. The combination in curly brackets could further be written as
\begin{equation}
\begin{aligned}
    &\left(\nabla_{k_i}\varepsilon_{+,\bm{k}} +s e_*(\bm{k}) \nabla_{k_i}\varepsilon_{-,\bm{k}}\right)   \left(\nabla_{k_j}\varepsilon_{+,\bm{k}} +s e_*(\bm{k}) \nabla_{k_j}\varepsilon_{-,\bm{k}}\right)\\
    &=\frac{1}{4}\Big[(1+s e_*(\bm{k})))\nabla_{k_i}\xi_{\bm{k}}+(1-s e_*(\bm{k})))\nabla_{k_i}\xi_{-\bm{k}}\Big]\Big[(1+s e_*(\bm{k})))\nabla_{k_j}\xi_{\bm{k}}+(1-s e_*(\bm{k})))\nabla_{k_j}\xi_{-\bm{k}}\Big]
    \end{aligned}
\end{equation}
We recall that $1+e_*(\bm{k})=|u_{\bm{k}}|^2$ and $1-e_*(\bm{k})=|v_{\bm{k}}|^2$ are just the coherence factors. This allows for a simple interpretation. Each Bogoliubov quasiparticle is a superposition of an electron at $\bm{k}$ and a hole at $-\bm{k}$. With inversion breaking, those two components have different velocities, so the quasiparticle current is a coherence-factor weighted average of the two.

Again, focusing on the low-energy transitions within a given BFS pocket, we obtain
\begin{equation}
    \begin{aligned}
        G^{\rm intra}_{J^iJ^j}(\bs{q}, \Omega)
        &\approx \sum_{s = \pm } \frac{1}{2} \int \frac{d^2 \bs{k}}{(2\pi)^2} \frac{f\left[E_{s}(\bs{k}+\bs{q}/2)\right] - f\left[E_{s}(\bs{k}-\bs{q}/2)\right]}{i\Omega - E_{s}(\bs{k}+\bs{q}/2) + E_{s}(\bs{k}-\bs{q}/2)} \Big(\nabla_{k_i}\varepsilon_{+,\bm{k}} +s e_*(\bm{k}) \nabla_{k_i}\varepsilon_{-,\bm{k}}\Big)   \Big(\nabla_{k_j}\varepsilon_{+,\bm{k}} +s e_*(\bm{k}) \nabla_{k_j}\varepsilon_{-,\bm{k}}\Big) \;.
    \end{aligned}
\end{equation}
For small $\Omega$ and $q$ but $\Omega/q$ fixed, we find
\begin{equation}
     G^{\rm intra}_{J^iJ^j}(\bm{q},\Omega) \approx \chi_J \delta_{ij} -\frac{i\Omega}{ q}  \Gamma^{ij}_{\hat{\bm{q}}}\left(\frac{\Omega}{q}\right) \;.\label{eq:Pi_JJ_full}
\end{equation}
Here $\chi_J$ is the static contribution from the BFS given by
\begin{equation}
     \chi_J = \frac{1}{2}\sum_{s = \pm} \int\frac{d^2\bm{k} }{(2\pi)^2}\Big(\nabla_{k_1}\varepsilon_{+,\bm{k}} +s e_*(\bm{k}) \nabla_{k_1}\varepsilon_{-,\bm{k}}\Big)^2    \delta(E_s(\bm{k}))\;.
\end{equation}
Here we set $i=j=1$ for the velocity components because this tensor is symmetric and diagonal due to the $C_3$ symmetry. The function $\Gamma_{\hat{\bm{q}}}\left(q/\Omega\right)$ is given by
\begin{equation}
\Gamma^{ij}_{\hat{\bm{q}}}\left(\Omega/q\right) = \frac{1}{2}\sum_{s = \pm} \int\frac{d^2\bm{k} }{(2\pi)^2} \frac{ \delta(E_s(\bm{k})) }{i\Omega/q-\bm{v}_s\cdot \hat{\bm{q}} }\Big(\nabla_{k_i}\varepsilon_{+,\bm{k}} +s e_*(\bm{k}) \nabla_{k_i}\varepsilon_{-,\bm{k}}\Big)   \Big(\nabla_{k_j}\varepsilon_{+,\bm{k}} +s e_*(\bm{k}) \nabla_{k_j}\varepsilon_{-,\bm{k}}\Big) \;.\label{eq:Gamma_IJ}
\end{equation}
Projection to longitudinal and transverse components gives 
\begin{equation}
    \Gamma^{LL}_{\hat{\bs{q}}} = \hat q_i \Gamma^{ij}_{\hat{\bs{q}}} \hat q_j \,, \quad \Gamma^{LT}_{\hat{\bs{q}}} = \hat q_i \Gamma^{ij}_{\hat{\bs{q}}} (\hat{z} \times \hat q)_j \,, \quad \Gamma^{TT}_{\hat{\bs{q}}} = (\hat{z} \times \hat q)_i \Gamma^{ij}_{\hat{\bs{q}}} (\hat{z} \times \hat q)_j  \,. 
\end{equation}
Taking the $\Omega \rightarrow 0$ limit in $\Gamma^{TT}_{\hat{\bs{q}}}$ produces the right panel of Fig.~\ref{fig:fig_Gamma}. 

Similarly to Eq.~\eqref{eq:interband}, interband processes contribute another constant and higher order corrections analytic in $q^2$ and $\Omega^2$
\begin{equation}
   (\text{regular interband contributions}) \approx \frac{1}{2}\int \frac{d^2 \bs{k}}{(2\pi)^2}\Big(f\left[E_{-}(\bs{k})\right] - f\left[E_{+}(\bs{k})\right]\Big)\frac{|\Delta_{\bm{k}}|^2}{E_0^3(\bm{k})} \nabla_{k_i}\varepsilon_{-,\bm{k}}\nabla_{k_j}\varepsilon_{-,\bm{k}}+\mathcal{O}(\Omega^2, \; q^2) 
\end{equation}
The constant term here is again proportional to $\delta_{ij}$ due to the $C_3$ symmetry, and thus it should be added to $\chi_J$ (we denote their sum as $\tilde{\chi}_J$). The full answer is therefore
\begin{equation}\label{eq:GJJfull}
    G_{J^i J^j}(\bs{q},\Omega) = \tilde \chi_J - \frac{i\Omega}{q} \Gamma^{ij}_{\hat{\bs{q}}} \left(\frac{\Omega}{q}\right) + \mathcal{O}(\Omega^2, q^2) \,. 
\end{equation}

Using Eq.~\eqref{eq:diamag_BdG} in combination with the results above, we find the following expression for the static superfluid stiffness 
 \begin{equation}
\begin{aligned}   \rho_s(\Omega=0,\bm{q}\rightarrow 0)&= \frac{1}{4}(k_{\rm diam} - \tilde{\chi}_J)=\frac{1}{8}\sum\limits_{s=\pm} \int \frac{d^2\bm{k} }{(2\pi)^2} \Big\{\frac{s|\Delta_{\bm{k}}|^2}{E_0^3(\bm{k})} f\left[E_{s}(\bs{k})\right] \left((\nabla_{k_1}\varepsilon_{-,\bm{k}})^2 -\nabla_{k_1}\varepsilon_{+,\bm{k}}\nabla_{k_1}\xi(\bm{k}) \right)\\
   &+\left[(1+se_*(\bm{k}))\nabla_{k_1}\xi(\bm{k})\nabla_{k_1}E_s(\bm{k})-\Big(\nabla_{k_1}\varepsilon_{+,\bm{k}} +s e_*(\bm{k}) \nabla_{k_1}\varepsilon_{-,\bm{k}}\Big)^2  \right]  \delta(E_s(\bm{k}))\Big\}\;.\label{eq:rho_s_full_exp}
   \end{aligned}
\end{equation}
Here we integrated by parts once in the diamagnetic contribution. For a specific dispersion in Eq.~\eqref{eq:xi_C3}, the calculation of these integrals can be performed using the explicit formulas presented in Sec.~\ref{subapp:BdG_rhorho} for the density-density correlator.

\subsection{Density-current correlation function}\label{subapp:BdG_rhoJ}
Finally, we consider the current-density correlator. The corresponding matrix element at $\bm{q}=0$ reads as
\begin{equation}
    \Tr \left[ \hat{P}_s(\bm{k})\left(\nabla_{k_i}\varepsilon_{+,\bm{k}}\hat{\tau}_0+\nabla_{k_i}\varepsilon_{-,\bm{k}}\hat{\tau}_z\right)\hat{P}_{s'}(\bm{k})\hat{\tau}_z\right]=s\delta_{s',s}\Big\{\nabla_{k_i}\varepsilon_{+,\bm{k}}+s e_*(\bm{k}) \nabla_{k_i}\varepsilon_{-,\bm{k}}\Big\}e_*(\bm{k})    +\frac{\delta_{s',-s} |\Delta_{\bm{k}}|^2}{E_0^2(\bm{k})} \nabla_{k_i}\varepsilon_{-,\bm{k}}\;.
\end{equation}
The intraband contribution with $s'=s$ reads as
\begin{equation}
    \begin{aligned}
        G^{\rm intra}_{J^i \rho}(\bs{q}, \Omega)
        &\approx \sum_{s = \pm } \frac{s}{2}\int \frac{d^2 \bs{k}}{(2\pi)^2} \frac{f\left[E_{s}(\bs{k}+\bs{q}/2)\right] - f\left[E_{s}(\bs{k}-\bs{q}/2)\right]}{i\Omega - E_{s}(\bs{k}+\bs{q}/2) + E_{s}(\bs{k}-\bs{q}/2)} \Big(\nabla_{k_i}\varepsilon_{+,\bm{k}} +s e_*(\bm{k}) \nabla_{k_i}\varepsilon_{-,\bm{k}}\Big)   e_*(\bm{k}) \;.
    \end{aligned}
\end{equation}
Thus, for small $\Omega$ and $q$ but $\Omega/q$ fixed, we find
\begin{equation}
     G_{J^i\rho}(\bm{q},\Omega) \approx -\frac{i\Omega}{ q}  \Gamma^{i}_{\hat{\bm{q}}}\left(\frac{\Omega}{q}\right) + \mathcal{O}(\Omega^2, q^2) \;,\label{eq:GrhoJ_full}
\end{equation}
where the subleading $\mathcal{O}(\Omega^2, q^2)$ terms come from regular interband contributions and $\Gamma^i_{\hat{\bs{q}}}$ is defined as
\begin{equation}
\Gamma^{i}_{\hat{\bm{q}}}\left(\Omega/q\right) = \frac{1}{2}\sum_{s = \pm} \int\frac{d^2\bm{k} }{(2\pi)^2} \frac{ \delta(E_s(\bm{k})) }{i\Omega/q-\bm{v}_s\cdot \hat{\bm{q}} }\Big(s\nabla_{k_i}\varepsilon_{+,\bm{k}} + e_*(\bm{k}) \nabla_{k_i}\varepsilon_{-,\bm{k}}\Big) e_*(\bm{k}) \;.
\end{equation}

\subsection{Alternative formulation of the Ginzburg-Landau theory: reviving gapless Bogoliubov fermions}\label{subapp:LG_withBFS}

In principle, the phase-only effective action \eqref{eq:LG_phaseonly_app} allows us to compute all gauge-invariant response functions. However, the compact representation \eqref{eq:LG_phaseonly_app} has an important drawback: since fermions are completely integrated out, singular effects of the gapless BFS are hidden in some subtle non-analytic coefficients of \eqref{eq:LG_phaseonly_app}. To make the gapless BFS more transparent, it is thus useful to have an extended Ginzburg-Landau action in which the gapless quasiparticles are included explicitly. 

Towards that end, we return to the full action that involves both $f$ and $\phi$. Adopting a Wilsonian approach, we can first integrate out momentum modes of $f$ that are separated from the mean-field BFS by a momentum cutoff $\Lambda$. Since these modes are gapped, they generate analytic contributions to the effective action 
\begin{equation}
    S_{\rm analytic}[\phi] \approx \frac{i \bar \rho_f}{2} \int d \tau d^2\bs{r} \partial_{\tau} \phi + \frac{1}{8} \int d^3 x \, \partial_{\mu} \phi \, \Pi^{\mu\nu}_{\rm BdG, > \Lambda}  \,\partial_{\nu} \phi \,,
\end{equation}
where $\Pi^{\mu\nu}_{\rm BdG, > \Lambda}$ are reduced correlation functions in the mean-field BdG state in which all momentum loop integrals avoid a ring of width $\Lambda$ around the BFS. By excluding the most singular gapless quasiparticles near the BFS, $\Pi^{\mu\nu}_{\rm BdG, > \Lambda}$ capture regular renormalizations of the compressibility and superfluid stiffness that are analytic in $\bs{q}$ and $\omega$ (see for example the constant shifts to $G_{\rho\rho}$ and $G_{J^iJ^j}$ in \eqref{eq:Grhorhofull} and \eqref{eq:GJJfull}).

The remaining fermionic action is $S_{\rm BFS, <\Lambda}[f]$, which only includes momentum modes of $f$ that are within distance $\Lambda$ of the BFS. When we turn on the background gauge field $A$, the minimally coupled action can be constructed by replacing $\partial_{\mu}$ with covariant derivatives $\partial_{\mu} - i A_{\mu}$ in the fermionic action $S_{\rm BFS, <\Lambda}[f]$ and replacing $\partial_{\mu} \phi$ with $\partial_{\mu} \phi - 2 A_{\mu}$ in the phase action. The full theory can therefore be written as
\begin{equation}\label{eq:LG_withBFS_app}
    S_{\rm eff}[f, \phi, A] = S_{\rm BFS, < \Lambda}[f, A] + S_{\rm analytic}[\phi, 2A] \,. 
\end{equation}
Upon further integrating out momentum modes of $f$ within distance $\Lambda$ of the BFS, we recover the phase-only action \eqref{eq:LG_phaseonly_app}. This extended action allows us to address several important questions:
\begin{enumerate}
    \item The Bogoliubov fermions themselves generically have some residual interactions beyond the mean-field limit. These interactions could dress the Bogoliubov quasiparticles with a renormalized energy and/or a finite lifetime. Both of these effects can be treated in a unified fashion by adding terms quartic in $f$ to \eqref{eq:LG_withBFS_app}. Note that even though the $U(1)$ symmetry is broken in the mean-field state, we don't generate new quadratic terms like $f^\dagger f^\dagger$ in the effective theory since these terms are eliminated by an appropriate choice of the saddle point (i.e. an appropriate Bogoliubov rotation in the Hamiltonian language).
    \item When the gauge field $A$ is dynamical and contains additional Chern-Simons terms (as is the case in the CBFL phase), retaining the fermion $f$ allows us to identify global topological properties of the system that are not accessible in the phase-only formulation. 
    \item At quantum phase transitions out of the BFS, quantum critical fluctuations could lead to singular damping effects of the gapless Bogoliubov quasiparticles. In these cases, directly integrating out the BFS generates an infinite series of non-analytic corrections to the phase mode effective action, which cannot be truncated at quadratic order. Retaining the Bogoliubov fermions explicitly in the action allows us to study such quantum phase transitions in a more controlled manner. 
\end{enumerate}

\section{Low energy observables in the CBFL phase}\label{app:CBFL_properties}

In this section, we use our results obtained for a Bogoliubov Fermi surface of electrons in Sec.~\ref{app:LGtheory} to calculate physical response functions and other observables in the Composite Bogoliubov Fermi liquid (CBFL). To this end, we shift the external gauge field $A$ in Eq.~\eqref{eq:CBFL_response_function} to be $A+a$, where $a$ is the emergent $U(1)$ Chern-Simons gauge field, and add a short-ranged density-density repulsion $v(r)$. We will be working in the Coulomb gauge. The full action becomes
\begin{equation}\label{eq:ACFL_app}
\begin{aligned}
    S_{\rm ACFL} &= \int_{\tau, \bs{r}} \bar f(\bs{r},\tau) [\partial_{\tau}-i (a_0+A_0)  + \xi_{-i \nabla -\bm{a}-\bm{A}}] f(\bs{r},\tau) - \frac{1}{4}\int_{\tau, \bs{r}, \bs{r}'} V(\bs{r} - \bs{r}') \bar f(\bs{r}, \tau) \bar f(\bs{r}', \tau) f(\bs{r}',\tau) f(\bs{r},\tau) \\
    &+ \frac{i}{4\pi}\int_{\tau, \bs{r}} a_0 (\bm{\nabla} \times \bm{a})_z+\frac{1}{2}\int_{\tau, \bs{r}, \bs{r}'} v(\bs{r} - \bs{r}') \bar f(\bs{r}', \tau) f(\bs{r}', \tau) \bar{f}(\bs{r},\tau) f(\bs{r},\tau)\;.
    \end{aligned}
\end{equation} 
Our assignment of signs for the gauge field is consistent with our convention in Eq.~\eqref{eq:IRaction_app_withfermion}. The equation of motion for $a_0$ implements the usual flux attachment constraint $\rho_f = \frac{1}{4\pi} (\bm{\nabla} \times \bm{a})_z$. It will be convenient to separate the transverse and longitudinal components of the gauge fields as
\begin{equation}
    \quad \bm{a}(\bs{q}) = a_L(\bs{q}) \hat{\bm{q}} + a_T(\bs{q}) \hat z \times \hat{\bm{q}}\;,\quad \quad a_{L/T}^*(\bs{q})=  -a_{L/T}(-\bs{q})\;.
\end{equation}
In the Coulomb gauge $ a_L(\bs{q})=0$, and thus we find
\begin{gather}
L_{\rm CBFL}= \frac{1}{2} (A^*_{\alpha}(\bs{q}, \Omega)+a^*_{\alpha}(\bs{q}, \Omega)) \Pi_{\rm BFL}^{\alpha\beta}(\bs{q}, \Omega)( A_{\beta}(\bs{q}, \Omega)+a_{\beta}(\bs{q}, \Omega))+
    \frac{1}{2} a^*_{\alpha}(\bs{q}, \Omega) \Pi_{a}^{\alpha\beta}(\bs{q}, \Omega)a_{\beta}(\bs{q}, \Omega)\;,\notag\\
  \Pi_{ a}^{\alpha\beta} =\begin{pmatrix}
     0   & \frac{q}{4\pi}\\
     - \frac{q}{4\pi}  & \frac{q^2 v_0}{(4\pi)^2}
    \end{pmatrix},\quad \quad \quad  \Pi_{\rm BFL}^{\alpha\beta}=   \begin{pmatrix}
   \Pi_{\rm BdG}^{00}   & \Pi_{\rm BdG}^{T0}\\
  \Pi_{\rm BdG}^{T0} & \Pi_{\rm BdG}^{TT}
    \end{pmatrix} -\notag \\
    -\frac{1}{\Omega^2 \Pi_{\rm BdG}^{00}+q^2 \Pi_{\rm BdG}^{LL}+2q\Omega \Pi_{\rm BdG}^{L0}}\begin{pmatrix}
     \left[\Omega\Pi_{\rm BdG}^{00}+ q \Pi_{\rm BdG}^{L0}\right]^2   & \left[\Omega \Pi_{\rm BdG}^{T0}+q\Pi_{\rm BdG}^{LT}\right]\left[\Omega \Pi_{\rm BdG}^{00}+q \Pi_{\rm BdG}^{L0}\right]\\
    \left[\Omega \Pi_{\rm BdG}^{T0}+q\Pi_{\rm BdG}^{LT}\right]\left[\Omega \Pi_{\rm BdG}^{00}+q \Pi_{\rm BdG}^{L0}\right] & [\Omega\Pi_{\rm BdG}^{T0}+q\Pi_{\rm BdG}^{LT}]^2
    \end{pmatrix} 
    \label{eq:CBFL_response_function}
\end{gather}
where $\alpha,\beta=0, T$. Also, $v_0$ is the  repulsion strength, and we suppressed the frequency and momentum arguments of all correlators. The $2$x$2$ matrix of the mean-field correlators $\Pi_{\rm BdG}^{\alpha\beta}$ is obtained from the $3$x$3$ matrix $\Pi_{\rm BdG}^{\mu\nu}$ given in Eq.~\eqref{eq:Pi_MF_munu} by projecting its current operators onto the transverse direction. Note that in the inversion-broken system, the current-density correlators $\Pi_{\rm BdG}^{T0}$ and $\Pi_{\rm BdG}^{L0}$ are finite at finite frequency, and so is the transverse/longitudinal current correlator. However, in the static limit these correlators vanish (e.g. the static term in Eq.~\eqref{eq:Pi_JJ_full} is diagonal and does not couple transverse and longitudinal currents, etc.), and so we recover $\Pi_{\rm BFL}^{TT}(\bm{q}\rightarrow 0, \Omega=0)=\Pi_{\rm BdG}^{TT}(\bm{q}\rightarrow 0, \Omega=0)$, as in a fully gapped superconductor.

Integrating out $a$ then gives the physical response theory
\begin{equation}
    L_{\rm CBFL} = \frac{1}{2} A_{\alpha}^* \Pi^{\alpha\beta}_{\rm CBFL} A_{\beta} \,, \quad 
    \Pi_{\rm CBFL} = \Pi_{\rm a} (\Pi_{\rm BFL} + \Pi_{\rm a})^{-1} \Pi_{\rm BFL} \,.\label{eq:Pi_CBFL_full}
\end{equation}
Inverting this relation gives the Ioffe-Larkin composition rule~\cite{Ioffe1989_rule} 
\begin{equation}
    \Pi_{\rm CBFL}^{-1} = \Pi_{\rm a}^{-1} + \Pi_{\rm BFL}^{-1} \,.\label{eq:Ioffe-Larkin_2} 
\end{equation}
Let us now pick $\hat{\bm{q}}=\hat{e}_x$, $\hat{z}\times \hat{\bm{q}}=\hat{e}_y$, and so $A_{T}=A_y$. The conductivity tensor is obtained as
\begin{equation}
    \sigma(\bm{q},\Omega)= \frac{1}{\Omega}\begin{pmatrix}
        \Pi_{\rm CBFL}^{LL}(\bm{q},\Omega) &  \Pi_{\rm CBFL}^{LT}(\bm{q},\Omega)\\
        \Pi_{\rm CBFL}^{TL}(\bm{q},\Omega)&  \Pi_{\rm CBFL}^{TT}(\bm{q},\Omega)
    \end{pmatrix}=\frac{1}{\Omega}\begin{pmatrix}
        -\Omega/q &0\\
        0& 1
    \end{pmatrix}\begin{pmatrix}
        \Pi_{\rm CBFL}^{00}(\bm{q},\Omega) &  \Pi_{\rm CBFL}^{0T}(\bm{q},\Omega)\\
        \Pi_{\rm CBFL}^{T0}(\bm{q},\Omega)&  \Pi_{\rm CBFL}^{TT}(\bm{q},\Omega)
    \end{pmatrix}\begin{pmatrix}
        -\Omega/q &0\\
        0& 1
    \end{pmatrix}\;,\label{eq:conductivity_q_omega}
\end{equation}
where we expressed the longitudinal current components using the relation in  Eq.~\eqref{eq:gauge_inv}. More generally, for any gauge invariant response function  (including $\Pi_{\rm CBFL}$ and $\Pi_{\rm BFL}$) we have
\begin{equation}\label{eq:LLto00}
    \Pi_{}^{LL}(\bm{q},\Omega) = \frac{\Omega^2}{q^2}\Pi_{}^{00}(\bm{q},\Omega) ,\quad\quad  \Pi_{}^{LT}(\bm{q},\Omega)= -\frac{\Omega}{q}\Pi_{}^{0T}(\bm{q},\Omega),\quad\quad  \Pi_{}^{TL}(\bm{q},\Omega)= -\frac{\Omega}{q}\Pi_{}^{T0}(\bm{q},\Omega) \,.
\end{equation}
We note in passing that these relations are consistent with the continuity equation following from Eq.~\eqref{eq:IRaction_app_withfermion}
\begin{equation}\label{eq:continuity_eq}
     J_L(\bm{q},\Omega)= -\frac{i\Omega}{q}\rho(\bm{q},\Omega)\;,\quad J_L(-\bm{q},-\Omega)= -\frac{i\Omega}{q}\rho(-\bm{q},-\Omega)\;.
 \end{equation}
 To see that, we need to recall our sign convention in $\Pi^{\mu\nu}$ Eq.~\eqref{eq:Pi_MF_munu}. For example, using Eq.~\eqref{eq:continuity_eq} we obtain
 \begin{equation}
     \Pi_{\rm BFL}^{LT}(\bm{q},\Omega)=  -G_{J^L J^T}^{}(\bm{q},\Omega)=\frac{i\Omega}{q}G_{\rho J^T}^{}(\bm{q},\Omega)= -\frac{\Omega}{q}\Pi^{0T}_{\rm BFL}(\bm{q},\Omega)\;.
 \end{equation}
We also emphasize that the relations in Eq.~\eqref{eq:LLto00} and Eq.~\eqref{eq:continuity_eq} do not hold for correlators evaluated with the BdG mean-field Hamiltonian since those are not gauge invariant.

After performing analytic continuation in Eq.~\eqref{eq:conductivity_q_omega}, inverting it using Eq.~\eqref{eq:Ioffe-Larkin_2}, and taking the limit $\bm{q}\rightarrow 0$, we arrive at Eq.~\eqref{eq:Ioffe-Larkin} of the main text.

\subsection{Density-density correlator and equal-time structure factor}\label{subapp:density_CBFL}

The Landau damping associated with gapless Bogoliubov Fermi surfaces manifests itself in the physical density-density correlator for electrons, which is given by $G_{\rho_c\rho_c}^R(\bm{q},\omega) = \Pi_{\rm CBFL}^{00,R} (\bm{q},\omega)$. Using the results of Secs.~\ref{subapp:BdG_rhorho},\;\ref{subapp:BdG_JJ}, and \ref{subapp:BdG_rhoJ}, we see that for $q \ll k_F$ and $\omega\ll E_F$,  $\Pi_{\rm BdG}$ has the structure
\begin{equation}
  \Pi^R_{\rm BdG}(\bs{q}, \omega)
    \approx \begin{pmatrix}
       \tilde{\chi}_0 - \Gamma_{\hat{\bm{q}}} \omega/q & i\Gamma^T_{\hat{\bm{q}}}\omega/q \\ i\Gamma^T_{\hat{\bm{q}}}\omega/q & (k_{\rm diam}-\tilde{\chi}_J) + \Gamma^{TT}_{\hat{\bm{q}}} \omega/q  
    \end{pmatrix}\;,\label{eq:Pi_bdG_full}
\end{equation}
where $\tilde{\chi}_0$, $\tilde{\chi}_J$, and $k_{\rm diam}$  are real positive constants, and $\Gamma_{\hat{\bm{q}}}$, $\Gamma_{\hat{\bm{q}}}^T$, and $\Gamma_{\hat{\bm{q}}}^{TT}$ are complex functions of $\omega/q +i0^+$, which in the static limit $|\omega|\ll q$ approach constant complex values. We also recall that  in this static limit $\operatorname{Im}\Gamma_{\hat{\bm{q}}}<0$, $\operatorname{Im}\Gamma^{TT}_{\hat{\bm{q}}}<0$ (see also Fig.~\ref{fig:fig_Gamma}), and $\rho_s=(k_{\rm diam}-\tilde{\chi}_J)/4\;>0$ is the static superfluid stiffness (cf. Eq.~\eqref{eq:rho_s_full_exp}). The remaining correlators entering in the second term in Eq.~\eqref{eq:CBFL_response_function} can also be deduced in a similar way from the expressions given Secs.~\ref{subapp:BdG_rhorho},\;\ref{subapp:BdG_JJ}, and \ref{subapp:BdG_rhoJ} (note that one cannot use the continuity equation for $\Pi_{\rm BdG}$ because the correlators entering $\Pi_{\rm BdG}$ are computed in a mean-field state with broken $U(1)$ symmetry):
\begin{equation}
    \Pi_{\rm BdG}^{L0,R}(\bs{q}, \omega)=i  \Gamma^L_{\hat{\bm{q}}} \frac{\omega}{q},\quad  \quad\Pi_{\rm BdG}^{LL,R}(\bs{q}, \omega)=4\rho_s+\Gamma^{LL}_{\hat{\bm{q}}} \frac{\omega}{q},\quad \quad \Pi_{\rm BdG}^{LT,R}(\bs{q}, \omega)=\Gamma^{LT}_{\hat{\bm{q}}} \frac{\omega}{q}\;.
\end{equation}
We note in passing that in the normal state we have $\Gamma^{LL}_{\hat{\bm{q}}}=0$ and $\Gamma^{LT}_{\hat{\bm{q}}}=0$, as expected from the continuity equation.

Finally, using Eq.~\eqref{eq:Pi_CBFL_full} and expanding at small $\omega/q$ we obtain 
\begin{equation}
    G_{\rho_c\rho_c}^R(\bm{q},\omega) \approx \frac{q^2}{4(4\pi)^2\rho_s} \left(1 -\frac{\Gamma_{\hat{\bm{q}}}^{TT}}{4\rho_s}\;\frac{\omega}{q} \right),\;\quad |\omega|\ll v_F q\;.
    \label{eq:G_cc_omega_exp}
\end{equation}
We note that $ G_{\rho_c\rho_c}^R(\bm{q}\rightarrow 0,\omega=0) \sim q^2$, and therefore the CBFL state is incompressible.

Despite the fact the electron density propagator is massive, the appearance of the Landau damping still has consequences for the equal-time structure factor $ S(\bm{q})$, which is most conveniently obtained in the Euclidean representation as $ S(\bm{q})\equiv G_{\rho_c\rho_c}(\bm{q}\rightarrow 0,\tau=0)$. Specifically, we find
\begin{equation}
     S(\bm{q}) \equiv \int \frac{d\Omega}{2\pi} G_{\rho_c \rho_c}(\bs{q},\Omega)  \approx \frac{q^2}{ 32 \pi ^3} \int_{0}^\Lambda d\Omega \;\left(f_{\hat{\bm{q}}}\left(\frac{\Omega}{q}\right)+f_{\hat{\bm{q}}}\left(-\frac{\Omega}{q}\right)\right)\label{eq:S_q_full}\;,
\end{equation}
where the integral converges at small frequencies, and the UV cutoff was introduced because in this expression we only accounted for the non-analytic dynamical contributions (i.e. Landau damping) arising from the BFS. The explicit form of the function $f_{\hat{\bm{q}}}\left(\Omega/q\right)$ is
\begin{equation}
    f_{\hat{\bm{q}}}(\Omega/q)= \frac{(\Pi^{L0}_{\rm BdG})^2-\Pi^{00}_{\rm BdG} \Pi^{LL}_{\rm BdG}}{\Pi^{00}_{\rm BdG} \left((\Pi ^{LT}_{\rm BdG})^2-\Pi ^{LL}_{\rm BdG} \Pi ^{TT}_{\rm BdG}\right)+(\Pi ^{L0}_{\rm BdG})^2 \Pi ^{TT}_{\rm BdG}-2 \Pi ^{L0}_{\rm BdG} \Pi ^{LT}_{\rm BdG} \Pi ^{T0}_{\rm BdG}+\Pi ^{LL}_{\rm BdG} (\Pi ^{T0}_{\rm BdG})^2}\;.\label{eq:_def_f}
\end{equation}

At this point, we have not made any assumptions regarding the ratio $\Omega/q$, so all the coefficients such as $\Gamma^{TT}_{\hat{\bm{q}}}$, etc., in Eq.~\eqref{eq:_def_f} retain their full dependence on $\Omega/q$. Since we are interested in the small $q$ limit of Eq.~\eqref{eq:S_q_full}, we rescale $\Omega \equiv q\alpha $ and analyze the asymptotics of the remaining integral as a function of the cutoff. Unlike the low-frequency regime $\Omega \ll q$, where non-analytic dependence on frequency arises through $|\Omega|$, the polarization functions $\Pi_{\rm BdG}$ admit analytic expansions at large $\Omega$. Consequently, we find:
\begin{equation}
     f_{\hat{\bm{q}}}(\alpha) = f_{\hat{\bm{q}}}^{(0)}+\frac{f_{\hat{\bm{q}}}^{(1)}}{i \alpha}+\frac{f_{\hat{\bm{q}}}^{(2)}}{\alpha^2}+...
\end{equation}
The second term here is frequency-odd and so it does not contribute to Eq.~\eqref{eq:S_q_full}. As a result, we can rearrange the integral as follows
\begin{equation}
\begin{aligned}
   q\int_{0}^{\Lambda/q} d\alpha \;\left(f_{\hat{\bm{q}}}\left(\alpha\right)+f_{\hat{\bm{q}}}\left(-\alpha\right)\right) &= 2f_{\hat{\bm{q}}}^{(0)} \Lambda +q\int_0^{\Lambda/q} d\alpha (f_{\hat{\bm{q}}}(\alpha)+f_{\hat{\bm{q}}}(-\alpha)-2f_{\hat{\bm{q}}}^{(0)})\\
   &=2f_{\hat{\bm{q}}}^{(0)} \Lambda+q\int_0^{\infty} d\alpha (f_{\hat{\bm{q}}}(\alpha)+f_{\hat{\bm{q}}}(-\alpha)-2f_{\hat{\bm{q}}}^{(0)}) -q\int_{\Lambda/q}^{\infty} d\alpha (f_{\hat{\bm{q}}}(\alpha)+f_{\hat{\bm{q}}}(-\alpha)-2f_{\hat{\bm{q}}}^{(0)})
   \end{aligned}
\end{equation}
Here the second term provides a non-analytic dependence on $q=|\bm{q}|$ with a cutoff-independent prefactor, whereas 
the last term produces a series of negligible corrections starting with $\sim q^2/\Lambda$. After collecting all contributions, we obtain
\begin{equation}
     S(\bm{q})  \approx (\text{ cutoff-dependent prefactor})\times  q^2 + \frac{q^3}{32\pi^3}\int_0^{\infty} d\alpha \Big(f_{\hat{\bm{q}}}(\alpha)+f_{\hat{\bm{q}}}(-\alpha)-2f_{\hat{\bm{q}}}(\alpha\rightarrow \infty)\Big) +\mathcal{O}(q^4)\;.
\end{equation}
The prefactor of the $q^3$ term can, in principle, be computed explicitly from the known form of $f_{\hat{\bm{q}}}(\alpha)$. Importantly, it diverges as the superfluid stiffness approaches zero. This is expected, as gapless gauge field fluctuations in the normal state generate a stronger non-analyticity of the form $\sim q^3\ln (1/q)$.

To illustrate this general behavior, we consider the following toy integral, which captures the emergence of the $q^3$ term without any additional logarithmic corrections when the gauge field has both a finite mass $\rho_s$ and Landau damping (taken here to be the same as in the isotropic normal state):
\begin{equation}
     q^2\int_0^\Lambda \frac{d\Omega}{\rho_s+\frac{|\Omega|}{\sqrt{ q^2+\Omega^2}}}\approx \mathcal{O}(q^2) + q^3\left(\frac{2\arctan(\sqrt{(1-\rho_s)/(1+\rho_s)})}{(1-\rho_s)^{3/2}}-\frac{1}{1-\rho_s}\right)+...
\end{equation}
where the prefactor of the $q^3$ term diverges logarithmically $\sim \ln 1/\rho_s$ at small $\rho_s$. 

\subsection{Single-particle spectral function}\label{subapp:monopoles}

In this section, we compute the electron Green's function in the CBFL phase. Recall that the basic composite fermion Lagrangian at $\nu = 1/2$ takes the general form 
\begin{equation}
    L[f, a, \alpha, A] = L_{\rm CF}[f, a] - \frac{2i}{4\pi} \alpha d \alpha + \frac{i}{2\pi} \alpha d (A-a) \,.
\end{equation}
For the electron Green's function on the infinite plane, the global topological properties of the Chern-Simons term do not play an important role and we can safely integrate out the gauge field $\alpha$ in favor of $a$. The resulting Lagrangian takes the more familiar form
\begin{equation}
    L[f,a,A] = L_{\rm CF}[f,a] + \frac{i}{8\pi} a\, da + \frac{i}{8\pi} A \, d A - \frac{i}{4\pi} a \, d A \,. 
\end{equation}
The mutual Chern-Simons term between $a$ and $A$ implies that the monopole operator $\mathcal{M}_a^{\dagger,2}$ that inserts $4\pi$ flux of $a$ carries electric charge $-1$ under $A$. On the other hand, the self Chern-Simons term for $a$ implies that $\mathcal{M}_a^{\dagger,2}$ carries gauge charge $1$ under $a$. Therefore, the gauge invariant electron operator that carries $+1$ charge under $A$ and is neutral under $a$ can be identified with $c^{\dagger} = f^{\dagger} \mathcal{M}_a^2$. 

Now let us enter the CBFL phase. At the mean-field level, the gauge flux of $a$ vanishes and the $f$ sector forms a gapless Bogoliubov Fermi liquid. To compute the electron Green's function, let us introduce a Wilson line operator 
\begin{equation}
    \mathcal{W}_a[\gamma] = e^{i\int_{\gamma} a_{\mu} dx^{\mu}} \,, \label{eq:W_Wilson}
\end{equation}
where $\gamma$ is the straight-line path connecting $(0,0)$ to $(\bs{x},\tau)$ where $\tau$ is the Euclidean time. In terms of the Wilson line, we can decompose the electron Green's function as 
\begin{equation}
    G_c(\bs{x}, \tau) \equiv \ev{c^{\dagger}(\bs{x}, \tau) c(0,0)} = \ev{f^{\dagger}(\bs{x},\tau) \mathcal{W}_a[\gamma] f(0,0) \mathcal{M}_a^2(\bs{x},\tau)\mathcal{W}^{\dagger}_a[\gamma]\mathcal{M}_a^{\dagger,2}(0,0)} \,. 
\end{equation}
Since the formation of the CF superconductor Higgses the dynamical gauge field $a$, fluctuations of $a$ are strongly suppressed in the low energy limit. This suppression allows us to approximately factorize the electron Green's function as a product of two gauge-invariant correlation functions 
\begin{equation}
    G_c(\bs{x}, \tau) \approx G_f(\bs{x},\tau) G_{\mathcal{M}_a^2}(-\bs{x},-\tau) \,, 
\end{equation}
where 
\begin{equation}
    G_f(\bs{x},\tau) = \ev{f^{\dagger}(\bs{x},\tau) \mathcal{W}_a[\gamma]f(0,0)} \,, \quad G_{\mathcal{M}_a^2}(\bs{x},\tau) = \ev{\mathcal{M}_a^{\dagger,2}(\bs{x},\tau) \mathcal{W}^{\dagger}_a[\gamma] \mathcal{M}_a^2(0,0)} \,. \label{eq:correlator_monopoles_general}
\end{equation}

Now let us study these two correlation functions in turn. At the mean-field level, the gapless BFS gives a power-law decaying Green's function for $f$. Since $a$ is Higgsed, the added Wilson line does not contribute in the low energy limit and $G_f(\bs{x},\tau)$ inherits the power law decay. The more subtle object is the monopole correlation function $G_{\mathcal{M}_a^2}(\bs{x}, \tau)$. Upon integrating out $f$ and setting the background gauge field $A$ to zero, we have the effective Lagrangian
\begin{equation}
    L_{\rm CBFL} = L[\phi, 2a] + \frac{i}{8\pi} a_{\mu} \epsilon_{\mu\nu\lambda} \partial_{\nu} a_{\lambda} + \frac{1}{2e^2} (\epsilon_{\mu\nu\lambda} \partial_{\nu} a_{\lambda})^2 \,, \quad L[\phi, 2a] = \frac{1}{8} (\partial_{\mu} \phi - 2a_{\mu}) \Pi^{\mu\nu}_{\rm BdG} (\partial_{\nu} \phi - 2a_{\nu})  \,,
\end{equation}
where a Maxwell term for $a$ has been added to regularize the theory. Our goal is to determine the monopole correlation functions of this theory. To that end, we begin by defining a more general class of $U(1)$ gauge theories 
\begin{equation}
    L = \frac{1}{2} (\partial \phi - 2a) D(\partial \phi - 2a) + \frac{i\kappa}{2} a E a + \frac{1}{2e^2} (E a)^2 \,, \quad E_{\mu\nu} = \epsilon_{\mu\lambda\nu} \partial_{\lambda} \,, 
\end{equation}
where all spacetime indices have been omitted to avoid notational clutter. Results of interest to us can be recovered by taking $D^{\mu\nu} = \Pi^{\mu\nu}_{\rm BdG}/4$, $\kappa = 1/4\pi$. Since the Lagrangian is quadratic in $\phi$, it is tempting to integrate out $\phi$ and obtain an effective Lagrangian for $a$. While such a procedure would give the correct density and current correlators (as in SM Sec.~\ref{app:LGtheory}), it is not sufficient for monopole correlation functions which are sensitive to the compactness of $\phi$. For that reason, we will now turn to a dual BF description.

The first step in our derivation is to decouple the phase Lagrangian using a Hubbard-Stratanovich field $C_{\mu}$ 
\begin{equation}\label{eq:spectral_lag_nomono}
    L[C, a] = \frac{1}{2} C D^{-1} C + i C (\partial \phi - 2a) + \frac{i\kappa}{2} a E a + \frac{1}{2e^2} (E a)^2 \,.
\end{equation}
Integrating over the $2\pi$-periodic compact scalar $\phi$ enforces a constraint $\partial_{\mu} C_{\mu} = 0$. This means that we can make a change of variables $C_{\mu} = \frac{1}{2\pi} \epsilon_{\mu\nu\lambda} \partial_{\nu} b_{\lambda}$, where $b$ is a new $U(1)$ gauge field with a redundancy $b_{\mu} \rightarrow b_{\mu} + \partial_{\mu} \alpha$. In terms of this new $U(1)$ gauge field, we can rewrite the Lagrangian as
\begin{equation}
    L[b,a] = \frac{1}{2(2\pi)^2} b E D^{-1} E b  - \frac{2i}{2\pi} b E a + \frac{i\kappa}{2} a E a + \frac{1}{2e^2} (Ea)^2 \,.
\end{equation}
Now let us introduce monopoles. Following Dirac's procedure, a $2\pi n$-monopole at spacetime location $x_m$ corresponds to a singular gauge field configuration in which 
\begin{equation}
    \epsilon_{\mu\nu\lambda}\partial_{\nu} a_{\lambda} = B_{\mu} - \tilde J_{\mu} \,, 
\end{equation}
where $B_{\mu}$ is the smooth magnetic field generated by the monopole and $\tilde J$ is a Dirac string that satisfies $\partial_{\mu} \tilde J^{\mu} = 2\pi n \delta^3(x-x_m)$. For a monopole-antimonopole pair at locations $x_1$ and $x_2$, we modify the string such that $\partial_{\mu} \tilde J^{\mu} = 2\pi n \delta^3(x-x_1) - 2\pi n \delta^3(x - x_2)$. Note that the Dirac string is chosen so that $\partial_{\mu} \epsilon_{\mu\nu\lambda} \partial_{\nu} a_{\lambda} = 0$, which is consistent with $a$ being a well-defined connection on a $U(1)$ principal bundle. 

When bare monopoles are gauge-invariant, the correct Lagrangian in the presence of monopoles can be obtained by replacing $Ea$ with $Ea + \tilde J$ everywhere in \eqref{eq:spectral_lag_nomono}. However, from the Chern-Simons terms, we see that a $2\pi n$-monopole of $a$ is not gauge invariant, but rather carries charge $2\pi \kappa n$ under $a$ and charge $-2n$ under $b$. Therefore, upon introducing the monopole current $\tilde J$, we should also include a Wilson line for $a$ with current $J_a = -\kappa \tilde J$ and a Wilson line for $b$ with current $J_b = 2/(2\pi) \tilde J$ to cancel the gauge charge. At this point we recall that for our case of interest $\kappa = 1/4\pi$, $n=2$, and $n\kappa=1/2\pi$, which precisely corresponds to the Wilson line $\mathcal{W}^\dagger_a[\gamma]$ that we inserted in Eq.~\eqref{eq:correlator_monopoles_general}. After adding these appropriate Wilson lines and replacing $Ea$ with $Ea + \tilde J$, we land on the correct monopole Lagrangian
\begin{equation}\label{eq:spectral_lag_mono}
    \begin{aligned}
    L[b, a, \tilde J] &= \frac{1}{2(2\pi)^2} b E D^{-1} E b - \frac{2i}{2\pi} b (E a + \tilde J) + i b J_b + \frac{1}{2e^2} (Ea + \tilde J)^2 + \frac{i\kappa}{2} a Ea + i \kappa a \tilde J + i a J_a\\
    &= \frac{1}{2(2\pi)^2} b E D^{-1} E b - \frac{2i}{2\pi} b E a + \frac{1}{2e^2} (Ea + \tilde J_T)^2 + \frac{1}{2e^2} |\tilde J_L|^2 + \frac{i\kappa}{2} a Ea \,,
    \end{aligned}
\end{equation}
where in the last line, we decomposed the monopole current $\tilde J$ into transverse $\tilde J_T$ and longitudinal $\tilde J_L$ parts which are orthogonal to each other. 
At this stage, we see that $b$ only couples to the transverse part of $a$. Thus, we can integrate out $b$ safely and obtain a Lagrangian entirely in terms of $a$ and $\tilde J$ (using the identity $E^2 = q^2 P_T$)
\begin{equation}
    L[a, \tilde J] = \frac{1}{2} a (4P_TDP_T) a + \frac{1}{2e^2} (Ea)^2 + \frac{i\kappa}{2} a Ea + \frac{1}{e^2} \tilde J_T Ea + \frac{1}{2e^2} |\tilde J|^2  \,.
\end{equation}
Finally, now that $\tilde J$ only couples to the transverse part of $a$, we can integrate out $a$ to obtain an effective Lagrangian for $\tilde J$ (from now on, we work in momentum space so that $\tilde J^*$ refers to $\tilde J(q)^*$)
\begin{equation}
    L[\tilde J] = \frac{1}{2e^2} |\tilde J|^2 - \frac{1}{2e^4} \tilde J_T^* E \left[q^2/e^2 P_T + 4D_T +  i\kappa E\right]^{-1} E \tilde J_T \,,
\end{equation}
where $P_{T,\mu\nu} = \delta_{\mu\nu} - q_{\mu} q_{\nu}/q^2$ is the transverse projector and $D_T = P_T D P_T$. Since the operator inside the inverse is entirely transverse, we can use a general inversion formula valid in the transverse subspace 
\begin{equation}
    K^{-1} = \frac{P_T \Tr K - K}{\det' K} \,,\label{eq:fency_inversion_formula}
\end{equation}
where $\det'$ is taken over the transverse subspace (i.e. $\det' K = [(\Tr K)^2-\Tr K^2]/2$ is a product of all non-zero eigenvalues of $K$). The inversion formula in Eq.~\eqref{eq:fency_inversion_formula} can also be derived by introducing a regulator term $\varepsilon P_L$, which renders $K$ invertible in the conventional sense. After computing the inverse using the Cayley–Hamilton formula and acting with $P_T$ on both sides, we take the limit $\varepsilon \rightarrow 0$, recovering Eq.~\eqref{eq:fency_inversion_formula}. Applying this formula to $K = q^2/ e^2 P_T + 4D_T + i\kappa E$, we find 
\begin{equation}
    K^{-1} = \frac{e^4\left[\left(\frac{q^2}{e^2} + 4  \Tr D_T - 4D_T\right) P_T - i \kappa E\right]}{q^4+ 4 e^2 q^2 \Tr D_T+16 e^4 \det ' D_T +e^4\kappa \left(\kappa q^2 -4i\Tr[D_T E]\right)} \,. 
\end{equation}

Plugging this back into the monopole Lagrangian gives
\begin{equation}
    \begin{aligned}
    L[\tilde J] &= \frac{1}{2e^2} |\tilde J|^2 - \frac{1}{2e^2} \tilde J_T^* \frac{q^2 \left(q^2 + 4 e^2 \Tr D_T\right) P_T - 4 e^2 ED_TE - i \kappa e^2 q^2 E}{q^4+ 4 e^2 q^2 \Tr D_T+16 e^4 \det ' D_T +e^4\kappa^2 q^2 -4ie^4\kappa\Tr[D_T E]}  \tilde J_T \\
    &= \frac{1}{2e^2} |\tilde J|^2 - \frac{1}{2e^2} \tilde J_T^* \frac{q^2 + 4 e^2 \Tr D_T - 4 e^2/q^2 ED_TE}{q^2+ 4 e^2 \Tr D_T+16 e^4/q^2 \det ' D_T +e^4\kappa^2  -4ie^4\kappa/q^2\Tr[D_T E]}  \tilde J_T \,,
    \end{aligned}
\end{equation}
where we used the identity $E^2 = q^2 P_T$ and the fact that $\tilde J^*_T E \tilde J_T = 0$ (we assume that $J$ is real) to simplify various terms. Note that $D_T$ is generally not a symmetric tensor and $\Tr ED_T$ need not vanish. 

Finally, to make the physics more transparent, we can decompose the Lagrangian into a term that depends on the full string $\tilde J$ and another term depending only on $\partial_{\mu} \tilde J^{\mu}$ which encodes interactions between the string endpoints (i.e. monopoles) 
\begin{equation}\label{eq:mono_general}
    L[\tilde J] = L_{\rm string}[\tilde J] + L_{\rm end}[\partial_{\mu} \tilde J^{\mu}] \,,
\end{equation}
with
\begin{equation}
    L_{\rm string}[\tilde J] = \frac{1}{2e^2} \tilde J^* \frac{16 e^4/q^2 \det' D_T + 4 e^2/q^2 ED_TE +e^4\kappa^2  -4ie^4\kappa/q^2\Tr[D_T E]}{q^2 + 4e^2 \Tr D_T + 16 e^4/q^2 \det' D_T + e^4\kappa^2  -4ie^4\kappa/q^2\Tr[D_T E]}  \tilde J \,,
\end{equation}
\begin{equation}
    L_{\rm end}[\partial_{\mu} \tilde J^{\mu}] = \frac{1}{2e^2} \frac{|\partial_{\mu} \tilde J^{\mu}|^2}{q^2} \frac{q^2 + 4e^2 \Tr D_T}{q^2 + 4e^2 \Tr D_T + 16 e^4/q^2 \det' D_T +e^4\kappa^2  -4ie^4\kappa/q^2\Tr[D_T E]} 
\end{equation}
As a sanity check, let us match this formula with existing results in the literature:
\begin{enumerate}
    \item If $D = 0$, the formulae simplify to
    \begin{equation}
        L_{\rm string}[\tilde J] = \frac{1}{2e^2} \tilde J^* \frac{\kappa^2 e^4}{q^2 + \kappa^2 e^4} \tilde J \,, \quad L_{\rm end}[\partial_{\mu} \tilde J^{\mu}] = \frac{1}{2e^2} \frac{|\partial_{\mu} \tilde J^{\mu}|^2}{q^2 + \kappa^2 e^4} \,. 
    \end{equation}
    The string term captures the UV logarithmically divergent string tension and the endpoint term describes an exponentially screened interaction between monopoles mediated by the massive photon. This answer is in exact agreement with Ref.~\cite{Diamantini1993_CS_monopole}. 
    \item If $\kappa = 0$ and $4D_T = m P_T$ is the superfluid stiffness of the relativistic Abelian Higgs model, then $4 \Tr D_T = 2 m$ and $16 \det D_T = m^2$. Thus, the formulae simplify to
    \begin{equation}
        L_{\rm string}[\tilde J] = \frac{1}{2e^2} \tilde J^* \frac{m e^2}{q^2 + m e^2} \tilde J \,, \quad L_{\rm end}[\partial_{\mu} \tilde J^{\mu}] = \frac{1}{2e^2} \frac{|\partial_{\mu} \tilde J^{\mu}|^2}{q^2 + m e^2} \,. 
    \end{equation}
    These results are again in exact agreement with Ref.~\cite{Einhorn1978}. 
\end{enumerate}
Given these sanity checks, we can apply the general formula \eqref{eq:mono_general} to the CBFL phase, where $4D = \Pi_{\rm BdG}$. In the static limit, $\Pi_{\rm BdG,\mu\nu} $ is a diagonal matrix with entries are $\tilde{\chi}_0$ for $\mu=\nu=0$ and $\rho_s$ for $\mu=\nu=1,2$  (cf. Eq.~\eqref{eq:Pi_bdG_full}). In this case
\begin{equation}
    \operatorname{Tr}D_T=\frac{1}{4q^2} \left[2\rho_s q_0^2+(\rho_s+\tilde{\chi}_0)|\bm{q}|^2\right],\quad     \operatorname{det}'D_T=\frac{\rho_s}{16q^2} \left(\rho_s q_0^2+\tilde{\chi}_0|\bm{q}|^2\right),\quad (ED_TE)_{00} = \rho_s |\bs{q}|^2/4 \,. 
\end{equation}
At large $|\bm{q}|$ these expressions simplify to $\operatorname{Tr}D_T\approx (\rho_s+\tilde{\chi}_0)/4$,  $\operatorname{det}'D_T\approx \rho_s\tilde{\chi}_0/16$, and $(ED_TE)_{00}\sim \rho_s |\bs{q}|^2/4$. \color{black}If we use this approximation and choose a temporally separated monopole-antimonopole pair
\begin{equation}
    \tilde J_{\mu}(x) = 2 \pi n \delta_{\mu 0} [\theta(x_0 - \tau/2) - \theta(x_0 + \tau/2)] \delta^2(\bs{x}) \,, \quad \tilde J_{\mu}(q) = 2\pi n\delta_{\mu0} \frac{-2 \sin [q_0\tau/2]}{q_0} \,, 
\end{equation}
then the monopole action (obtained from integrating the monopole Lagrangian $L[\tilde J]$ over spacetime) takes the form
\begin{equation}
    \begin{aligned}
    S[\tilde J] &\equiv - \ln \ev{\mathcal{M}_a^{\dagger,n} \mathcal{M}_a^n}(0, \tau) \\
    &= \frac{(2\pi n)^2}{2} \int \frac{d^3 q}{(2\pi)^3} \frac{ e^2 \left[\kappa ^2 q^4+\rho_s \left(\tilde{\chi}_0|\bm{q}|^2  +\rho_s q_0^2 \right)\right] + \rho_s e^2 |\bs{q}|^2 q^2}{e^4 \kappa ^2 q^4+e^4 \rho_s\left(\tilde{\chi}_0|\bm{q}|^2  +\rho_sq_0^2 \right)+e^2 q^2 \left[|\bm{q}|^2 (\rho_s+\tilde{\chi}_0)+2\rho_s q_0^2 \right]+q^6} \frac{4 \sin [q_0 \tau/2]^2}{q_0^2} \\
    &+ \frac{(2\pi n)^2}{2e^2} \int \frac{d^3 q}{(2\pi)^3} 4 \sin [q_0 \tau/2]^2 \frac{e^2 \left[|\bm{q}|^2 (\rho_s+\tilde{\chi}_0)+2 \rho_s q_0^2 \right]+q^4}{e^4 \kappa ^2 q^4+e^4 \rho_s \left(\tilde{\chi}_0|\bm{q}|^2  + \rho_s q_0^2\right)+e^2 q^2 \left[|\bm{q}|^2 (\rho_s+\tilde{\chi}_0 )+2 \rho_s q_0^2 \right]+q^6}  \,.
    \end{aligned}
\end{equation}
\color{black}
The second term gives a contribution which decays exponentially with $|\tau|$ and can be neglected at large $\tau$ (formally, there is also a $|\tau|$-independent UV divergent contribution from a monopole self-interaction which can be ignored due to proper normalization of the monopole correlation function). As for the first term, we find that the integral over $q_0$ produces a linear-in-$|\tau|$ dependence, whereas the integral over spatial momenta contributes a logarithmically divergent prefactor. This is the familiar divergent string tension that we have seen in Maxwell Chern-Simons and Abelian Higgs models~\cite{Einhorn1978,Diamantini1993_CS_monopole}. Specifically, focusing on the large $|\bs{q}|$ behavior of the integrand, we can easily extract the leading logarithm as
\begin{equation}
    \begin{aligned}
    - \ln \ev{\mathcal{M}_a^{\dagger,n} \mathcal{M}_a^n}(0, \tau) &\approx \frac{(2\pi n)^2}{2e^2 (2\pi)^2} \int d q_0 \frac{4 \sin [q_0 \tau/2]^2}{q_0^2} \int^{\Lambda_{\rm UV}} q dq \frac{\rho_s e^2 + \kappa^2 e^4}{q^2} \\
    &= \frac{(2\pi n)^2 (\rho_s e^2 + \kappa^2 e^4)}{e^2 (2\pi)^2} \int d q_0 \frac{\sin [q_0 \tau/2]^2}{q_0^2} \log \left[\frac{\Lambda^2_{\rm UV}}{\rho_s e^2 + \kappa^2 e^4}\right] \,. 
    \end{aligned}
\end{equation}
Here $\Lambda_{\rm UV}$ is associated with the short-distance regularization of the Dirac string, which in the Higgs case is related to the coherence length. 
Further integrating over $q_0$, we indeed obtain a $|\tau|$-linear monopole action 
\begin{equation}\label{eq:mono_CBFL_final}
    - \ln \ev{\mathcal{M}_a^{\dagger,n} \mathcal{M}_a^n}(0, \tau) \approx  \frac{n^2 \pi}{2} (\rho_s + \kappa^2 e^2) \log \left[\frac{\Lambda^2_{\rm UV}}{\rho_s e^2 + \kappa^2 e^4}\right] |\tau| \,. 
\end{equation}
Taking $n = 2$ and $\kappa = 1/(4\pi)$, we recover a monopole correlation function which decays exponentially in $|\tau|$.  

Finally, let us reintroduce the Landau damping terms. After the same rescaling of spacetime coordinates, we can put $\Pi_{\rm BdG}$ into the form 
\begin{equation}
    4D_{T,\mu\nu} = (P_T \Pi_{\rm BdG} P_T)_{\mu\nu} = m P_{T,\mu\nu} + \Gamma^{\mu\nu} \frac{|q_0|}{|\bs{q}|} \,. 
\end{equation}
Plugging this form into \eqref{eq:mono_general} gives the exact monopole correlation function. While the full integral is tricky, we observe that the leading string tension derived in \eqref{eq:mono_CBFL_final} only depends on the leading logarithmic divergence in the integral over spatial momenta $|\bs{q}|$. In this limit, the Landau damping is suppressed by an additional factor of $1/|\bs{q}|$, which means that the corrections due to Landau damping do not have a UV divergence in $|\bs{q}|$ and do not contribute to the string tension.  

\subsection{Thermodynamics of the CBFL}\label{subapp:thermodynamics}

In this section, we compute the free energy of the CBFL state at nonzero temperature. In SM Sec.~\ref{subapp:LG_withBFS}, we went from $S_{\rm BFL}[f, \phi, a]$ in \eqref{eq:IRaction_app_withfermion} to $S_{\rm eff}[\phi, 2a]$ in \eqref{eq:IRaction_app} by integrating out the fermion fluctuations and dropping terms independent of $\phi$. This procedure is legitimate for computing correlation functions but not for thermodynamics because the free energy receives contributions both from the BFS and from the gauge field sector. The former contribution is manifested as an extra fermion determinant in the partition function
\begin{equation}
    Z = \int D \phi\, D a \exp{- \frac{1}{2} \int_{\tau,\bm{r}} a_{\alpha} (\Pi^{\alpha\beta}_{\rm BFL}+\Pi^{\alpha\beta}_{a}) a_{\beta}} \mathrm{Pf}[\partial_{\tau} + \hat{\mathcal{H}}_{\rm BdG}] \,, \label{eq:Z}
\end{equation}
where $\Pi^{\alpha\beta}_{\rm BFL}$ and $\Pi^{\alpha\beta}_{a}$ are given in Eq.~\eqref{eq:CBFL_response_function}. The Pfaffian is related to the fermion path integral
\begin{equation}
    \mathrm{Pf}[\partial_{\tau} + \hat{\mathcal{H}}_{\rm BdG}] = \int D f \, \exp{- \frac{1}{2} \int_{\tau, \bs{r}} \bar \Psi [\partial_{\tau} + \hat{\mathcal{H}}_{\rm BdG}] \Psi} \,. 
\end{equation}
By diagonalizing the BdG Hamiltonian, we know that the Pfaffian factorizes into a product of contributions from different Bogoliubov Fermi pockets. At the quadratic level, each pocket is simply a free Fermi gas. The free energy is therefore 
\begin{equation}
    F_{\rm BFL}(T) = E_{\rm BFL} - T\sum\limits_{s=\pm}\int \frac{d^2\bm{k}}{(2\pi)^2}\ln\left(1+e^{-\beta |E_{s}(\bm{k})|}\right) \approx E_{\rm BFL} - \frac{\pi^2}{6} T^2\sum\limits_{s=\pm} \int\frac{d^2\bm{k}}{(2\pi)^2} \delta(E_{s}(\bm{k})) \,, 
\end{equation}
The remaining momentum integral corresponds to the thermodynamic density of states at the Bogoliubov Fermi surfaces. Note that this quantity does not contain any coherence factors, and thus, it is distinct from $\chi_0$ which appeared in the calculation of the density-density correlator in Eq.~\eqref{eq:def-Gamma}. Beyond the quadratic level, there is no singular infrared contributions to the entropy beyond RPA since the dynamical gauge field is Higgsed. Therefore, the temperature scaling of the specific heat computed from the quadratic approximation will be robust. Taking a derivative with respect to $T$, we obtain the BFS contribution to the entropy
\begin{equation}
    S_{\rm BFL}(T) = - \frac{\partial F_{\rm BFL}}{\partial T} = \frac{\pi^2 T}{3} \sum\limits_{s=\pm} \int\frac{d^2\bm{k}}{(2\pi)^2} \delta(E_{s}(\bm{k}))\,. 
\end{equation}

As for the gauge field sector, upon integrating out $a$ in Eq.~\eqref{eq:Z}, we obtain the following contribution to the free energy
\begin{equation}
    F_{\rm gauge}(T) = \frac{T}{2} \int_{\bs{q}} \sum_{\omega_n} \ln \det \left[\Pi_{\rm BFL}(\bs{q}, i\omega_n)+\Pi_{a}(\bs{q}, i\omega_n)\right] = \int_{\omega, \bs{q}} \frac{1}{e^{\beta \omega} - 1} \tan^{-1} \left(\frac{\Im \det \left[\Pi_{\rm BFL}^R(\bs{q}, \omega)+\Pi_{a}^R(\bs{q}, \omega)\right]}{\Re \det \left[\Pi_{\rm BFL}^R(\bs{q}, \omega)+\Pi_{a}^R(\bs{q}, \omega)\right]} \right) . 
\end{equation}
The determinant can be evaluated using Eq.~\eqref{eq:Pi_bdG_full} and Eq.~\eqref{eq:CBFL_response_function}.  Combining everything together, we infer that 
\begin{equation}
\begin{aligned}
    \Im \det [\Pi^R_{\rm BFL}(\bs{q}, \omega)+ \Pi^R_{a}(\bs{q}, \omega)] &\approx \Im\Big[\tilde{\chi}_0 \Gamma^{TT}_{\hat{\bm{q}}} - 4 \rho_s \Gamma_{\hat{\bm{q}}}\Big] \omega/q \,, \\
    \Re \det [\Pi^R_{\rm BFL}(\bs{q}, \omega)+ \Pi^R_{a}(\bs{q}, \omega)] &\approx 4 \tilde{\chi}_0 \rho_s +\frac{q^2}{(4\pi)^2}\left(1+\tilde{\chi}_0 v_0\right) + \mathcal{O}(\omega/q) \,,
    \end{aligned}
\end{equation}
where we recall that $\rho_s = (k_{\rm diag}-\tilde{\chi}_J)/4$ is the superfluid stiffness of the BFS superconductor. It turns out that the second term in Eq.~\eqref{eq:CBFL_response_function} contributes only to the sub-leading corrections. Given these approximations, we can obtain a simplified formula for $F_{\rm gauge}(T)$
\begin{equation}
    F_{\rm gauge}(T) = \int_{-\infty}^{\infty} \frac{d \omega}{2\pi} \int\frac{d\phi_{\bm{q}}}{2\pi}\int_0^{\infty} \frac{q dq}{2\pi} \frac{1}{e^{\beta \omega} - 1} \tan^{-1} \left(\frac{Z_{\hat{\bm{q}}} \omega}{q(1+q^2/\Lambda^2)}\right) \,, \quad Z_{\hat{\bm{q}}} \equiv \frac{\tilde{\chi}_0 \Im\Gamma^{TT}_{\hat{\bm{q}}} - 4 \rho_s \Im\Gamma_{\hat{\bm{q}}}}{4\tilde{\chi}_0 \rho_s} \,. 
\end{equation}
Here $\Lambda^2= 4^3  \pi^2 \tilde{\chi}_0 \rho_s/(1+\tilde{\chi}_0 v_0)$. We recall that both $\tilde \chi_0 \Im \Gamma^{TT}_{\hat{\bs{q}}}$ and $\rho_s \Im \Gamma_{\hat{\bs{q}}}$ are negative. So the sign of $Z_{\hat{\bm{q}}}$ is not fixed a priori. However, for small $\rho_s$, we know that the first term dominates over the second term and $Z_{\hat{\bs{q}}}$ is negative. Since a positive $Z_{\hat{\bs{q}}}$ signals an instability in the gauge sector, we conjecture that the sign change in $Z_{\hat{\bs{q}}}$ arises only when the BFS disappears across a quantum phase transition. It would be interesting to check this conjecture through a more detailed calculation in the future.

The entropy can be extracted by taking a temperature derivative of $F_{\rm gauge}(T)$
\begin{equation}
    S_{\rm gauge}(T) = - \frac{\partial F_{\rm gauge}(T)}{\partial T} = \int_{-\infty}^{\infty} \frac{\omega d \omega}{2\pi} \int\frac{d\phi_{\bm{q}}}{2\pi} \int_0^{\infty} \frac{q dq}{2\pi} \frac{e^{\beta \omega}}{(e^{\beta \omega}-1)^2 T^2} \tan^{-1} \left(\frac{|Z_{\hat{\bm{q}}}| \omega}{q(1+q^2/\Lambda^2)}\right) \,. 
\end{equation}
The leading temperature dependence of this integral originates from the small $q$ regime so we can neglect the $q^2/\Lambda^2$ term in the denominator and instead impose a large-momentum cutoff $\Lambda$. We note that in principle, there is also another natural UV cutoff associated with the fact that the Landau damping form of $\Pi_{\rm BFL}$ that we used only holds for $q$ much smaller than the linear size of the BFS. We expect that for small $\Delta$, the scale $\Lambda$ is smaller because it is proportional to the superfluid stiffness, while the size of the Bogoliubov pockets is  the largest at small $\Delta$. The integral over $q$ then can be approximated as
\begin{equation}
    \int_0^{\Lambda} \frac{q dq}{2\pi} \tan^{-1} \left(\frac{|Z_{\hat{\bm{q}}}| \omega}{q}\right) \approx \frac{\Lambda |Z_{\hat{\bm{q}}}|}{2\pi} \omega + \mathcal{O}(\omega^2) \,. 
\end{equation}
This means that
\begin{equation}
    S_{\rm gauge}(T) \approx \frac{\Lambda}{T^2} \int \frac{d\phi_{\bm{q}}}{2\pi}  |Z_{\hat{\bm{q}}}|\int_{-\infty}^{\infty} \frac{\omega^2 d \omega}{(2\pi)^2} \frac{e^{\beta \omega}}{(e^{\beta \omega}-1)^2} \,. 
\end{equation}
After defining a dimensionless variable $x = \beta \omega$, we can rewrite the integral as
\begin{equation}
      S_{\rm gauge}(T) \approx \frac{\Lambda T}{(2\pi)^2} \int \frac{d\phi_{\bm{q}}}{2\pi}  |Z_{\hat{\bm{q}}}|\int_{-\infty}^{\infty} x^2 dx \, \frac{e^x}{(e^{x}-1)^2} = \frac{\Lambda T}{6}   \int \frac{d\phi_{\bm{q}}}{2\pi}  |Z_{\hat{\bm{q}}}|\,. 
\end{equation}
Comparing the scaling form of $S_{\rm gauge}(T)$ and $S_{\rm BFL}(T)$, we see that both terms contribute a $T$-linear specific heat at low temperature. 

\section{Ground state degeneracy at half-filling}\label{app:degeneracy}
In this section, we elaborate on the torus ground state degeneracy (GSD) of the ACFL and the Pfaffian state. 

By a simple change of variables $b = A-a$, we can rewrite the ACFL Lagrangian in \eqref{eq:ACFL} as
\begin{equation}
    L_{\rm ACFL} = L[f,A-b] - \frac{2i}{4\pi} \alpha d \alpha + \frac{i}{2\pi} \alpha d b \,. 
\end{equation}
Even when $f$ is gapped, integrating over $b$ sets $\alpha = 0$ and gives a trivial gapped insulator. A gapless Fermi surface of $f$ cannot induce an extra topological order. Therefore, the ACFL has a unique ground state on a spatial torus. Note that, in contrast, the CFL in a half-filled Landau level has a two-fold torus GSD protected by the continuum magnetic translation symmetry~\cite{Geraedts2016_DiracCFL,Dong2023_ACFL}. This degeneracy is lifted by an arbitrarily weak periodic potential and is not associated with any intrinsic topological order in the system.

When the CFs pair up into a fully gapped $p+ip$ superconductor, the resulting electronic phase has Pfaffian topological order
\begin{equation}
    \mathrm{Pfaffian} = \frac{\mathrm{Ising} \times U(1)_8}{\mathbb{Z}_2} \,,
\end{equation}
where the Ising sector contains three anyons $\{1, \sigma, \psi\}$ and the $U(1)_8$ sector is generated by a single Abelian anyon $a$ with $a^8 = 1$. The $\mathbb{Z}_2$ quotient imposes the restriction that the $\{1, \psi\}$ anyons in the Ising sector must be bound to an even power of $a$, while the non-Abelian $\sigma$ anyon in the Ising sector must be bound to an odd power of $a$~\cite{Moore1991_nonAbelian,Read1999_pair}. Additionally, $\psi a^4$ is identified with the microscopic electron, which is a local excitation. Therefore, the full anyon set is
\begin{equation}
    \mathrm{Pfaffian} = \{a^{2n}, \sigma a^{2n+1}, \psi a^{2n}\} \,, \quad n \in \{0,1,2,3\} \,.
\end{equation}
When the system is placed on a spatial torus, this anyon content implies a six-fold ground state degeneracy, with finite-size splittings that decay exponentially with the system size.

\section{Emergent 1-form symmetry in the CBFL}\label{app:topology}

In this section, we give a more formal derivation of the torus ground state degeneracy of the CBFL state from the perspective of generalized symmetries~\cite{Gaiotto2015_gensym}. We begin with the effective Euclidean Lagrangian for the CBFL phase presented in the main text
\begin{equation}
    L_{\rm CBFL} = L_{\rm BFL}[\psi, A-b] + \frac{2i}{2\pi} \beta d (A-b) - \frac{2i}{4\pi} \alpha d \alpha + \frac{i}{2\pi} \alpha d b \,. 
\end{equation}
Through a shift of variables $\alpha \rightarrow \alpha + 2 \beta$, we can rewrite the Lagrangian in a more convenient form
\begin{equation}
    L_{\rm BFL} = L[\psi,A-b] - \frac{2i}{4\pi} \alpha d \alpha - \frac{4i}{2\pi} \alpha d \beta - \frac{8i}{4\pi} \beta d \beta + \frac{i}{2\pi} \alpha d b + \frac{2i}{2\pi} \beta d A \,.
\end{equation}
When the term involving $\psi$ is neglected, this Lagrangian enjoys an emergent $\mathbb{Z}_8$ 1-form symmetry generated by
\begin{equation}
    X: b \rightarrow b + \lambda/2 \,, \quad \beta \rightarrow \beta + \lambda/8 \,,
\end{equation}
where $\lambda$ is a flat 1-form. However, in the presence of a gapless $\psi$ sector, the shift $b \rightarrow b + \lambda/2$ no longer leaves $L[\psi, A-b]$ invariant. The surviving 1-form symmetry is therefore a $\mathbb{Z}_4$ subgroup generated by 
\begin{equation}
    \tilde X = X^2: \beta \rightarrow \beta + \frac{\lambda}{4} \,. 
\end{equation}
We can couple the system to a background $\mathbb{Z}_4$ 2-form gauge field $B$ by substituting $d \beta$ by $d \beta - B/4$ with the properly quantized 2-form field $B$ transforming as $B \rightarrow B + d \lambda$. The resulting Lagrangian is
\begin{equation}
    L_{\rm BFL}[B] = L[\psi,A-b] - \frac{2i}{4\pi} \alpha d \alpha - \frac{4i}{2\pi} \alpha d \beta - \frac{8i}{4\pi} \beta d \beta + \frac{i}{2\pi} \alpha B + \frac{2i}{2\pi} \beta B + \frac{i}{2\pi} \alpha d b - \frac{i}{4\pi} A B \,.
\end{equation}
Under a $\mathbb{Z}_4$ gauge transformation $\beta \rightarrow \beta + \lambda/4$, the mixed Chern-Simons term between $\beta$ and $B$ shifts by
\begin{equation}
    \Delta L = \frac{i}{4\pi} \lambda \, B \,. 
\end{equation}
This final equation implies a self-anomaly of the $\mathbb{Z}_4$ 1-form symmetry with coefficient $k = 2$. If we put the theory on a torus, the Wilson loops of $\beta$ along the two noncontractable cycles (labeled $x$ and $y$) fail to commute
\begin{equation}
    W^x_{\beta} W_{\beta}^y = W_{\beta}^y W_{\beta}^x e^{2\pi i k/4} = - W_{\beta}^y W_{\beta}^x \,. 
\end{equation}
This projective commutation relation between $W^x_{\beta}$ and $W^y_{\beta}$ enforces a 2-fold ground state degeneracy (with an exponentially suppressed finite-size energy splitting) which contrasts with both the fully gapped quantum Hall state and the ACFL state. Therefore, the CBFL state, despite having an infinite number of gapless modes, remains topologically nontrivial.

\section{Preliminary results on daughter states of the CBFL}\label{app:daughter}

In this section, we present some preliminary results on the effect of doping/additional external magnetic field on the CBFL phase realized in a half-filled Chern band. Precisely at $\nu = 1/2$, the CBFL phase is a paired state of CFs with a gapless Bogoliubov Fermi surface. Upon doping away from $\nu = 1/2$/turning on an external magnetic field $B$, the CFs start to see a small effective magnetic field $\delta B$. While the CF superconductor likely forms a vortex lattice at small $\delta B$, the vortex lattice may melt into translation-invariant neighboring FQH states at larger values of $\delta B$. The sequence of such FQH states are usually referred to as the daughter states of the parent phase at $\nu = 1/2$. 

If we started with an ACFL at $\nu = 1/2$, the natural daughter states would be fermionic IQH states in which the CFs fill $p$ Landau levels defined by the effective magnetic field $\delta B$ they see. These states are the famous Jain states that occur at electronic filling fractions $\nu = p/(2p+1)$ for $p \in \mathbb{Z}$. If we started with the Pfaffian state at $\nu = 1/2$, the natural daughter states would be the Levin-Halperin sequence~\cite{Levin2009_daughter}. The two members in this sequence closest to $\nu = 1/2$ occur at $\nu = 8/17$ and $\nu = 7/13$ respectively. Note that although these filling fractions also fit into the Jain sequence, they differ from the Jain states by a stacked $\mathrm{E}_8$ bosonic phase which has chiral central charge $c_- = 8$ but no Hall conductance~\cite{Kitaev2005_anyons,Lu2012_bosoninvertible}. 

Now let us study the CBFL. Following the main text, we conjecture that a moderate value of $\delta B$ destroys the phase coherence in the CF superconductor while preserving the pairing gap for a fraction $\epsilon$ of CFs in the parent ACFL. This means that the effective Lagrangian can be written as
\begin{equation}
    L = L[\Phi, 2a] + L[f, a] - \frac{2i}{4\pi} \alpha d \alpha + \frac{i}{2\pi} \alpha d (A-a) \,, \quad \rho_f = (1-\epsilon) \rho_c \,, \quad \rho_{\Phi} = \frac{\epsilon}{2} \rho_c \,. 
\end{equation}
Here, $f$ denotes the unpaired CFs while $\Phi \sim f^2$ denotes the paired CFs, which see twice the effective magnetic field but has half the density. Upon doping to $\rho_c = 1/2 + \delta$, the internal gauge flux must adjust itself so that 
\begin{equation}
    \frac{1}{2} + \delta = \frac{1}{2} \cdot \frac{1}{2\pi} \nabla \times (\bs{A} - \bs{a}) \quad \rightarrow \quad \frac{1}{2\pi} \nabla \times \bs{a} = -2 \delta \,. 
\end{equation}
This means that the effective Landau level filling of $\Phi$ and $f$ are 
\begin{equation}
    \nu_f = \frac{2\pi \rho_f}{\nabla \times \bs{a}} = - \frac{(1/2 + \delta)(1-\epsilon)}{2 \delta} \,, \quad \nu_{\Phi} = \frac{2\pi \rho_{\Phi}}{2\nabla \times \bs{a}} = - \frac{\epsilon}{8 \delta} (\frac{1}{2} + \delta) \,. 
\end{equation}
For general $\delta, \epsilon$, both of these filling fractions are irrational and gapless phases/very complicated fractionalized phases are allowed by filling constraints. These states are likely very fragile. Much more natural are states in which $\epsilon$ adjusts itself so that $\nu_f$ is an integer and $\nu_{\Phi}$ is an even integer. If this occurs, $f$ and $\Phi$ can form gapped fermionic and bosonic IQH phases which have an enhanced stability. 

As a sanity check, let us verify some specific limits. When $\epsilon = 0$, the entire CF sector is unpaired and 
\begin{equation}
    \frac{1/2 + \delta}{2 \delta} = - p \quad \rightarrow \quad \delta = -\frac{1}{2(2p+1)} \,.
\end{equation}
These conditions are precisely satisfied by states in the Jain sequence $\nu = p/(2p+1)$. In the opposite limit $\epsilon = 1$, the solutions are
\begin{equation}
    - \frac{1/2 + \delta}{8\delta} = 2n \quad \rightarrow \quad \delta = -\frac{1}{2(16n+1)} \,. 
\end{equation}
These are precisely the daughter states of the strong-pairing $U(1)_8$ phase at $\nu = 1/2$~\cite{Yutushui2024_daughter}. 

For intermediate values of $\epsilon$, the most general integer solution $\nu_f = p$ and $\nu_{\Phi} = 2n$ gives a condition
\begin{equation}
    -p = \frac{(1/2+\delta)(1-\epsilon)}{2\delta} \,, \quad -2n = \frac{\epsilon(1/2+\delta)}{8\delta} \,. 
\end{equation}
Solving for $\delta, \epsilon$ simultaneously gives
\begin{equation}
    \delta = - \frac{1}{2(1+2p+16n)} \,, \quad \epsilon = \frac{8n}{8n+p} \,. 
\end{equation}
Therefore, in addition to the daughter states of ACFL and $U(1)_8$, we get new daughter states at more exotic filling fractions.\footnote{This family of states has previously appeared in Refs.~\cite{Yutushui2024_daughter,Zheltonozhskii2024_daughter} as tools for constructing daughter states of gapped phases at $\nu = 1/2$. However, these earlier works do not give a physical interpretation of these states for general $p, n$ as they do not naturally arise if the parent state is fully gapped.}  For the simplest choice of bosonic sector with $n = 1$, the allowed daughter states show up at a sequence of fractional $\epsilon$
\begin{equation}
    \delta = - \frac{1}{2(17+2p)} \,, \quad \epsilon = \frac{8}{8+p} \,. 
\end{equation}
For $p > 1$, we get a sequence of daughter states with decreasing paired fractions. This should be a reasonable approximation when the parent CBFL at $\nu = 1/2$ has small Bogoliubov Fermi pockets.

Now, let us assume that a particular physical system indeed realizes the scenario with some arbitrary $p, n$. Then the effective field theory of the daughter state takes the form
\begin{equation}
    L = \frac{p i}{4\pi} a da + 2p i \mathrm{CS}_g + \frac{n i}{4\pi} (2a) d (2a) - \frac{2i}{4\pi} \alpha d \alpha + \frac{i}{2\pi} \alpha d (A-a) = \frac{(p+8n)i}{4\pi} a da + 2p i \mathrm{CS}_g - \frac{2i}{4\pi} \alpha d \alpha + \frac{i}{2\pi} \alpha d (A-a) \,. 
\end{equation}
This TQFT is equivalent to the Jain state at $\nu = \frac{(p+8n)}{2(p+8n) + 1}$, up to stacking with $n$ copies of the $E_8$ state (invertible bosonic phase with $c_- = 8$).

\end{document}